\pdfoutput=1
\documentclass[11pt,twoside,a4paper,cmspaper,final,collab]{cms-tdr}
\def\svnVersion{ecbbf64}\def\svnDate{2025/11/16}\def\cmsCernNoTag{CERN-EP-2025-202}\def\cmsCernDate{\today}\def\cmsMessage{Published in Physical Review Letters as \href{http://dx.doi.org/10.1103/9nwb-splk}{\doi{10.1103/9nwb-splk}.}}
\begin{document}\cmsNoteHeader{HIG-24-018}

\ifthenelse{\boolean{cms@external}}{%
    \providecommand{\cmsLeft}{upper\xspace}
    \providecommand{\cmsRight}{lower\xspace}
    \providecommand{\cmsNarrowTable}[1]{{\scriptsize#1}}
}{
    \providecommand{\cmsLeft}{left\xspace}
    \providecommand{\cmsRight}{right\xspace}
    \providecommand{\cmsNarrowTable}[1]{#1}
}
\providecommand{\cmsTable}[1]{\resizebox{\textwidth}{!}{#1}}
\providecommand{\cmsWideTable}[1]{\resizebox{\textwidth}{!}{#1}}
\newlength\cmsTabSkip\setlength{\cmsTabSkip}{1ex}
\newcommand{\cmsNormalSep}{\hspace{\tabcolsep}\extracolsep{0pt}}
\newcommand{\cmsExtraSep}{\hspace{6\tabcolsep}\extracolsep{\fill}}

\newlength{\cmsFigWidth}
\ifthenelse{\boolean{cms@external}}{\setlength{\cmsFigWidth}{\columnwidth}}{\setlength{\cmsFigWidth}{0.6\textwidth}}

\ifthenelse{\boolean{cms@external}}{%
    \newcommand{\appname}{End Matter}
    \def\appendixname{\appname}
    \providecommand{\EndMatter}{ in the \appname\xspace}
    \providecommand{\suppMaterial}[1]{Supplemental Material~\cite{suppMatBib}\xspace}
}{
    \providecommand{\EndMatter}{ in the \nameref{endmatter}\xspace}
    \providecommand{\suppMaterial}[1]{Supplemental Material (Section~\ref{#1})\nocite{suppMatBib}\xspace}
}

\newcommand{\ttH}{\ensuremath{\ttbar\PH}\xspace}
\newcommand{\hcc}{\ensuremath{\PH{\to}\ccbar}\xspace}
\newcommand{\hbb}{\ensuremath{\PH{\to}\bbbar}\xspace}
\newcommand{\BR}{\ensuremath{\mathcal{B}}\xspace}
\newcommand{\brHcc}{\ensuremath{\BR(\PH{\to}\ccbar)}\xspace}
\newcommand{\brHbb}{\ensuremath{\BR(\PH{\to}\bbbar)}\xspace}
\newcommand{\sigmaBrHcc}{\ensuremath{\sigma(\ttbar\PH)\brHcc}\xspace}

\newcommand{\yc}{\ensuremath{y_{\PQc}}\xspace}
\newcommand{\kappaC}{\ensuremath{\kappa_{\PQc}}\xspace}
\newcommand{\kappaB}{\ensuremath{\kappa_{\PQb}}\xspace}

\newcommand{\ttZ}{\ensuremath{\ttbar\PZ}\xspace}
\newcommand{\ttW}{\ensuremath{\ttbar\PW}\xspace}
\newcommand{\tWZ}{\ensuremath{\PQt\PW\PZ}\xspace}
\newcommand{\VH}{\ensuremath{\PV\PH}\xspace}

\newcommand{\zcc}{\ensuremath{\PZ{\to}\ccbar}\xspace}
\newcommand{\zbb}{\ensuremath{\PZ{\to}\bbbar}\xspace}
\newcommand{\ttHbb}{\ensuremath{\ttH(\hbb)}\xspace}
\newcommand{\ttHcc}{\ensuremath{\ttH(\hcc)}\xspace}
\newcommand{\ttZbb}{\ensuremath{\ttZ(\zbb)}\xspace}
\newcommand{\ttZcc}{\ensuremath{\ttZ(\zcc)}\xspace}

\newcommand{\muHcc}{\ensuremath{\mu_{\ttHcc}}\xspace}
\newcommand{\muJustHcc}{\ensuremath{\mu_{\hcc}}\xspace}
\newcommand{\muHbb}{\ensuremath{\mu_{\ttHbb}}\xspace}
\newcommand{\muZcc}{\ensuremath{\mu_{\ttZcc}}\xspace}
\newcommand{\muZbb}{\ensuremath{\mu_{\ttZbb}}\xspace}

\newcommand{\ttjets}{\ensuremath{\ttbar{+}\text{jets}}\xspace}

\newcommand{\ttbbbar}{\ensuremath{\ttbar\bbbar}\xspace}

\newcommand{\ttbjets}{\ensuremath{\ttbar{+}{\geq}1\PQb}\xspace}
\newcommand{\ttcjets}{\ensuremath{\ttbar{+}{\geq}1\PQc}\xspace}
\newcommand{\ttlight}{\ensuremath{\ttbar{+}\text{light}}\xspace}

\newcommand{\ttbb}{\ensuremath{\ttbar{+}{\geq}2\PQb}\xspace}
\newcommand{\ttbj}{\ensuremath{\ttbar{+}\PQb}\xspace}

\newcommand{\ttcc}{\ensuremath{\ttbar{+}{\geq}2\PQc}\xspace}
\newcommand{\ttcj}{\ensuremath{\ttbar{+}\PQc}\xspace}
\newcommand{\PQj}{{\HepParticle{j}{}{}}\xspace}
\newcommand{\ttDoubleC}{\ensuremath{\smash[b]{\ttbar{+}\PQj_{\PQc\PQc}}}\xspace}
\newcommand{\ttDoubleB}{\ensuremath{\smash[b]{\ttbar{+}\PQj_{\PQb\PQb}}}\xspace}

\newcommand{\POWHEGBOX}{{\POWHEG}\textsc{-box-res}\xspace}
\newcommand{\OPENLOOPS}{\textsc{OpenLoops}\xspace}
\newcommand{\fourFS}{\ensuremath{\text{4FS}}\xspace}
\newcommand{\fiveFS}{\ensuremath{\text{5FS}}\xspace}

\newcommand{\deepJet}{\textsc{DeepJet}\xspace}
\newcommand{\pNet}{\textsc{ParticleNet}\xspace}
\newcommand{\parT}{\textsc{ParT}\xspace}
\newcommand{\parTlong}{\textsc{Particle Transformer}\xspace}

\newcommand{\FH}{\ensuremath{\text{0L}}\xspace}
\newcommand{\SL}{\ensuremath{\text{1L}}\xspace}
\newcommand{\DL}{\ensuremath{\text{2L}}\xspace}

\newcommand{\PQC}{{\HepParticle{C}{}{}}\xspace}
\newcommand{\pBpC}{\ensuremath{p_{\PB{+}\PQC}}\xspace}
\newcommand{\pBvC}{\ensuremath{p_{\PB\text{vs}\PQC}}\xspace}
\newcommand{\tagBin}[1]{\texttt{#1}\xspace}
\newcommand{\Wjets}{\ensuremath{\PW{+}\text{jets}}\xspace}
\newcommand{\Zjets}{\ensuremath{\PZ{+}\text{jets}}\xspace}

\newcommand{\score}[1]{\ensuremath{\mathcal{D}_{#1}}\xspace}
\newcommand{\ttX}{\ensuremath{\ttbar\PX}\xspace}
\newcommand{\Xbb}{\ensuremath{\PX{\to}\bbbar}\xspace}
\newcommand{\Xcc}{\ensuremath{\PX{\to}\ccbar}\xspace}

\newcommand{\pmasym}[2]{\ensuremath{^{#1}_{#2}}}
\newcommand{\pb}{\unit{pb}}

\newcommand{\gbb}{\ensuremath{\Pg{\to}\bbbar}\xspace}
\newcommand{\gcc}{\ensuremath{\Pg{\to}\ccbar}\xspace}
\newcommand{\abseta}{\ensuremath{\abs{\eta}}\xspace}
\newcommand{\pp}{\ensuremath{\Pp\Pp}\xspace}

\cmsNoteHeader{HIG-24-018}
\title{Simultaneous probe of the charm and bottom quark Yukawa couplings using \texorpdfstring{\ttH}{ttH} events}

\date{\today}

\abstract{A search for the standard model Higgs boson decaying to a charm quark-antiquark pair, \hcc, produced in association with a top quark-antiquark pair (\ttH) is presented. The search is performed with data from proton-proton collisions at $\sqrt{s}=13\TeV$, corresponding to an integrated luminosity of 138\fbinv. Advanced machine learning techniques are employed for jet flavor identification and event classification. The Higgs boson decay to a bottom quark-antiquark pair is measured simultaneously and the observed \ttHbb event rate relative to the standard model expectation is 0.91\pmasym{+0.26}{-0.22}. The observed (expected) upper limit on the product of production cross section and branching fraction \sigmaBrHcc is 0.11 (0.13)\pb at 95\% confidence level, corresponding to 7.8 (8.7) times the standard model prediction. When combined with the previous search for \hcc via associated production with a \PW or \PZ boson, the observed (expected) 95\% confidence interval on the Higgs-charm Yukawa coupling modifier, \kappaC, is $\abs{\kappaC}<3.5$ (2.7), the most stringent constraint to date.}

\hypersetup{%
pdfauthor={CMS Collaboration},%
pdftitle={Simultaneous probe of the charm and bottom quark Yukawa couplings using ttH events},%
pdfsubject={CMS},%
pdfkeywords={CMS, Higgs boson, charm quark, bottom quark, ttH, jet tagging}}

\maketitle

The discovery of a Higgs boson (\PH) by the ATLAS~\cite{Aad:2012tfa} and CMS~\cite{CMS:HIG-12-028,CMS:HIG-12-036} experiments was a landmark achievement in understanding electroweak (EW) symmetry breaking.
With a measured mass of $125.38\pm0.14\GeV$~\cite{CMS:HIG-19-004}, the Higgs boson's observed interactions with gauge bosons and third-generation fermions~\cite{Aad:2014eha,CMS:HIG-13-001,Aad:2014eva,CMS:HIG-13-002,ATLAS:2014aga,Aad:2015ona,CMS:HIG-13-023,Aad:2015vsa,CMS:HIG-13-004,CMS:HIG-16-043,CMS:HIG-16-042}, as well as all its measured properties~\cite{Aad:2015gba,CMS:HIG-14-009,CMS:HIG-12-041,Aad:2013xqa,CMS:HIG-15-002,CMS:HIG-14-042,CMS:HIG-16-041,Aaboud:2018ezd,Aaboud:2018wps}, align with the standard model (SM) predictions.
Following the first evidence of Higgs boson decays to muons, \ie, second-generation leptons~\cite{CMS:HIG-17-019}, another important milestone is to observe its couplings to second-generation quarks.

The charm quark Yukawa coupling, \yc, can be significantly modified in the presence of physics beyond the SM~\cite{Delaunay:2013pja,Perez:2015aoa,Perez:2015lra,Ghosh:2015gpa,Botella:2016krk,Harnik:2012pb,Altmannshofer:2016zrn,Erdelyi:2024sls,Giannakopoulou:2024unn}.
Searches for Higgs boson decays to a charm quark-antiquark pair, \ccbar, provide direct access to \yc.
To date, the most sensitive approach at the LHC exploits associated production of a Higgs boson with a \PV (\PW or \PZ) boson.
Using proton-proton (\pp) collision data at 13\TeV, corresponding to an integrated luminosity of about 140\fbinv, the ATLAS~\cite{ATLAS:2024yzu} and CMS~\cite{CMS:HIG-21-008} Collaborations reported observed (expected) 95\% confidence level (\CL) intervals on $\kappaC=\yc/\yc^{\,\text{SM}}$ of $\abs{\kappaC}<4.2$ (4.1) and $1.1<\abs{\kappaC}<5.5$ ($\abs{\kappaC}<3.4$), respectively.
The search in the dominant gluon fusion production mode yields lower sensitivity because of the overwhelming \ccbar background from quantum chromodynamics (QCD) multijet production~\cite{CMS:HIG-21-012}.
Despite these advancements, projections for the High-Luminosity LHC~\cite{ATL-PHYS-PUB-2022-018} indicate that current approaches may prove insufficient to achieve evidence of \hcc decay within the experiment's lifetime.

In this Letter, we present a new approach: to search for \hcc via associated production of a Higgs boson with a top quark-antiquark pair (\ttH) and to measure simultaneously the \hbb decay. We use \pp collision data at $\sqrt{s}=13\TeV$, collected with the CMS detector in 2016--2018, and corresponding to an integrated luminosity of 138\fbinv~\cite{CMS:LUM-17-003,CMS:LUM-17-004,CMS:LUM-18-002}.
The analogous processes \ttZcc and \ttZbb are measured to validate the analysis strategy.
For \ttHcc and \ttHbb, the presence of multiple jets, including several \PQb and \PQc jets, poses significant challenges in identifying the jet origins and in reconstructing the top quark and Higgs boson decays.
To address this, we employ advanced machine learning algorithms, \pNet~\cite{Qu:2019gqs} for jet flavor identification and \parTlong (\parT)~\cite{Qu:2022mxj} for event classification.
The resulting sensitivity is comparable to the best achieved in the \VH channel~\cite{CMS:HIG-21-008}.

The CMS apparatus~\cite{CMS:2008xjf,CMS:2023gfb} is a multipurpose, nearly hermetic detector, designed to trigger on~\cite{CMS:2020cmk,CMS:2016ngn,CMS:2024aqx} and identify electrons, muons, photons, and hadrons~\cite{CMS:2020uim,CMS:2018rym,CMS:2014pgm}.
A global ``particle-flow'' algorithm~\cite{CMS:2017yfk} aims to reconstruct all individual particles in an event, combining information provided by the all-silicon inner tracker and by the crystal electromagnetic and brass-scintillator hadron calorimeters, operating inside a 3.8\unit{T} superconducting solenoid, with data from the gas-ionization muon detectors embedded in the flux-return yoke outside the solenoid.
The reconstructed particles are used to build jets, employing the anti-\kt algorithm~\cite{Cacciari:2008gp, Cacciari:2011ma} with a distance parameter $R=0.4$, and to compute missing transverse momentum~\cite{CMS:2016lmd,CMS:2019ctu}.

Signal and background processes are simulated using various Monte Carlo event generators.
The \ttH signal process is generated at next-to-leading order (NLO) accuracy in QCD using \POWHEG v2~\cite{Nason:2004rx,Frixione:2007vw,Alioli:2010xd}, with the Higgs boson and top quark masses set to 125 and 172.5\GeV, respectively.
The main background is \ttbar production with additional jets (\ttjets), particularly when these additional jets originate from the hadronization of \PQb or \PQc quarks, as such events closely resemble the signal topology.
This background is categorized based on the flavor of the additional jets, \ie, those that do not originate from top quark decays.
Jets containing at least one \PQb hadron (at least one \PQc hadron and no \PQb hadron) are defined as \PQb (\PQc) jets, following the ghost association procedure~\cite{CMS:BTV-16-002}. The remaining jets are labeled as light jets.
Events containing one extra \PQb jet are labeled as \ttbj, while those with two or more are labeled as \ttbb.
Similarly, events with no \PQb jets but one (two or more) extra \PQc jet are labeled as \ttcj (\ttcc).
The remaining \ttjets events, containing no \PQb or \PQc jets other than those from top quark decays, are labeled \ttlight.
To achieve the highest available accuracy in modeling the \ttjets components, a dedicated \ttbbbar sample is used to predict the \ttbj and \ttbb backgrounds.
This sample is generated at NLO in QCD in the four-flavor scheme using \POWHEGBOX~\cite{Jezo:2015aia,Jezo:2018yaf} and \OPENLOOPS~\cite{Buccioni:2019sur}, explicitly including additional \PQb quarks in the matrix element calculation.
Contributions to \ttbj and \ttbb from double parton scattering---amounting to $\approx$15\% of the \ttbbbar production cross section---are not included in the \ttbbbar sample and are modeled separately.
The remaining components, \ttcj, \ttcc, and \ttlight, are modeled using an inclusive \ttbar sample simulated with \POWHEG v2 at NLO in QCD in the five-flavor scheme~\cite{Frixione:2007nw}.
Measurements of \ttbjets and \ttcjets production~\cite{CMS:TOP-22-009,ATLAS:2024aht,CMS:TOP-20-003,ATLAS:2024plw} indicate moderate mismodeling in simulation, necessitating corrections of the background estimates based on control samples in data.

Single top quark production in the $t$ channel ($s$ channel) and in association with a \PW boson is simulated at NLO accuracy with \POWHEG v2~\cite{Re:2010bp,Frederix:2012dh,Alioli:2009je} (\MGvATNLO v2.6.5~\cite{Alwall:2014hca}).
The production cross sections for the \ttbar and single top quark samples are computed at next-to-NLO (NNLO)~\cite{Czakon:2011xx, Kidonakis:2013zqa}.
For \ttbar, the differential cross section as a function of top quark \pt is corrected to the NNLO QCD + NLO EW prediction~\cite{Czakon:2017wor}.
The \ttW and \ttZ (\tWZ) processes are simulated at NLO (leading order, LO) accuracy in QCD using \MGvATNLO.
The matching of jets from matrix element calculations and those from parton showers is done with the FxFx~\cite{Frederix:2012ps} (MLM~\cite{Alwall:2007fs}) prescription for NLO (LO) samples.
For all samples, the proton structure is described by the NNLO NNPDF3.1 parton distribution function (PDF) set~\cite{Ball:2014uwa}.
Parton showering and hadronization are handled by \PYTHIA v8.240~\cite{Sjostrand:2014zea} with the CP5 underlying event tune~\cite{CMS:GEN-17-001}.
Additional \pp interactions within the same or nearby bunch crossings (pileup) are simulated with \PYTHIA and added to the hard-scattering process, with events reweighted to match the pileup profile observed in data.
The detector response is modeled with \GEANTfour~\cite{GEANT4}.

The analysis is carried out in three mutually exclusive channels targeting the fully hadronic (\FH), single-lepton (\SL), and dilepton (\DL) decays of the top quark-antiquark pair.
Events are collected using triggers based on high jet and \PQb jet multiplicities, or on the presence of one or two well-identified and isolated leptons (electrons or muons).
In the offline selection, only jets with transverse momentum $\pt>25\GeV$ and pseudorapidity $\abseta<2.4$ are considered.
In the \FH channel, events must contain at least seven jets, of which at least six have $\pt>40\GeV$, and have a jet \pt scalar sum exceeding 500\GeV to satisfy trigger requirements.
The \SL channel selects events with one isolated electron or muon with a minimum \pt of 26 to 30\GeV, depending on the lepton flavor and the trigger requirements in each data-taking year, along with at least five jets.
The \DL channel selects events with two oppositely charged leptons, at least one with $\pt>25\GeV$, and at least four jets.
To suppress the Drell--Yan background, events with a dielectron or dimuon invariant mass below 20\GeV or within the \PZ boson mass window (76--106\GeV) are excluded.
Across all channels, events must contain at least three jets tagged as either \PQb or \PQc jets, with at least one tagged as a \PQb jet, using criteria corresponding to tagging efficiencies of approximately 70\% (50\%) for \PQb (\PQc) jets.
Additional details of the event selection criteria are provided in \ifthenelse{\boolean{cms@external}}{\suppMaterial{supp:baseline-selection}, which includes the references after Ref.~[99]}{\suppMaterial{supp:baseline-selection}}.

\begin{figure}[!tp]
\centering
\includegraphics[width=\cmsFigWidth]{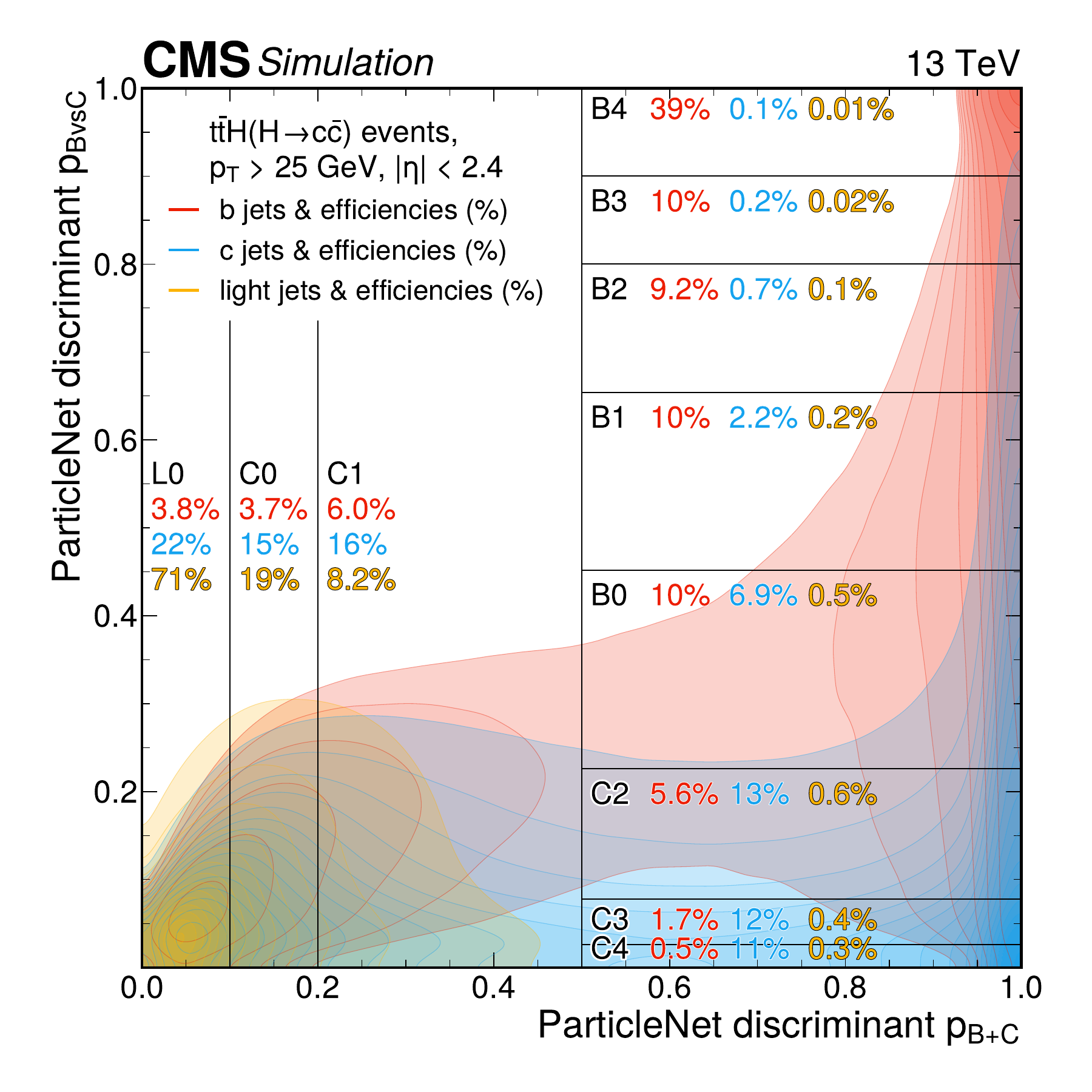}
\caption{%
    Distribution of \PQb, \PQc, and light jets in the two-dimensional \pNet discriminant plane.
    The vertical and horizontal lines correspond to the edges of the tagging categories.
    The numbers in each bin correspond to the tagging efficiencies for \PQb (red), \PQc (blue), and light (yellow) jets, evaluated on a sample of simulated \ttHcc events.
    The contour lines represent constant density values for each jet type in steps of 5\%.
}
\label{fig:ftag-performance}
\end{figure}

A key challenge in this analysis is accurately identifying jet flavor---not only efficiently tagging heavy-flavor jets initiated by \PQb or \PQc quarks but also distinguishing between them. To address this, \pNet~\cite{Qu:2019gqs}, a dynamic graph convolutional neural network~\cite{Wang:2018nkf}, is employed to classify jet flavors by leveraging spatial and kinematic correlations of individual particles and secondary vertices associated with the jets.
Two discriminants are constructed from the \pNet outputs: \pBpC, which differentiates heavy-flavor from light jets, and \pBvC, which distinguishes \PQb from \PQc jets.
Based on these, 11 mutually exclusive tagging categories are defined as shown in Fig.~\ref{fig:ftag-performance}: five \PQb-tagging categories (\tagBin{B0}--\tagBin{B4}), five \PQc-tagging categories (\tagBin{C0}--\tagBin{C4}), and one untagged category (\tagBin{L0}) enriched in light jets.
Compared to the previously used \deepJet algorithm~\cite{Bols:2020bkb}, \pNet improves background jet rejection by up to a factor of two at the same signal jet efficiency, demonstrating superior performance in distinguishing jet flavors.

To correct for potential mismodeling of tagging efficiencies in simulation, flavor-dependent scale factors are derived using data samples targeting the production of \Wjets, dileptonic \ttbar, and \Zjets, which are enriched in \PQc, \PQb, and light jets, respectively.
These scale factors, defined as data-to-simulation efficiency ratios for each jet type in the 11 tagging categories as functions of jet \pt, are extracted via a simultaneous profile likelihood fit to all three data samples.
Further details about the tagger performance and calibration are provided in \suppMaterial{supp:flavor-tagging}.
The scale factors typically range from 0.75 to 1.5, with uncertainties of up to $\approx$15\% for \PQb jets and up to $\approx$50\% for \PQc and light jets.

A further challenge involves distinguishing the \ttH signal from the \ttjets background. Both processes produce numerous jets, including multiple \PQb and \PQc jets, which may originate from top quark or Higgs boson decays, or from additional quark or gluon radiation.
This complexity hinders precise jet origin identification and full decay chain reconstruction.
To overcome this, a multiclass event classifier based on \parT~\cite{Qu:2022mxj} is developed to classify events directly from final-state objects---jets, leptons, and missing transverse momentum---without requiring explicit reconstruction of the top quarks or the Higgs boson.
For every object, the classifier receives kinematic inputs in terms of $\ln(\pt/\GeVns)$, $\ln(E/\GeVns)$, and $\eta$.
Jet flavor is encoded via 10 boolean values corresponding to the tagging categories \tagBin{B0}--\tagBin{B4} and \tagBin{C0}--\tagBin{C4}.
For each lepton, a flag is included to distinguish between electrons and muons.
To better capture event kinematic values and object correlations, \parT constructs pairwise features (\eg, angular separation and invariant mass) for all object pairs using their four-momenta.
The classifier is trained to assign likelihood scores across 10 (9) classes in the \FH (\SL, \DL) channel:
\begin{itemize}
    \item Two \ttH classes: \ttHcc, \ttHbb.
    \item Two \ttZ classes: \ttZcc, \ttZbb.
    \item Five \ttjets classes: \ttbj, \ttbb, \ttcj, \ttcc, \ttlight.
    \item (Only in the \FH channel) one class for QCD multijet.
\end{itemize}

The \parT classifier is used to refine event selection and categorization.
In the \FH channel, a stringent requirement on the QCD multijet discriminant suppresses this background by nearly four orders of magnitude, allowing it to be neglected relative to \ttjets and enabling a uniform event categorization strategy across all three channels, as depicted in Fig.~\ref{fig:event-categorization}\EndMatter.
Only events with a high \ttH or \ttZ likelihood---defined as $\score{\ttX}>0.6$, where \score{\ttX} is the sum of all four \ttH and \ttZ discriminants---are retained for signal extraction and background estimation.
A requirement on the \ttlight discriminant, $\score{\ttlight}<0.05$ (0.02) in the \FH and \DL (\SL) channels, further reduces \ttlight contamination.
Events passing these criteria are categorized into four signal regions (SRs) with $\score{\ttX}>0.85$, each enriched in \ttHcc, \ttHbb, \ttZcc, and \ttZbb, along with five control regions (CRs) with $0.6<\score{\ttX}<0.85$, which are used to estimate the normalizations of various \ttjets background components in a phase space similar to that of the SRs.
Additional requirements on heavy-flavor jet multiplicity enhance purity: at least three \PQb jets are required in the \PQb-enriched regions (\ttHbb, \ttZbb, \ttbb, and \ttbj), at least two \PQc jets in the \PQc-enriched \ttHcc, \ttZcc, and \ttcc regions, and at least one \PQc jet in the \ttcj region.
A signal-depleted sideband region, defined by $0.4<\score{\ttX}<0.6$ with an analogous SR and CR structure, is used to validate the background estimation strategy.

Production rates of the signal processes are determined via a binned profile likelihood fit to data.
The fitted variable is the \parT classifier discriminant for each category.
For each of the \ttHcc, \ttHbb, \ttZcc, and \ttZbb processes, the expected yield is scaled by an independent signal strength modifier $\mu$, defined as $(\sigma\BR)_{\text{obs}}/(\sigma\BR)_{\text{SM}}$. Here $\sigma$ is the production cross section and \BR is the branching fraction, which is allowed to freely float in the fit, following the procedure in Ref.~\cite{CMS:HIG-14-009}.
The dominant background, \ttjets, is estimated by including CRs in the fit, with the normalizations of the \ttcj, \ttcc, \ttbj, \ttbb, and \ttlight components allowed to float independently.
The normalization factors for each \ttjets component are shared between the \SL and \DL channels but not with the \FH channel, to account for potential phase space differences due to the requirement of more energetic jets in the \FH case.
Minor backgrounds, such as single top quark production, \ttW, and \tWZ, are estimated directly from simulations assuming SM production rates.
Results are obtained using \textsc{combine}~\cite{CMS:2024onh}, the CMS statistical analysis tool based on the \textsc{RooFit}~\cite{Verkerke:2003ir} and \textsc{RooStats}~\cite{Moneta:2010pm} frameworks.

Systematic uncertainties affecting normalizations and shapes of fitted variables are incorporated via nuisance parameters that encode the appropriate correlations across channels.
The contributions of each uncertainty source to the total uncertainty in the fitted \muHcc and \muHbb are summarized in Table~\ref{tab:syst-breakdown}\EndMatter.
For \muHcc, the leading uncertainty is statistical because of the limited number of events in the SRs as well as in the CRs used to extract background normalizations.
The main systematic uncertainty arises from the theoretical modeling of the \ttcj and \ttlight processes, contributing $\approx$32\% and $\approx$29\% of the total uncertainty, respectively.
For \muHbb, the largest uncertainties are theoretical, including those associated with the renormalization and factorization scales, flavor scheme, and parton shower modeling in the \ttbbbar simulation, representing $\approx$60\% of the total uncertainty.
Theoretical uncertainties in the \ttH signal simulation contribute an additional $\approx$47\%.
For both measurements, the primary experimental uncertainty is associated with jet flavor identification efficiencies, representing $\approx$39\% ($\approx$28\%) of the total uncertainty in \muHcc (\muHbb).

The analysis is validated by measuring the \ttZ signal strengths:
\begin{equation}\begin{aligned}
    \muZcc & = 1.02\pmasym{+0.79}{-0.84}, \\
    \muZbb & = 1.47\pmasym{+0.45}{-0.41},
\end{aligned}\end{equation}
which are consistent with the SM prediction within uncertainties and with measurements in the leptonic decay channel~\cite{CMS:TOP-18-009,CMS:TOP-23-004,ATLAS:2023eld}.
The significance of the excess over the background-only hypothesis is computed using the asymptotic distribution of a test statistic based on the profile likelihood ratio~\cite{CMS:NOTE-2011-005,Cowan:2010js}.
The observed (expected) significance is 1.2 (1.3) standard deviations for \ttZcc and 3.5 (2.4) standard deviations for \ttZbb.

\begin{figure}[!ht]
\centering
\includegraphics[width=0.48\textwidth]{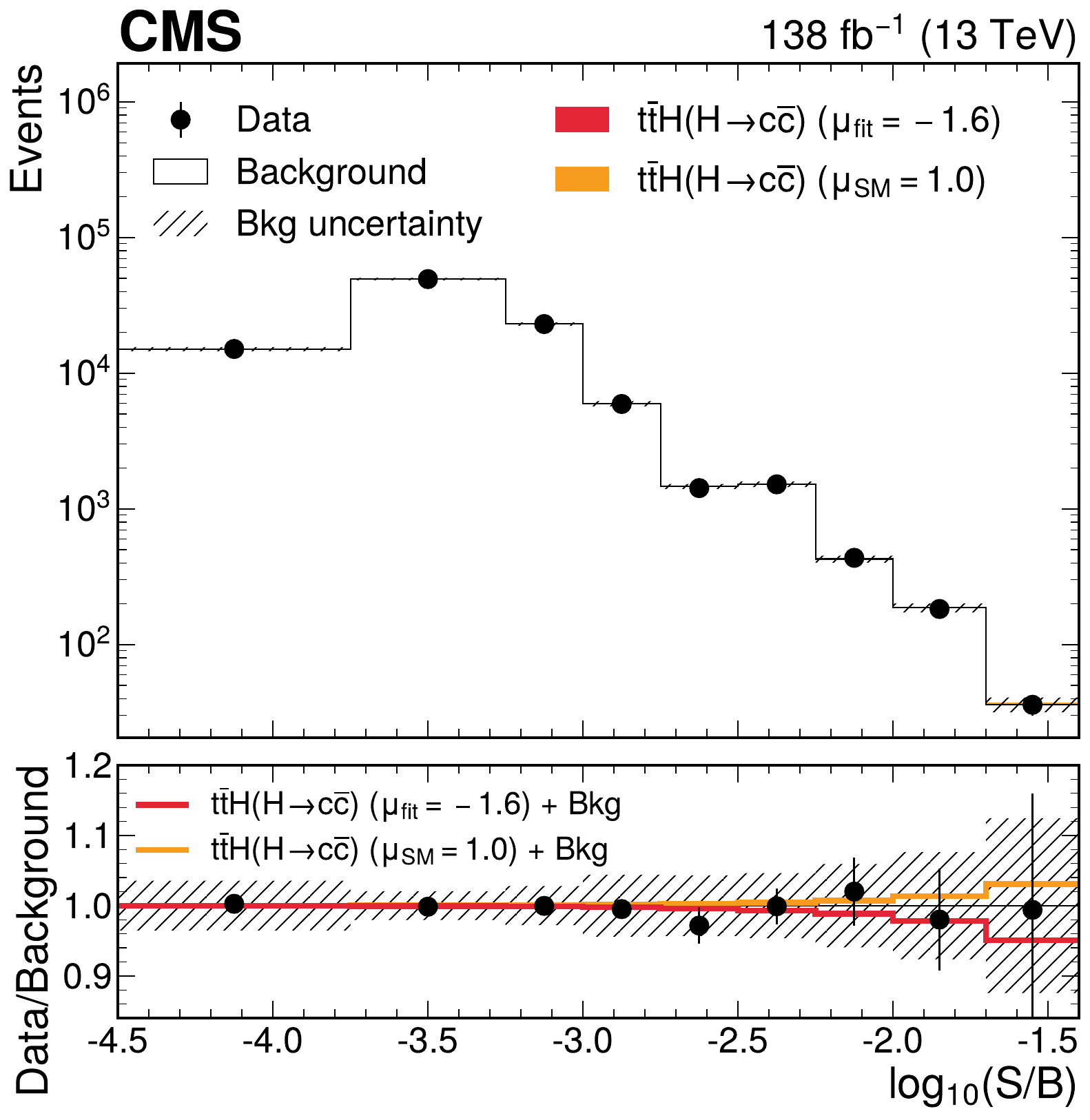}%
\hfill%
\includegraphics[width=0.48\textwidth]{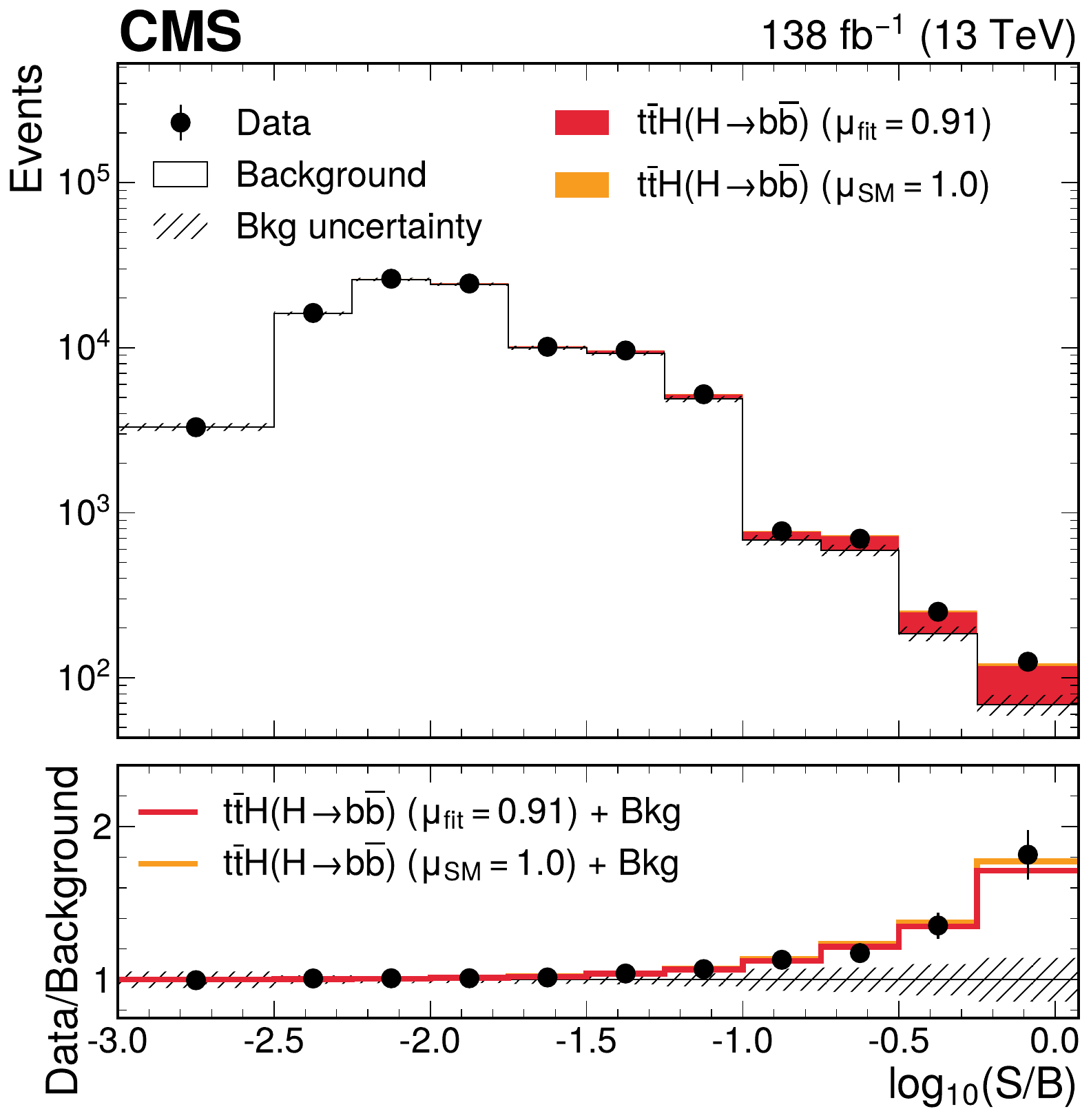}
\caption{%
    Observed and expected event yields from all SRs and CRs as a function of $\log_{10}(S/B)$, where $S$ are the expected \ttHcc (\cmsLeft) and \ttHbb (\cmsRight) yields, and $B$ are the post-fit total background yields.
    Signal contributions are shown for the best fit signal strength (red) and for the SM prediction, $\mu=1$ (orange).
    The lower panel shows the ratio of the data to the post-fit background predictions, compared to the signal-plus-background predictions.
}
\label{fig:results:soverb}
\end{figure}

\begin{figure}[!t]
\centering
\includegraphics[width=\cmsFigWidth]{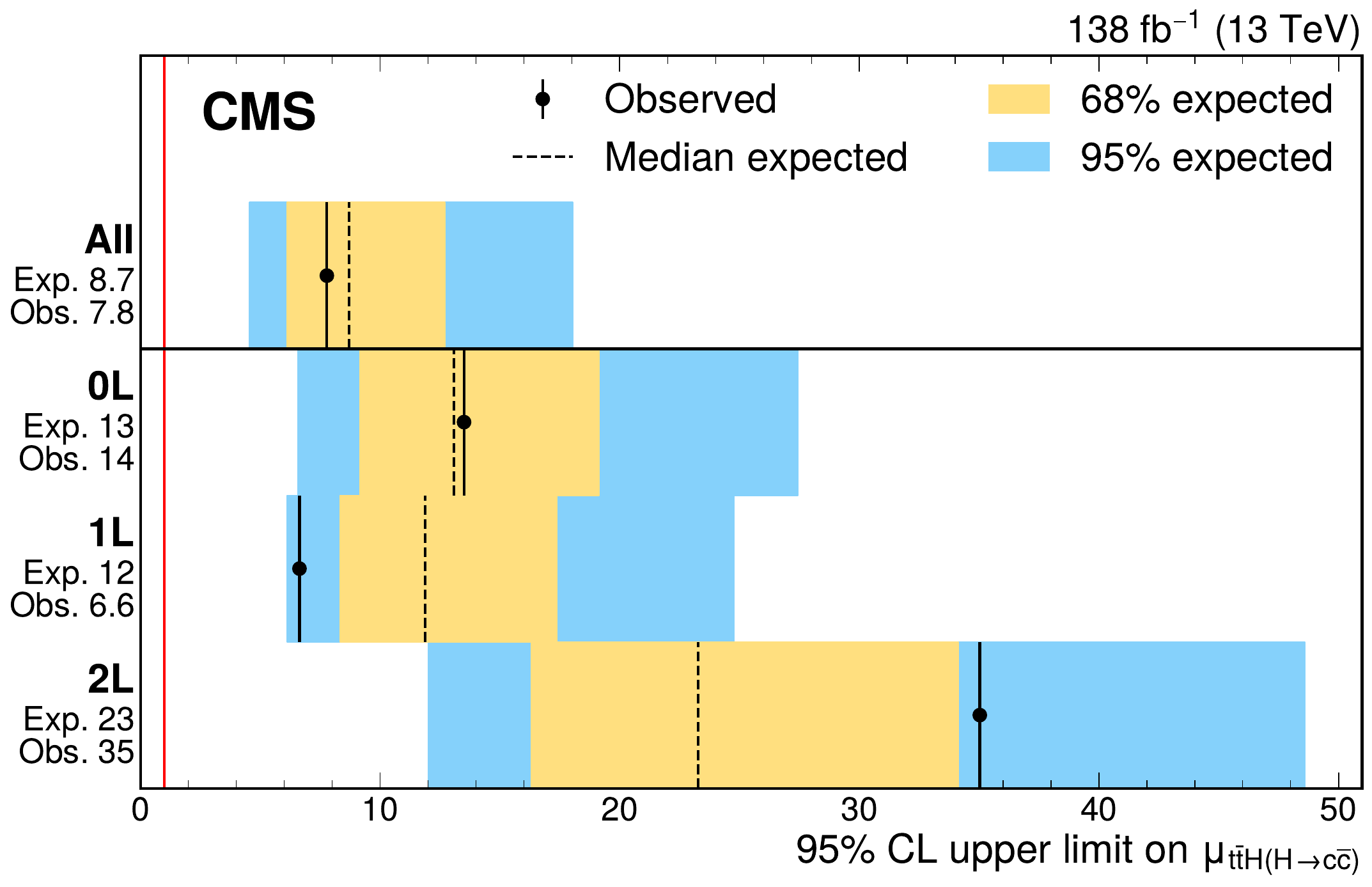}
\caption{%
    The 95\% \CL upper limits on \muHcc.
    The yellow and blue bands indicate the expected 68\% and 95\% \CL regions, respectively, under the background-only hypothesis.
    The vertical red line indicates the SM value $\muHcc=1$.
}
\label{fig:results:limits}
\end{figure}

Figure~\ref{fig:results:soverb} shows the observed and expected event yields from all CRs and SRs as a function of $\log_{10}(S/B)$, the logarithm of the ratio of \ttHcc (or \ttHbb) and background yields.
The best fit signal strengths for \ttH production are
\begin{equation}\begin{aligned}
    \muHcc & = -1.6\pm4.5, \\
    \muHbb & = 0.91\pmasym{+0.26}{-0.22},
\end{aligned}\end{equation}
{\tolerance=800
with an observed (expected) significance of 4.4 (4.5) standard deviations for the \ttHbb process.
The best fit \muHbb value is closer to the SM prediction than was the previous measurement~\cite{CMS:HIG-19-011}, with a compatibility $p$-value of 10\% (3\%) between the two measurements, assuming uncorrelated (50\% correlated) systematic uncertainties. Because of a very different background model, a more inclusive event selection, and a low correlation of the neural network scores, the total systematic uncertainty can be considered mostly uncorrelated.
No excess over the background-only hypothesis is observed in the search for \ttHcc.
An upper limit on \muHcc is extracted using the \CLs criterion~\cite{Junk:1999kv,Read:2002hq}.
The test statistic is the profile likelihood ratio modified for upper limits~\cite{CMS:NOTE-2011-005}, with the asymptotic approximation~\cite{Cowan:2010js} used in the limit setting procedure.
The observed (expected) 95\% \CL upper limit on \muHcc is 7.8 (8.7), corresponding to an observed (expected) upper limit on \sigmaBrHcc of 0.11 (0.13)\pb.
The contributions from the individual channels are summarized in Fig.~\ref{fig:results:limits}.
Tabulated results are provided in the HEPData record for this analysis~\cite{hepdata}.
\par}

\begin{figure}[!tp]
\centering
    \includegraphics[width=\cmsFigWidth]{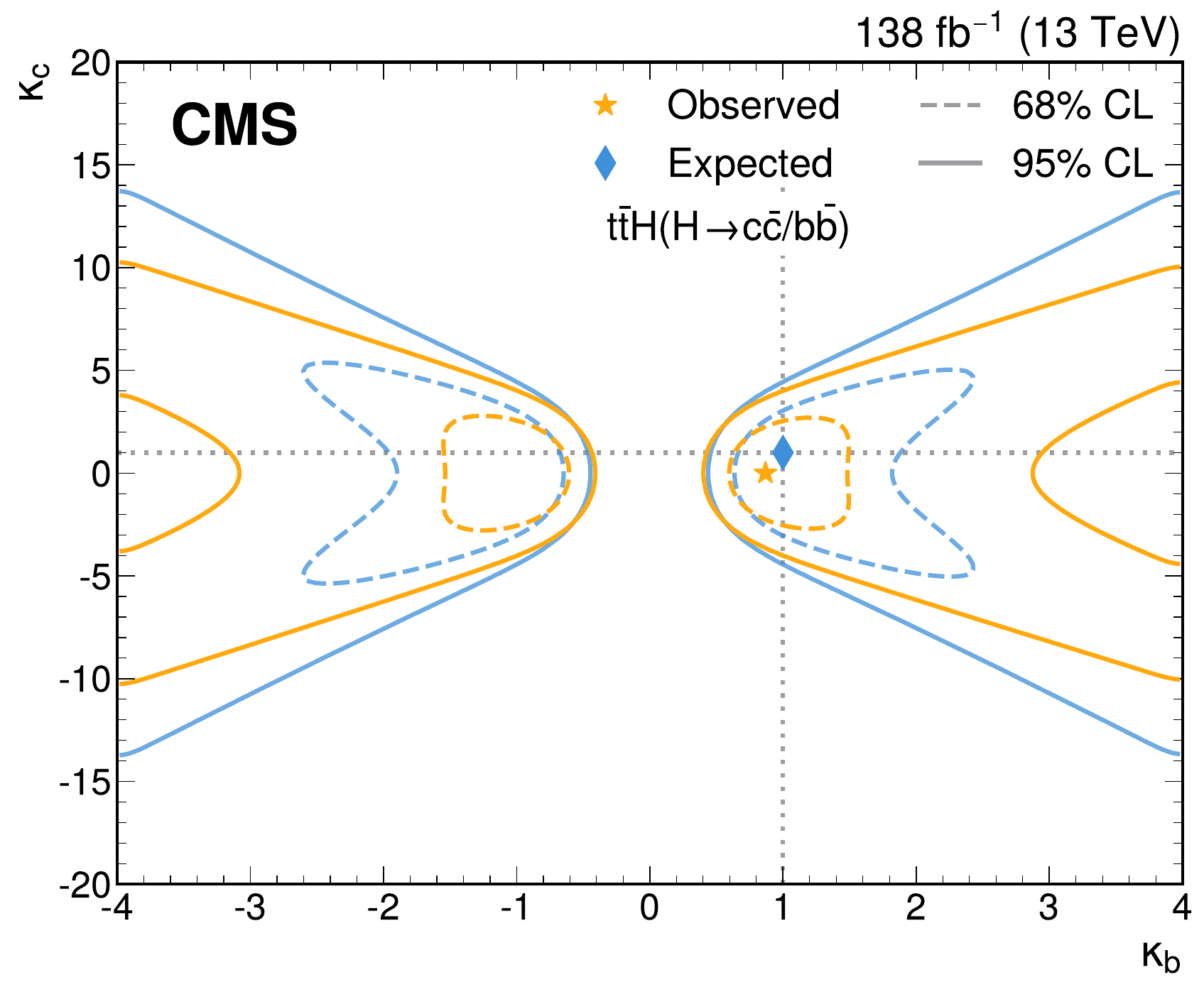}
\caption{%
    Constraints on the Higgs boson coupling modifiers \kappaC and \kappaB.
    The 68\% (95\%) \CL intervals are indicated by the dashed (solid) lines.
    The observed (expected) best fit values are shown by the orange (blue) markers.
}
\label{fig:results:kappa}
\end{figure}

{\tolerance=800
The result is interpreted in the $\kappa$-framework~\cite{Heinemeyer:2013tqa,deFlorian:2016spz} by reparameterizing \brHcc and \brHbb in terms of the charm and bottom quark Yukawa coupling modifiers \kappaC and \kappaB, assuming that only the Higgs boson decay widths are altered:
\begin{equation}\begin{aligned}
    \brHcc & = \frac{\kappaC^2 \, \BR_{\text{SM}}^{\hcc}}{1 + (\kappaC^2 -1) \, \BR_{\text{SM}}^{\hcc} + (\kappaB^2 -1) \, \BR_{\text{SM}}^{\hbb}}, \\
    \brHbb & = \frac{\kappaB^2 \, \BR_{\text{SM}}^{\hbb}}{1 + (\kappaC^2 -1) \, \BR_{\text{SM}}^{\hcc} + (\kappaB^2 -1) \, \BR_{\text{SM}}^{\hbb}}.
\end{aligned}\end{equation}}%
Figure~\ref{fig:results:kappa} shows the two-dimensional profile likelihood scan of \kappaC and \kappaB.
When fixing \kappaB to the SM expectation, the observed (expected) 95\% \CL interval is $\abs{\kappaC}<3.0$ (3.3).

A combined analysis with the previous search in the \VH channel~\cite{CMS:HIG-21-008} is performed. Common experimental uncertainties are correlated, except for jet flavor tagging, which differs in algorithms and calibration methods.
Background modeling uncertainties are uncorrelated because of differing dominant backgrounds, while theoretical uncertainties in Higgs boson production and decay for the same processes are correlated.
The expected 95\% \CL upper limit on \muJustHcc, assuming SM production rates for \ttH and \VH, is 5.6, representing a 36\% (26\%) improvement over the \ttH (\VH) channel alone.
The observed limit is 9.3, driven by a small upward statistical fluctuation in the \VH channel.
For \kappaC, the combination improves the expected 95\% \CL interval to $\abs{\kappaC}<2.7$, while the observed is $\abs{\kappaC}<3.5$.

In summary, a search for the SM Higgs boson decaying to a charm quark-antiquark pair via \ttH production is presented, alongside a simultaneous measurement of the Higgs boson decay to a bottom quark-antiquark pair.
Novel jet flavor identification tools and event classification techniques using advanced machine learning algorithms are developed for this analysis.
The observed \ttH signals relative to the SM predictions are $\muHcc=-1.6\pm4.5$ and $\muHbb=0.91\pmasym{+0.26}{-0.22}$, with an observed (expected) significance of 4.4 (4.5) standard deviations for the \ttHbb process.
The observed (expected) upper limit on \sigmaBrHcc is 0.11 (0.13)\pb, corresponding to 7.8 (8.7) times the theoretical prediction for an SM Higgs boson mass of 125.38\GeV.
When combined with the previous search for \hcc via associated production with a \PW or \PZ boson, the observed (expected) 95\% \CL interval on the charm quark Yukawa coupling modifier, \kappaC, is $\abs{\kappaC}<3.5$ (2.7).
This represents the most stringent constraint on \kappaC to date.

\begin{acknowledgments}
We congratulate our colleagues in the CERN accelerator departments for the excellent performance of the LHC and thank the technical and administrative staffs at CERN and at other CMS institutes for their contributions to the success of the CMS effort. In addition, we gratefully acknowledge the computing centers and personnel of the Worldwide LHC Computing Grid and other centers for delivering so effectively the computing infrastructure essential to our analyses. Finally, we acknowledge the enduring support for the construction and operation of the LHC, the CMS detector, and the supporting computing infrastructure provided by the following funding agencies: SC (Armenia), BMBWF and FWF (Austria); FNRS and FWO (Belgium); CNPq, CAPES, FAPERJ, FAPERGS, and FAPESP (Brazil); MES and BNSF (Bulgaria); CERN; CAS, MoST, and NSFC (China); MINCIENCIAS (Colombia); MSES and CSF (Croatia); RIF (Cyprus); SENESCYT (Ecuador); ERC PRG, TARISTU24-TK10 and MoER TK202 (Estonia); Academy of Finland, MEC, and HIP (Finland); CEA and CNRS/IN2P3 (France); SRNSF (Georgia); BMFTR, DFG, and HGF (Germany); GSRI (Greece); NKFIH (Hungary); DAE and DST (India); IPM (Iran); SFI (Ireland); INFN (Italy); MSIT and NRF (Republic of Korea); MES (Latvia); LMTLT (Lithuania); MOE and UM (Malaysia); BUAP, CINVESTAV, CONACYT, LNS, SEP, and UASLP-FAI (Mexico); MOS (Montenegro); MBIE (New Zealand); PAEC (Pakistan); MES, NSC, and NAWA (Poland); FCT (Portugal); MESTD (Serbia); MICIU/AEI and PCTI (Spain); MOSTR (Sri Lanka); Swiss Funding Agencies (Switzerland); MST (Taipei); MHESI (Thailand); TUBITAK and TENMAK (T\"{u}rkiye); NASU (Ukraine); STFC (United Kingdom); DOE and NSF (USA).
\end{acknowledgments}

\bibliography{auto_generated}

\providecommand{\href}[2]{#2}\begingroup\raggedright\begin{thebibliography}{100}%
\makeatletter
\providecommand{\hrefCMSnoop }[0]{\@secondoftwo}%
\makeatother
\providecommand{\doi}{\texttt{doi:}\begingroup \urlstyle{tt}\Url}

\bibitem{Aad:2012tfa}
\hrefCMSnoop {}{{ATLAS Collaboration}, ``Observation of a new particle in the
  search for the standard model {Higgs} boson with the {ATLAS} detector at the
  {LHC}'',} \textit{ Phys. Lett. B} \textbf{ 716} (2012) 1,
  \href{http://dx.doi.org/10.1016/j.physletb.2012.08.020}{\doi{10.1016/j.physletb.2012.08.020}}.

\bibitem{CMS:HIG-12-028}
\hrefCMSnoop {}{{CMS Collaboration}, ``Observation of a new boson at a mass of
  {125\GeV} with the {CMS} experiment at the {LHC}'',} \textit{ Phys. Lett. B}
  \textbf{ 716} (2012) 30,
  \href{http://dx.doi.org/10.1016/j.physletb.2012.08.021}{\doi{10.1016/j.physletb.2012.08.021}}.

\bibitem{CMS:HIG-12-036}
\hrefCMSnoop {}{{CMS Collaboration}, ``Observation of a new boson with mass
  near {125\GeV} in ${\Pp\Pp}$ collisions at $\sqrt{s}=7$ and {8\TeV}'',}
  \textit{ J. High Energy Phys.} \textbf{ 2013} (2013), no.~06, 081,
  \href{http://dx.doi.org/10.1007/JHEP06(2013)081}{\doi{10.1007/JHEP06(2013)081}}.

\bibitem{CMS:HIG-19-004}
\hrefCMSnoop {}{{CMS Collaboration}, ``A measurement of the {Higgs} boson mass
  in the diphoton decay channel'',} \textit{ Phys. Lett. B} \textbf{ 805}
  (2020) 135425,
  \href{http://dx.doi.org/10.1016/j.physletb.2020.135425}{\doi{10.1016/j.physletb.2020.135425}}.

\bibitem{Aad:2014eha}
\hrefCMSnoop {}{{ATLAS Collaboration}, ``Measurement of {Higgs} boson
  production in the diphoton decay channel in ${\Pp\Pp}$ collisions at
  center-of-mass energies of 7 and {8\TeV} with the {ATLAS} detector'',}
  \textit{ Phys. Rev. D} \textbf{ 90} (2014) 112015,
  \href{http://dx.doi.org/10.1103/PhysRevD.90.112015}{\doi{10.1103/PhysRevD.90.112015}}.

\bibitem{CMS:HIG-13-001}
\hrefCMSnoop {}{{CMS Collaboration}, ``Observation of the diphoton decay of the
  {Higgs} boson and measurement of its properties'',} \textit{ Eur. Phys. J. C}
  \textbf{ 74} (2014) 3076,
  \href{http://dx.doi.org/10.1140/epjc/s10052-014-3076-z}{\doi{10.1140/epjc/s10052-014-3076-z}}.

\bibitem{Aad:2014eva}
\hrefCMSnoop {}{{ATLAS Collaboration}, ``Measurements of {Higgs} boson
  production and couplings in the four-lepton channel in ${\Pp\Pp}$ collisions
  at center-of-mass energies of 7 and {8\TeV} with the {ATLAS} detector'',}
  \textit{ Phys. Rev. D} \textbf{ 91} (2015) 012006,
  \href{http://dx.doi.org/10.1103/PhysRevD.91.012006}{\doi{10.1103/PhysRevD.91.012006}}.

\bibitem{CMS:HIG-13-002}
\hrefCMSnoop {}{{CMS Collaboration}, ``Measurement of the properties of a
  {Higgs} boson in the four-lepton final state'',} \textit{ Phys. Rev. D}
  \textbf{ 89} (2014) 092007,
  \href{http://dx.doi.org/10.1103/PhysRevD.89.092007}{\doi{10.1103/PhysRevD.89.092007}}.

\bibitem{ATLAS:2014aga}
\hrefCMSnoop {}{{ATLAS Collaboration}, ``Observation and measurement of {Higgs}
  boson decays to ${\PW\PW^\ast}$ with the {ATLAS} detector'',} \textit{ Phys.
  Rev. D} \textbf{ 92} (2015) 012006,
  \href{http://dx.doi.org/10.1103/PhysRevD.92.012006}{\doi{10.1103/PhysRevD.92.012006}}.

\bibitem{Aad:2015ona}
\hrefCMSnoop {}{{ATLAS Collaboration}, ``Study of ${(\PW\!/\PZ)\PH}$ production
  and {Higgs} boson couplings using ${\PH\to\PW\PW^\ast}$ decays with the
  {ATLAS} detector'',} \textit{ J. High Energy Phys.} \textbf{ 2015} (2015),
  no.~08, 137,
  \href{http://dx.doi.org/10.1007/JHEP08(2015)137}{\doi{10.1007/JHEP08(2015)137}}.

\bibitem{CMS:HIG-13-023}
\hrefCMSnoop {}{{CMS Collaboration}, ``Measurement of {Higgs} boson production
  and properties in the ${\PW\PW}$ decay channel with leptonic final states'',}
  \textit{ J. High Energy Phys.} \textbf{ 2014} (2014), no.~01, 096,
  \href{http://dx.doi.org/10.1007/JHEP01(2014)096}{\doi{10.1007/JHEP01(2014)096}}.

\bibitem{Aad:2015vsa}
\hrefCMSnoop {}{{ATLAS Collaboration}, ``Evidence for the {Higgs}-boson
  {Yukawa} coupling to tau leptons with the {ATLAS} detector'',} \textit{ J.
  High Energy Phys.} \textbf{ 2015} (2015), no.~04, 117,
  \href{http://dx.doi.org/10.1007/JHEP04(2015)117}{\doi{10.1007/JHEP04(2015)117}}.

\bibitem{CMS:HIG-13-004}
\hrefCMSnoop {}{{CMS Collaboration}, ``Evidence for the {125\GeV} {Higgs} boson
  decaying to a pair of {\PGt} leptons'',} \textit{ J. High Energy Phys.}
  \textbf{ 2014} (2014), no.~05, 104,
  \href{http://dx.doi.org/10.1007/JHEP05(2014)104}{\doi{10.1007/JHEP05(2014)104}}.

\bibitem{CMS:HIG-16-043}
\hrefCMSnoop {}{{CMS Collaboration}, ``Observation of the {Higgs} boson decay
  to a pair of {\PGt} leptons'',} \textit{ Phys. Lett. B} \textbf{ 779} (2018)
  283,
  \href{http://dx.doi.org/10.1016/j.physletb.2018.02.004}{\doi{10.1016/j.physletb.2018.02.004}}.

\bibitem{CMS:HIG-16-042}
\hrefCMSnoop {}{{CMS Collaboration}, ``Measurements of properties of the
  {Higgs} boson decaying to a {\PW} boson pair in ${\Pp\Pp}$ collisions at
  $\sqrt{s}={13\TeV}$'',} \textit{ Phys. Lett. B} \textbf{ 791} (2019) 96,
  \href{http://dx.doi.org/10.1016/j.physletb.2018.12.073}{\doi{10.1016/j.physletb.2018.12.073}}.

\bibitem{Aad:2015gba}
\hrefCMSnoop {}{{ATLAS Collaboration}, ``Measurements of the {Higgs} boson
  production and decay rates and coupling strengths using ${\Pp\Pp}$ collision
  data at $\sqrt{s}=7$ and {8\TeV} in the {ATLAS} experiment'',} \textit{ Eur.
  Phys. J. C} \textbf{ 76} (2016) 6,
  \href{http://dx.doi.org/10.1140/epjc/s10052-015-3769-y}{\doi{10.1140/epjc/s10052-015-3769-y}}.

\bibitem{CMS:HIG-14-009}
\hrefCMSnoop {}{{CMS Collaboration}, ``Precise determination of the mass of the
  {Higgs} boson and tests of compatibility of its couplings with the standard
  model predictions using proton collisions at 7 and {8\TeV}'',} \textit{ Eur.
  Phys. J. C} \textbf{ 75} (2015) 212,
  \href{http://dx.doi.org/10.1140/epjc/s10052-015-3351-7}{\doi{10.1140/epjc/s10052-015-3351-7}}.

\bibitem{CMS:HIG-12-041}
\hrefCMSnoop {}{{CMS Collaboration}, ``Study of the mass and spin-parity of the
  {Higgs} boson candidate via its decays to {\PZ} boson pairs'',} \textit{
  Phys. Rev. Lett.} \textbf{ 110} (2013) 081803,
  \href{http://dx.doi.org/10.1103/PhysRevLett.110.081803}{\doi{10.1103/PhysRevLett.110.081803}}.

\bibitem{Aad:2013xqa}
\hrefCMSnoop {}{{ATLAS Collaboration}, ``Evidence for the spin-0 nature of the
  {Higgs} boson using {ATLAS} data'',} \textit{ Phys. Lett. B} \textbf{ 726}
  (2013) 120,
  \href{http://dx.doi.org/10.1016/j.physletb.2013.08.026}{\doi{10.1016/j.physletb.2013.08.026}}.

\bibitem{CMS:HIG-15-002}
\hrefCMSnoop {}{{ATLAS and CMS Collaborations}, ``Measurements of the {Higgs}
  boson production and decay rates and constraints on its couplings from a
  combined {ATLAS} and {CMS} analysis of the {LHC} ${\Pp\Pp}$ collision data at
  $\sqrt{s}=7$ and {8\TeV}'',} \textit{ J. High Energy Phys.} \textbf{ 2016}
  (2016), no.~08, 045,
  \href{http://dx.doi.org/10.1007/JHEP08(2016)045}{\doi{10.1007/JHEP08(2016)045}}.

\bibitem{CMS:HIG-14-042}
\hrefCMSnoop {}{{ATLAS and CMS Collaborations}, ``Combined measurement of the
  {Higgs} boson mass in ${\Pp\Pp}$ collisions at $\sqrt{s}=7$ and {8\TeV} with
  the {ATLAS} and {CMS} experiments'',} \textit{ Phys. Rev. Lett.} \textbf{
  114} (2015) 191803,
  \href{http://dx.doi.org/10.1103/PhysRevLett.114.191803}{\doi{10.1103/PhysRevLett.114.191803}}.

\bibitem{CMS:HIG-16-041}
\hrefCMSnoop {}{{CMS Collaboration}, ``Measurements of properties of the
  {Higgs} boson decaying into the four-lepton final state in ${\Pp\Pp}$
  collisions at $\sqrt{s}={13\TeV}$'',} \textit{ J. High Energy Phys.} \textbf{
  2017} (2017), no.~11, 047,
  \href{http://dx.doi.org/10.1007/JHEP11(2017)047}{\doi{10.1007/JHEP11(2017)047}}.

\bibitem{Aaboud:2018ezd}
\hrefCMSnoop {}{{ATLAS Collaboration}, ``Combined measurement of differential
  and total cross sections in the ${\PH\to\PGg\PGg}$ and the
  ${\PH\to\PZ\PZ^\ast\to4\Pell}$ decay channels at $\sqrt{s}={13\TeV}$ with the
  {ATLAS} detector'',} \textit{ Phys. Lett. B} \textbf{ 786} (2018) 114,
  \href{http://dx.doi.org/10.1016/j.physletb.2018.09.019}{\doi{10.1016/j.physletb.2018.09.019}}.

\bibitem{Aaboud:2018wps}
\hrefCMSnoop {}{{ATLAS Collaboration}, ``Measurement of the {Higgs} boson mass
  in the ${\PH\to\PZ\PZ^\ast\to4\Pell}$ and ${\PH\to\PGg\PGg}$ channels with
  $\sqrt{s}={13\TeV}$ ${\Pp\Pp}$ collisions using the {ATLAS} detector'',}
  \textit{ Phys. Lett. B} \textbf{ 784} (2018) 345,
  \href{http://dx.doi.org/10.1016/j.physletb.2018.07.050}{\doi{10.1016/j.physletb.2018.07.050}}.

\bibitem{CMS:HIG-17-019}
\hrefCMSnoop {}{{CMS Collaboration}, ``Search for the {Higgs} boson decaying to
  two muons in proton-proton collisions at $\sqrt{s}={13\TeV}$'',} \textit{
  Phys. Rev. Lett.} \textbf{ 122} (2019) 021801,
  \href{http://dx.doi.org/10.1103/PhysRevLett.122.021801}{\doi{10.1103/PhysRevLett.122.021801}}.

\bibitem{Delaunay:2013pja}
\hrefCMSnoop {}{C.~Delaunay, T.~Golling, G.~Perez, and Y.~Soreq, ``{Enhanced
  Higgs boson coupling to charm pairs}'',} \textit{ Phys. Rev. D} \textbf{ 89}
  (2014) 033014,
  \href{http://dx.doi.org/10.1103/PhysRevD.89.033014}{\doi{10.1103/PhysRevD.89.033014}}.

\bibitem{Perez:2015aoa}
\hrefCMSnoop {}{G.~Perez, Y.~Soreq, E.~Stamou, and K.~Tobioka, ``{Constraining
  the charm Yukawa and Higgs-quark coupling universality}'',} \textit{ Phys.
  Rev. D} \textbf{ 92} (2015) 033016,
\href{http://dx.doi.org/10.1103/PhysRevD.92.033016}{\doi{10.1103/PhysRevD.92.033016}}.

\bibitem{Perez:2015lra}
\hrefCMSnoop {}{G.~Perez, Y.~Soreq, E.~Stamou, and K.~Tobioka, ``{Prospects for
  measuring the Higgs boson coupling to light quarks}'',} \textit{ Phys. Rev.
  D} \textbf{ 93} (2016) 013001,
  \href{http://dx.doi.org/10.1103/PhysRevD.93.013001}{\doi{10.1103/PhysRevD.93.013001}}.

\bibitem{Ghosh:2015gpa}
\hrefCMSnoop {}{D.~Ghosh, R.~S. Gupta, and G.~Perez, ``{Is the Higgs mechanism
  of fermion mass generation a fact? A Yukawa-less first-two-generation
  model}'',} \textit{ Phys. Lett. B} \textbf{ 755} (2016) 504,
\href{http://dx.doi.org/10.1016/j.physletb.2016.02.059}{\doi{10.1016/j.physletb.2016.02.059}}.

\bibitem{Botella:2016krk}
\hrefCMSnoop {}{F.~J. Botella, G.~C. Branco, M.~N. Rebelo, and J.~I.
  Silva-Marcos, ``{What if the masses of the first two quark families are not
  generated by the standard model Higgs boson?}'',} \textit{ Phys. Rev. D}
  \textbf{ 94} (2016) 115031,
\href{http://dx.doi.org/10.1103/PhysRevD.94.115031}{\doi{10.1103/PhysRevD.94.115031}}.

\bibitem{Harnik:2012pb}
\hrefCMSnoop {}{R.~Harnik, J.~Kopp, and J.~Zupan, ``Flavor violating {Higgs}
  decays'',} \textit{ J. High Energy Phys.} \textbf{ 2013} (2013), no.~03, 026,
\href{http://dx.doi.org/10.1007/JHEP03(2013)026}{\doi{10.1007/JHEP03(2013)026}}.

\bibitem{Altmannshofer:2016zrn}
W.~Altmannshofer\hrefCMSnoop {}{ { et~al.}, ``Collider signatures of flavorful
  {Higgs} bosons'',} \textit{ Phys. Rev. D} \textbf{ 94} (2016) 115032,
\href{http://dx.doi.org/10.1103/PhysRevD.94.115032}{\doi{10.1103/PhysRevD.94.115032}}.

\bibitem{Erdelyi:2024sls}
\hrefCMSnoop {}{B.~A. Erdelyi, R.~Gr{\"o}ber, and N.~Selimovic, ``{How large
  can the light quark Yukawa couplings be?}'',} \textit{ J. High Energy Phys.}
  \textbf{ 2025} (2025), no.~05, 189,
  \href{http://dx.doi.org/10.1007/JHEP05(2025)189}{\doi{10.1007/JHEP05(2025)189}}.

\bibitem{Giannakopoulou:2024unn}
\hrefCMSnoop {}{A.~S. Giannakopoulou, P.~Meade, and M.~Valli, ``{How charming
  can the Higgs be?}'',} \textit{ J. High Energy Phys.} \textbf{ 2025} (2025),
  no.~02, 067,
  \href{http://dx.doi.org/10.1007/JHEP02(2025)067}{\doi{10.1007/JHEP02(2025)067}}.

\bibitem{ATLAS:2024yzu}
\hrefCMSnoop {}{{ATLAS Collaboration}, ``Measurements of ${\PW\PH}$ and
  ${\PZ\PH}$ production with {Higgs} boson decays into bottom quarks and direct
  constraints on the charm {Yukawa} coupling in {13\TeV} ${\Pp\Pp}$ collisions
  with the {ATLAS} detector'',} \textit{ J. High Energy Phys.} \textbf{ 2025}
  (2025), no.~04, 075,
  \href{http://dx.doi.org/10.1007/JHEP04(2025)075}{\doi{10.1007/JHEP04(2025)075}}.

\bibitem{CMS:HIG-21-008}
\hrefCMSnoop {}{{CMS Collaboration}, ``Search for {Higgs} boson decay to a
  charm quark-antiquark pair in proton-proton collisions at
  $\sqrt{s}={13\TeV}$'',} \textit{ Phys. Rev. Lett.} \textbf{ 131} (2023)
  061801,
  \href{http://dx.doi.org/10.1103/PhysRevLett.131.061801}{\doi{10.1103/PhysRevLett.131.061801}}.

\bibitem{CMS:HIG-21-012}
\hrefCMSnoop {}{{CMS Collaboration}, ``Search for boosted {Higgs} boson decay
  to a charm quark-antiquark pair in proton-proton collisions at
  $\sqrt{s}={13\TeV}$'',} \textit{ Phys. Rev. Lett.} \textbf{ 131} (2023)
  041801,
  \href{http://dx.doi.org/10.1103/PhysRevLett.131.041801}{\doi{10.1103/PhysRevLett.131.041801}}.

\bibitem{ATL-PHYS-PUB-2022-018}
\href {https://cds.cern.ch/record/2805993}{{ATLAS and CMS Collaborations},
  ``Snowmass white paper contribution: Physics with the {Phase-2} {ATLAS} and
  {CMS} detectors'',} CMS Physics Analysis Summary CMS-PAS-FTR-22-001,
  ATL-PHYS-PUB-2022-018, 2022.

\bibitem{CMS:LUM-17-003}
\hrefCMSnoop {}{{CMS Collaboration}, ``Precision luminosity measurement in
  proton-proton collisions at $\sqrt{s}={13\TeV}$ in 2015 and 2016 at {CMS}'',}
  \textit{ Eur. Phys. J. C} \textbf{ 81} (2021) 800,
  \href{http://dx.doi.org/10.1140/epjc/s10052-021-09538-2}{\doi{10.1140/epjc/s10052-021-09538-2}}.

\bibitem{CMS:LUM-17-004}
\href {https://cds.cern.ch/record/2621960}{{CMS Collaboration}, ``{CMS}
  luminosity measurement for the 2017 data-taking period at
  $\sqrt{s}={13\TeV}$'',} CMS Physics Analysis Summary CMS-PAS-LUM-17-004,
  2018.

\bibitem{CMS:LUM-18-002}
\href {https://cds.cern.ch/record/2676164}{{CMS Collaboration}, ``{CMS}
  luminosity measurement for the 2018 data-taking period at
  $\sqrt{s}={13\TeV}$'',} CMS Physics Analysis Summary CMS-PAS-LUM-18-002,
  2019.

\bibitem{Qu:2019gqs}
\hrefCMSnoop {}{H.~Qu and L.~Gouskos, ``Jet tagging via particle clouds'',}
  \textit{ Phys. Rev. D} \textbf{ 101} (2020) 056019,
  \href{http://dx.doi.org/10.1103/PhysRevD.101.056019}{\doi{10.1103/PhysRevD.101.056019}}.

\bibitem{Qu:2022mxj}
\href {https://proceedings.mlr.press/v162/qu22b.html}{H.~Qu, C.~Li, and
  S.~Qian, ``Particle transformer for jet tagging'',} in \textit{ {Proc. 39th
  International Conference on Machine Learning (ICML 2022): Baltimore MD, USA,
  July 17--23, 2022}}.
\newblock 2022.
\newblock
  \href{http://www.arXiv.org/abs/2202.03772}{\texttt{arXiv:2202.03772}}.
\newblock [PMLR 162 (2022) 18281].

\bibitem{CMS:2008xjf}
\hrefCMSnoop {}{{CMS Collaboration}, ``The {CMS} experiment at the {CERN}
  {LHC}'',} \textit{ J. Instrum.} \textbf{ 3} (2008) S08004,
  \href{http://dx.doi.org/10.1088/1748-0221/3/08/S08004}{\doi{10.1088/1748-0221/3/08/S08004}}.

\bibitem{CMS:2023gfb}
\hrefCMSnoop {}{{CMS Collaboration}, ``Development of the {CMS} detector for
  the {CERN} {LHC} \mbox{Run 3}'',} \textit{ J. Instrum.} \textbf{ 19} (2024)
  P05064,
  \href{http://dx.doi.org/10.1088/1748-0221/19/05/P05064}{\doi{10.1088/1748-0221/19/05/P05064}}.

\bibitem{CMS:2020cmk}
\hrefCMSnoop {}{{CMS Collaboration}, ``Performance of the {CMS} {\Lone} trigger
  in proton-proton collisions at $\sqrt{s}={13\TeV}$'',} \textit{ J. Instrum.}
  \textbf{ 15} (2020) P10017,
  \href{http://dx.doi.org/10.1088/1748-0221/15/10/P10017}{\doi{10.1088/1748-0221/15/10/P10017}}.

\bibitem{CMS:2016ngn}
\hrefCMSnoop {}{{CMS Collaboration}, ``The {CMS} trigger system'',} \textit{ J.
  Instrum.} \textbf{ 12} (2017) P01020,
  \href{http://dx.doi.org/10.1088/1748-0221/12/01/P01020}{\doi{10.1088/1748-0221/12/01/P01020}}.

\bibitem{CMS:2024aqx}
\hrefCMSnoop {}{{CMS Collaboration}, ``Performance of the {CMS} high-level
  trigger during {LHC} \mbox{Run 2}'',} \textit{ J. Instrum.} \textbf{ 19}
  (2024) P11021,
  \href{http://dx.doi.org/10.1088/1748-0221/19/11/P11021}{\doi{10.1088/1748-0221/19/11/P11021}}.

\bibitem{CMS:2020uim}
\hrefCMSnoop {}{{CMS Collaboration}, ``Electron and photon reconstruction and
  identification with the {CMS} experiment at the {CERN} {LHC}'',} \textit{ J.
  Instrum.} \textbf{ 16} (2021) P05014,
  \href{http://dx.doi.org/10.1088/1748-0221/16/05/P05014}{\doi{10.1088/1748-0221/16/05/P05014}}.

\bibitem{CMS:2018rym}
\hrefCMSnoop {}{{CMS Collaboration}, ``Performance of the {CMS} muon detector
  and muon reconstruction with proton-proton collisions at
  $\sqrt{s}={13\TeV}$'',} \textit{ J. Instrum.} \textbf{ 13} (2018) P06015,
  \href{http://dx.doi.org/10.1088/1748-0221/13/06/P06015}{\doi{10.1088/1748-0221/13/06/P06015}}.

\bibitem{CMS:2014pgm}
\hrefCMSnoop {}{{CMS Collaboration}, ``Description and performance of track and
  primary-vertex reconstruction with the {CMS} tracker'',} \textit{ J.
  Instrum.} \textbf{ 9} (2014) P10009,
  \href{http://dx.doi.org/10.1088/1748-0221/9/10/P10009}{\doi{10.1088/1748-0221/9/10/P10009}}.

\bibitem{CMS:2017yfk}
\hrefCMSnoop {}{{CMS Collaboration}, ``Particle-flow reconstruction and global
  event description with the {CMS} detector'',} \textit{ J. Instrum.} \textbf{
  12} (2017) P10003,
  \href{http://dx.doi.org/10.1088/1748-0221/12/10/P10003}{\doi{10.1088/1748-0221/12/10/P10003}}.

\bibitem{Cacciari:2008gp}
\hrefCMSnoop {}{M.~Cacciari, G.~P. Salam, and G.~Soyez, ``The anti-\kt jet
  clustering algorithm'',} \textit{ J. High Energy Phys.} \textbf{ 2008}
  (2008), no.~04, 063,
  \href{http://dx.doi.org/10.1088/1126-6708/2008/04/063}{\doi{10.1088/1126-6708/2008/04/063}}.

\bibitem{Cacciari:2011ma}
\hrefCMSnoop {}{M.~Cacciari, G.~P. Salam, and G.~Soyez, ``{\FASTJET} user
  manual'',} \textit{ Eur. Phys. J. C} \textbf{ 72} (2012) 1896,
  \href{http://dx.doi.org/10.1140/epjc/s10052-012-1896-2}{\doi{10.1140/epjc/s10052-012-1896-2}}.

\bibitem{CMS:2016lmd}
\hrefCMSnoop {}{{CMS Collaboration}, ``Jet energy scale and resolution in the
  {CMS} experiment in ${\Pp\Pp}$ collisions at {8\TeV}'',} \textit{ J.
  Instrum.} \textbf{ 12} (2017) P02014,
  \href{http://dx.doi.org/10.1088/1748-0221/12/02/P02014}{\doi{10.1088/1748-0221/12/02/P02014}}.

\bibitem{CMS:2019ctu}
\hrefCMSnoop {}{{CMS Collaboration}, ``Performance of missing transverse
  momentum reconstruction in proton-proton collisions at $\sqrt{s}={13\TeV}$
  using the {CMS} detector'',} \textit{ J. Instrum.} \textbf{ 14} (2019)
  P07004,
  \href{http://dx.doi.org/10.1088/1748-0221/14/07/P07004}{\doi{10.1088/1748-0221/14/07/P07004}}.

\bibitem{Nason:2004rx}
\hrefCMSnoop {}{P.~Nason, ``A new method for combining {NLO} {QCD} with shower
  {Monte Carlo} algorithms'',} \textit{ J. High Energy Phys.} \textbf{ 2004}
  (2004), no.~11, 040,
  \href{http://dx.doi.org/10.1088/1126-6708/2004/11/040}{\doi{10.1088/1126-6708/2004/11/040}}.

\bibitem{Frixione:2007vw}
\hrefCMSnoop {}{S.~Frixione, P.~Nason, and C.~Oleari, ``Matching {NLO} {QCD}
  computations with parton shower simulations: the {\POWHEG} method'',}
  \textit{ J. High Energy Phys.} \textbf{ 2007} (2007), no.~11, 070,
  \href{http://dx.doi.org/10.1088/1126-6708/2007/11/070}{\doi{10.1088/1126-6708/2007/11/070}}.

\bibitem{Alioli:2010xd}
\hrefCMSnoop {}{S.~Alioli, P.~Nason, C.~Oleari, and E.~Re, ``A general
  framework for implementing {NLO} calculations in shower {Monte Carlo}
  programs: the {\POWHEG} \textsc{box}'',} \textit{ J. High Energy Phys.}
  \textbf{ 2010} (2010), no.~06, 043,
  \href{http://dx.doi.org/10.1007/JHEP06(2010)043}{\doi{10.1007/JHEP06(2010)043}}.

\bibitem{CMS:BTV-16-002}
\hrefCMSnoop {}{{CMS Collaboration}, ``Identification of heavy-flavour jets
  with the {CMS} detector in ${\Pp\Pp}$ collisions at {13\TeV}'',} \textit{ J.
  Instrum.} \textbf{ 13} (2018) P05011,
  \href{http://dx.doi.org/10.1088/1748-0221/13/05/P05011}{\doi{10.1088/1748-0221/13/05/P05011}}.

\bibitem{Jezo:2015aia}
\hrefCMSnoop {}{T.~Je{\v{z}}o and P.~Nason, ``On the treatment of resonances in
  next-to-leading order calculations matched to a parton shower'',} \textit{ J.
  High Energy Phys.} \textbf{ 2015} (2015), no.~12, 065,
  \href{http://dx.doi.org/10.1007/JHEP12(2015)065}{\doi{10.1007/JHEP12(2015)065}}.

\bibitem{Jezo:2018yaf}
\hrefCMSnoop {}{T.~Je{\v{z}}o, J.~M. Lindert, N.~Moretti, and S.~Pozzorini,
  ``New {NLOPS} predictions for \ttbar{+}{\PQb}-jet production at the {LHC}'',}
  \textit{ Eur. Phys. J. C} \textbf{ 78} (2018) 502,
  \href{http://dx.doi.org/10.1140/epjc/s10052-018-5956-0}{\doi{10.1140/epjc/s10052-018-5956-0}}.

\bibitem{Buccioni:2019sur}
F.~Buccioni\hrefCMSnoop {}{ { et~al.}, ``\textsc{OpenLoops} 2'',} \textit{ Eur.
  Phys. J. C} \textbf{ 79} (2019) 866,
  \href{http://dx.doi.org/10.1140/epjc/s10052-019-7306-2}{\doi{10.1140/epjc/s10052-019-7306-2}}.

\bibitem{Frixione:2007nw}
\hrefCMSnoop {}{S.~Frixione, G.~Ridolfi, and P.~Nason, ``A positive-weight
  next-to-leading-order {Monte Carlo} for heavy flavour hadroproduction'',}
  \textit{ J. High Energy Phys.} \textbf{ 2007} (2007), no.~09, 126,
  \href{http://dx.doi.org/10.1088/1126-6708/2007/09/126}{\doi{10.1088/1126-6708/2007/09/126}}.

\bibitem{CMS:TOP-22-009}
\hrefCMSnoop {}{{CMS Collaboration}, ``Inclusive and differential cross section
  measurements of $\ttbar\bbbar$ production in the lepton+jets channel at
  $\sqrt{s}={13\TeV}$'',} \textit{ J. High Energy Phys.} \textbf{ 2024} (2024),
  no.~05, 042,
  \href{http://dx.doi.org/10.1007/JHEP05(2024)042}{\doi{10.1007/JHEP05(2024)042}}.

\bibitem{ATLAS:2024aht}
\hrefCMSnoop {}{{ATLAS Collaboration}, ``Measurement of \ttbar production in
  association with additional {\PQb}-jets in the ${\Pe\PGm}$ final state in
  proton-proton collisions at $\sqrt{s}={13\TeV}$ with the {ATLAS} detector'',}
  \textit{ J. High Energy Phys.} \textbf{ 2025} (2025), no.~01, 068,
  \href{http://dx.doi.org/10.1007/JHEP01(2025)068}{\doi{10.1007/JHEP01(2025)068}}.

\bibitem{CMS:TOP-20-003}
\hrefCMSnoop {}{{CMS Collaboration}, ``First measurement of the cross section
  for top quark pair production with additional charm jets using dileptonic
  final states in ${\Pp\Pp}$ collisions at $\sqrt{s}={13\TeV}$'',} \textit{
  Phys. Lett. B} \textbf{ 820} (2021) 136565,
  \href{http://dx.doi.org/10.1016/j.physletb.2021.136565}{\doi{10.1016/j.physletb.2021.136565}}.

\bibitem{ATLAS:2024plw}
\hrefCMSnoop {}{{ATLAS Collaboration}, ``Measurement of top-quark pair
  production in association with charm quarks in proton-proton collisions at
  $\sqrt{s}={13\TeV}$ with the {ATLAS} detector'',} \textit{ Phys. Lett. B}
  \textbf{ 860} (2024) 139177,
  \href{http://dx.doi.org/10.1016/j.physletb.2024.139177}{\doi{10.1016/j.physletb.2024.139177}}.

\bibitem{Re:2010bp}
\hrefCMSnoop {}{E.~Re, ``Single-top ${\PW\PQt}$-channel production matched with
  parton showers using the {\POWHEG} method'',} \textit{ Eur. Phys. J. C}
  \textbf{ 71} (2011) 1547,
  \href{http://dx.doi.org/10.1140/epjc/s10052-011-1547-z}{\doi{10.1140/epjc/s10052-011-1547-z}}.

\bibitem{Frederix:2012dh}
\hrefCMSnoop {}{R.~Frederix, E.~Re, and P.~Torrielli, ``Single-top $t$-channel
  hadroproduction in the four-flavors scheme with {\POWHEG} and {a\MCATNLO}'',}
  \textit{ J. High Energy Phys.} \textbf{ 2012} (2012), no.~09, 130,
  \href{http://dx.doi.org/10.1007/JHEP09(2012)130}{\doi{10.1007/JHEP09(2012)130}}.

\bibitem{Alioli:2009je}
\hrefCMSnoop {}{S.~Alioli, P.~Nason, C.~Oleari, and E.~Re, ``{NLO} single-top
  production matched with shower in {\POWHEG}: $s$- and $t$-channel
  contributions'',} \textit{ J. High Energy Phys.} \textbf{ 2009} (2009),
  no.~09, 111,
  \href{http://dx.doi.org/10.1088/1126-6708/2009/09/111}{\doi{10.1088/1126-6708/2009/09/111}}.
  [Erratum: \DOI{10.1007/JHEP02(2010)011}].

\bibitem{Alwall:2014hca}
J.~Alwall\hrefCMSnoop {}{ { et~al.}, ``The automated computation of tree-level
  and next-to-leading order differential cross sections, and their matching to
  parton shower simulations'',} \textit{ J. High Energy Phys.} \textbf{ 2014}
  (2014), no.~07, 079,
  \href{http://dx.doi.org/10.1007/JHEP07(2014)079}{\doi{10.1007/JHEP07(2014)079}}.

\bibitem{Czakon:2011xx}
\hrefCMSnoop {}{M.~Czakon and A.~Mitov, ``\textsc{top++}: a program for the
  calculation of the top-pair cross-section at hadron colliders'',} \textit{
  Comput. Phys. Commun.} \textbf{ 185} (2014) 2930,
  \href{http://dx.doi.org/10.1016/j.cpc.2014.06.021}{\doi{10.1016/j.cpc.2014.06.021}}.

\bibitem{Kidonakis:2013zqa}
\hrefCMSnoop {}{N.~Kidonakis, ``Top quark production'',} in \textit{ {Proc.
  Helmholtz International Summer School on Physics of Heavy Quarks and Hadrons
  (HQ 2013): Dubna, Russia, July 15--28, 2013}}.
\newblock 2013.
\newblock [DESY-PROC-2013-03].
  \href{http://dx.doi.org/10.3204/DESY-PROC-2013-03/Kidonakis}{\doi{10.3204/DESY-PROC-2013-03/Kidonakis}}.

\bibitem{Czakon:2017wor}
M.~Czakon\hrefCMSnoop {}{ { et~al.}, ``Top-pair production at the {LHC} through
  {NNLO} {QCD} and {NLO} {EW}'',} \textit{ J. High Energy Phys.} \textbf{ 2017}
  (2017), no.~10, 186,
  \href{http://dx.doi.org/10.1007/JHEP10(2017)186}{\doi{10.1007/JHEP10(2017)186}}.

\bibitem{Frederix:2012ps}
\hrefCMSnoop {}{R.~Frederix and S.~Frixione, ``Merging meets matching in
  {\MCATNLO}'',} \textit{ J. High Energy Phys.} \textbf{ 2012} (2012), no.~12,
  061,
  \href{http://dx.doi.org/10.1007/JHEP12(2012)061}{\doi{10.1007/JHEP12(2012)061}}.

\bibitem{Alwall:2007fs}
J.~Alwall\hrefCMSnoop {}{ { et~al.}, ``Comparative study of various algorithms
  for the merging of parton showers and matrix elements in hadronic
  collisions'',} \textit{ Eur. Phys. J. C} \textbf{ 53} (2008) 473,
  \href{http://dx.doi.org/10.1140/epjc/s10052-007-0490-5}{\doi{10.1140/epjc/s10052-007-0490-5}}.

\bibitem{Ball:2014uwa}
\hrefCMSnoop {}{{NNPDF} Collaboration, ``Parton distributions for the {LHC} run
  {II}'',} \textit{ J. High Energy Phys.} \textbf{ 2015} (2015), no.~04, 040,
  \href{http://dx.doi.org/10.1007/JHEP04(2015)040}{\doi{10.1007/JHEP04(2015)040}}.

\bibitem{Sjostrand:2014zea}
T.~Sj{\"o}strand\hrefCMSnoop {}{ { et~al.}, ``An introduction to
  {\PYTHIA8.2}'',} \textit{ Comput. Phys. Commun.} \textbf{ 191} (2015) 159,
  \href{http://dx.doi.org/10.1016/j.cpc.2015.01.024}{\doi{10.1016/j.cpc.2015.01.024}}.

\bibitem{CMS:GEN-17-001}
\hrefCMSnoop {}{{CMS Collaboration}, ``Extraction and validation of a new set
  of {CMS} {\PYTHIA8} tunes from underlying-event measurements'',} \textit{
  Eur. Phys. J. C} \textbf{ 80} (2020) 4,
  \href{http://dx.doi.org/10.1140/epjc/s10052-019-7499-4}{\doi{10.1140/epjc/s10052-019-7499-4}}.

\bibitem{GEANT4}
\hrefCMSnoop {}{{GEANT4} Collaboration, ``{\GEANTfour}---a simulation
  toolkit'',} \textit{ Nucl. Instrum. Meth. A} \textbf{ 506} (2003) 250,
  \href{http://dx.doi.org/10.1016/S0168-9002(03)01368-8}{\doi{10.1016/S0168-9002(03)01368-8}}.

\bibitem{suppMatBib}
\hrefCMSnoop {}{``Supplemental Material: More details on simulated samples, jet
  flavor identification, baseline event selection, neural network event
  classifier, systematic uncertainties, validation of the background model, and
  additional results''.}
\newblock [URL will be inserted by publisher].

\bibitem{Wang:2018nkf}
Y.~Wang\hrefCMSnoop {}{ { et~al.}, ``Dynamic graph {CNN} for learning on point
  clouds'',} \textit{ ACM Trans. Graph.} \textbf{ 38} (2018) 146,
  \href{http://dx.doi.org/10.1145/3326362}{\doi{10.1145/3326362}}.

\bibitem{Bols:2020bkb}
E.~Bols\hrefCMSnoop {}{ { et~al.}, ``Jet flavour classification using
  {DeepJet}'',} \textit{ J. Instrum.} \textbf{ 15} (2020) P12012,
  \href{http://dx.doi.org/10.1088/1748-0221/15/12/P12012}{\doi{10.1088/1748-0221/15/12/P12012}}.

\bibitem{CMS:2024onh}
\hrefCMSnoop {}{{CMS Collaboration}, ``The {CMS} statistical analysis and
  combination tool: \textsc{combine}'',} \textit{ Comput. Softw. Big Sci.}
  \textbf{ 8} (2024) 19,
  \href{http://dx.doi.org/10.1007/s41781-024-00121-4}{\doi{10.1007/s41781-024-00121-4}}.

\bibitem{Verkerke:2003ir}
\href
  {https://www.slac.stanford.edu/econf/C0303241/proc/papers/MOLT007.PDF}{W.~Verkerke
  and D.~Kirkby, ``The \textsc{RooFit} toolkit for data modeling'',} in
  \textit{ {Proc. 13th International Conference on Computing in High Energy and
  Nuclear Physics (CHEP 2003): La Jolla CA, United States, March 24--28,
  2003}}.
\newblock 2003.
\newblock
  \href{http://www.arXiv.org/abs/physics/0306116}{\texttt{arXiv:physics/0306116}}.
\newblock [eConf C0303241 (2003) MOLT007].

\bibitem{Moneta:2010pm}
L.~Moneta\hrefCMSnoop {}{ { et~al.}, ``The \textsc{RooStats} project'',} in
  \textit{ {Proc. 13th International Workshop on Advanced Computing and
  Analysis Techniques in Physics Research (ACAT 2010): Jaipur, India, February
  22--27, 2010}}.
\newblock 2010.
\newblock [PoS (ACAT2010) 057].
  \href{http://dx.doi.org/10.22323/1.093.0057}{\doi{10.22323/1.093.0057}}.

\bibitem{CMS:TOP-18-009}
\hrefCMSnoop {}{{CMS Collaboration}, ``Measurement of top quark pair production
  in association with a {\PZ} boson in proton-proton collisions at
  $\sqrt{s}={13\TeV}$'',} \textit{ J. High Energy Phys.} \textbf{ 2020} (2020),
  no.~03, 056,
  \href{http://dx.doi.org/10.1007/JHEP03(2020)056}{\doi{10.1007/JHEP03(2020)056}}.

\bibitem{CMS:TOP-23-004}
\hrefCMSnoop {}{{CMS Collaboration}, ``Measurements of inclusive and
  differential cross sections for top quark production in association with a
  {\PZ} boson in proton-proton collisions at $\sqrt{s}={13\TeV}$'',} \textit{
  J. High Energy Phys.} \textbf{ 2025} (2025), no.~02, 177,
  \href{http://dx.doi.org/10.1007/JHEP02(2025)177}{\doi{10.1007/JHEP02(2025)177}}.

\bibitem{ATLAS:2023eld}
\hrefCMSnoop {}{{ATLAS Collaboration}, ``Inclusive and differential
  cross-section measurements of ${\ttbar\PZ}$ production in ${\Pp\Pp}$
  collisions at $\sqrt{s}={13\TeV}$ with the {ATLAS} detector, including {EFT}
  and spin-correlation interpretations'',} \textit{ J. High Energy Phys.}
  \textbf{ 2024} (2024), no.~07, 163,
  \href{http://dx.doi.org/10.1007/JHEP07(2024)163}{\doi{10.1007/JHEP07(2024)163}}.

\bibitem{CMS:NOTE-2011-005}
\href {https://cds.cern.ch/record/1379837}{{ATLAS and CMS Collaborations, and
  LHC Higgs Combination Group}, ``Procedure for the {LHC} {Higgs} boson search
  combination in {Summer} 2011'',} Technical Report CMS-NOTE-2011-005,
  ATL-PHYS-PUB-2011-11, 2011.

\bibitem{Cowan:2010js}
\hrefCMSnoop {}{G.~Cowan, K.~Cranmer, E.~Gross, and O.~Vitells, ``Asymptotic
  formulae for likelihood-based tests of new physics'',} \textit{ Eur. Phys. J.
  C} \textbf{ 71} (2011) 1554,
  \href{http://dx.doi.org/10.1140/epjc/s10052-011-1554-0}{\doi{10.1140/epjc/s10052-011-1554-0}}.
  [Erratum: \DOI{10.1140/epjc/s10052-013-2501-z}].

\bibitem{CMS:HIG-19-011}
\hrefCMSnoop {}{{CMS Collaboration}, ``Measurement of the ${\ttbar\PH}$ and
  ${\PQt\PH}$ production rates in the ${\PH\to\bbbar}$ decay channel using
  proton-proton collision data at $\sqrt{s}={13\TeV}$'',} \textit{ J. High
  Energy Phys.} \textbf{ 2025} (2025), no.~02, 097,
  \href{http://dx.doi.org/10.1007/JHEP02(2025)097}{\doi{10.1007/JHEP02(2025)097}}.

\bibitem{Junk:1999kv}
\hrefCMSnoop {}{T.~Junk, ``Confidence level computation for combining searches
  with small statistics'',} \textit{ Nucl. Instrum. Meth. A} \textbf{ 434}
  (1999) 435,
  \href{http://dx.doi.org/10.1016/S0168-9002(99)00498-2}{\doi{10.1016/S0168-9002(99)00498-2}}.

\bibitem{Read:2002hq}
\hrefCMSnoop {}{A.~L. Read, ``Presentation of search results: The {\CLs}
  technique'',} \textit{ J. Phys. G} \textbf{ 28} (2002) 2693,
  \href{http://dx.doi.org/10.1088/0954-3899/28/10/313}{\doi{10.1088/0954-3899/28/10/313}}.

\bibitem{hepdata}
\hrefCMSnoop {}{``{HEPData} record for this analysis'',} 2025.
\newblock
  \href{http://dx.doi.org/10.17182/hepdata.159997}{\doi{10.17182/hepdata.159997}}.

\bibitem{Heinemeyer:2013tqa}
\hrefCMSnoop {}{{LHC Higgs Cross Section Working Group}, S.~Heinemeyer {
  et~al.}, ``Handbook of {LHC} {Higgs} cross sections: 3. {Higgs}
  properties'',} CERN Report CERN-2013-004, 2013.
\newblock
  \href{http://dx.doi.org/10.5170/CERN-2013-004}{\doi{10.5170/CERN-2013-004}}.

\bibitem{deFlorian:2016spz}
\hrefCMSnoop {}{{LHC Higgs Cross Section Working Group}, D.~de~Florian {
  et~al.}, ``Handbook of {LHC} {Higgs} cross sections: 4. {Deciphering} the
  nature of the {Higgs} sector'',} CERN Report CERN-2017-002-M, 2016.
\newblock
  \href{http://dx.doi.org/10.23731/CYRM-2017-002}{\doi{10.23731/CYRM-2017-002}}.

\bibitem{ATLAS:2018zbr}
\hrefCMSnoop {}{{ATLAS Collaboration}, ``Study of the hard double-parton
  scattering contribution to inclusive four-lepton production in ${\Pp\Pp}$
  collisions at $\sqrt{s}={8\TeV}$ with the {ATLAS} detector'',} \textit{ Phys.
  Lett. B} \textbf{ 790} (2019) 595,
  \href{http://dx.doi.org/10.1016/j.physletb.2019.01.062}{\doi{10.1016/j.physletb.2019.01.062}}.

\bibitem{Artoisenet:2012st}
\hrefCMSnoop {}{P.~Artoisenet, R.~Frederix, O.~Mattelaer, and R.~Rietkerk,
  ``Automatic spin-entangled decays of heavy resonances in {Monte Carlo}
  simulations'',} \textit{ J. High Energy Phys.} \textbf{ 2013} (2013), no.~03,
  015,
  \href{http://dx.doi.org/10.1007/JHEP03(2013)015}{\doi{10.1007/JHEP03(2013)015}}.

\bibitem{CMS:JME-18-001}
\hrefCMSnoop {}{{CMS Collaboration}, ``Pileup mitigation at {CMS} in {13\TeV}
  data'',} \textit{ J. Instrum.} \textbf{ 15} (2020) P09018,
  \href{http://dx.doi.org/10.1088/1748-0221/15/09/P09018}{\doi{10.1088/1748-0221/15/09/P09018}}.

\bibitem{Kulesza:2020nfh}
A.~Kulesza\hrefCMSnoop {}{ { et~al.}, ``Associated top quark pair production
  with a heavy boson: differential cross sections at {NLO+NNLL} accuracy'',}
  \textit{ Eur. Phys. J. C} \textbf{ 80} (2020) 428,
  \href{http://dx.doi.org/10.1140/epjc/s10052-020-7987-6}{\doi{10.1140/epjc/s10052-020-7987-6}}.

\bibitem{Mrenna:2016sih}
\hrefCMSnoop {}{S.~Mrenna and P.~Skands, ``Automated parton-shower variations
  in {\PYTHIA8}'',} \textit{ Phys. Rev. D} \textbf{ 94} (2016) 074005,
  \href{http://dx.doi.org/10.1103/PhysRevD.94.074005}{\doi{10.1103/PhysRevD.94.074005}}.

\bibitem{CMS:MLG-24-001}
\hrefCMSnoop {}{{CMS Collaboration}, ``Reweighting simulated events using
  machine-learning techniques in the {CMS} experiment'',} \textit{ Eur. Phys.
  J. C} \textbf{ 85} (2025) 495,
  \href{http://dx.doi.org/10.1140/epjc/s10052-025-14097-x}{\doi{10.1140/epjc/s10052-025-14097-x}}.

\bibitem{CMS:TOP-22-008}
\hrefCMSnoop {}{{CMS Collaboration}, ``Evidence for ${\PQt\PW\PZ}$ production
  in proton-proton collisions at $\sqrt{s}={13\TeV}$ in multilepton final
  states'',} \textit{ Phys. Lett. B} \textbf{ 855} (2024) 138815,
  \href{http://dx.doi.org/10.1016/j.physletb.2024.138815}{\doi{10.1016/j.physletb.2024.138815}}.

\bibitem{Buonocore:2023ljm}
L.~Buonocore\hrefCMSnoop {}{ { et~al.}, ``Precise predictions for the
  associated production of a {\PW} boson with a top-antitop quark pair at the
  {LHC}'',} \textit{ Phys. Rev. Lett.} \textbf{ 131} (2023) 231901,
  \href{http://dx.doi.org/10.1103/PhysRevLett.131.231901}{\doi{10.1103/PhysRevLett.131.231901}}.

\bibitem{ATLAS:2024gth}
\hrefCMSnoop {}{{ATLAS Collaboration}, ``Measurement of the associated
  production of a top-antitop-quark pair and a {Higgs} boson decaying into a
  \bbbar pair in ${\Pp\Pp}$ collisions at $\sqrt{s}={13\TeV}$ using the {ATLAS}
  detector at the {LHC}'',} \textit{ Eur. Phys. J. C} \textbf{ 85} (2025) 210,
  \href{http://dx.doi.org/10.1140/epjc/s10052-025-13740-x}{\doi{10.1140/epjc/s10052-025-13740-x}}.

\end{thebibliography}\endgroup

\appendix
\ifthenelse{\boolean{cms@external}}{\section{}}{\section{End Matter}\label{endmatter}}

\subsection{Analysis strategy and event selection}
\label{endmatter:selection}

\begin{figure}[!ht]
\centering
\includegraphics[width=\cmsFigWidth]{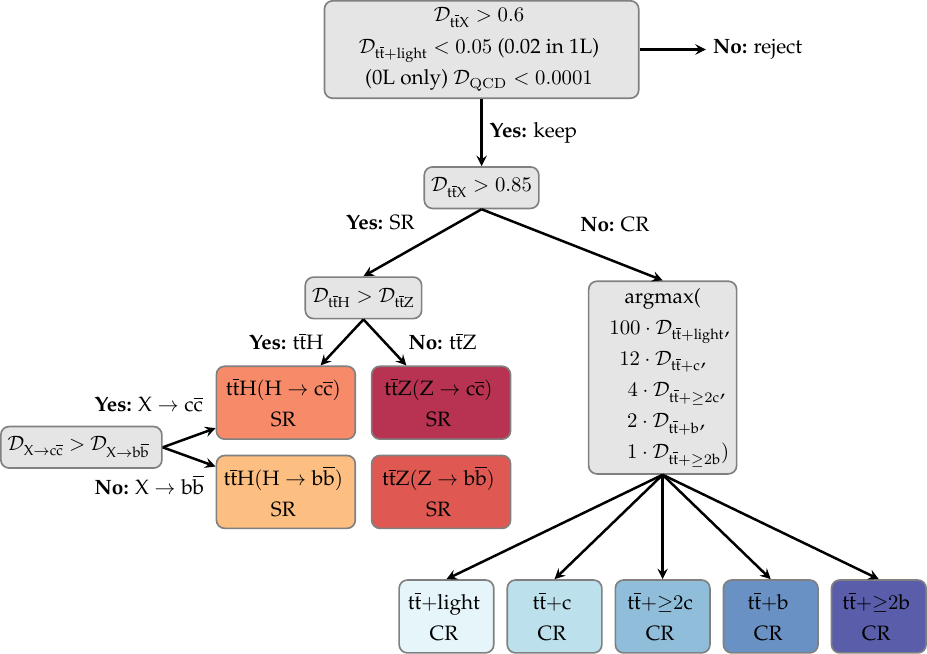}
\caption{Event categorization flowchart.}
\label{fig:event-categorization}
\end{figure}

{\tolerance=800
The \parT event classifier is used to select and categorize events, as illustrated in Fig.~\ref{fig:event-categorization}.
The classifier outputs 10 (9) likelihood scores that sum to unity for the \FH (\SL and \DL) channel.
These include four scores for the signal processes, \score{\ttHcc}, \score{\ttHbb}, \score{\ttZcc}, and \score{\ttZbb}, and six (five) for the background processes, \score{\ttlight}, \score{\ttcj}, \score{\ttcc}, \score{\ttbj}, \score{\ttbb}, and \score{QCD} (the last applies only to the \FH channel).
The sum of the four signal scores is defined as \score{\ttX}, while the sum of the five \ttjets background scores is defined as \score{\ttjets}.
\par}

In the \FH channel, an initial requirement of $\score{QCD}<0.0001$ suppresses the QCD multijet background to below 1\% of the \ttjets background, allowing the QCD multijet contribution to be neglected.
To reduce the \ttlight background, a requirement of $\score{\ttlight}<0.05$ (0.02) is applied in the \FH and \DL (\SL) channels.
Events with $\score{\ttX}>0.85$ are assigned to one of four SRs based on two pairs of signal discriminants, \score{\ttH} \vs \score{\ttZ} and \score{\Xcc} \vs \score{\Xbb}, where $\score{\ttH}=\score{\ttHcc}+\score{\ttHbb}$, $\score{\ttZ}=\score{\ttZcc}+\score{\ttZbb}$, $\score{\Xcc}=\score{\ttHcc}+\score{\ttZcc}$, and $\score{\Xbb}=\score{\ttHbb}+\score{\ttZbb}$.
Events with $0.6<\score{\ttX}<0.85$ are categorized into one of the five CRs, determined by the background class with the highest weighted score. The weights---100, 12, 4, 2 and 1 for \score{\ttlight}, \score{\ttcj}, \score{\ttcc}, \score{\ttbj}, \score{\ttbb}, respectively---are optimized to enhance the purity of each CR.

\begin{figure*}[!ht]
\centering
\includegraphics[width=\textwidth]{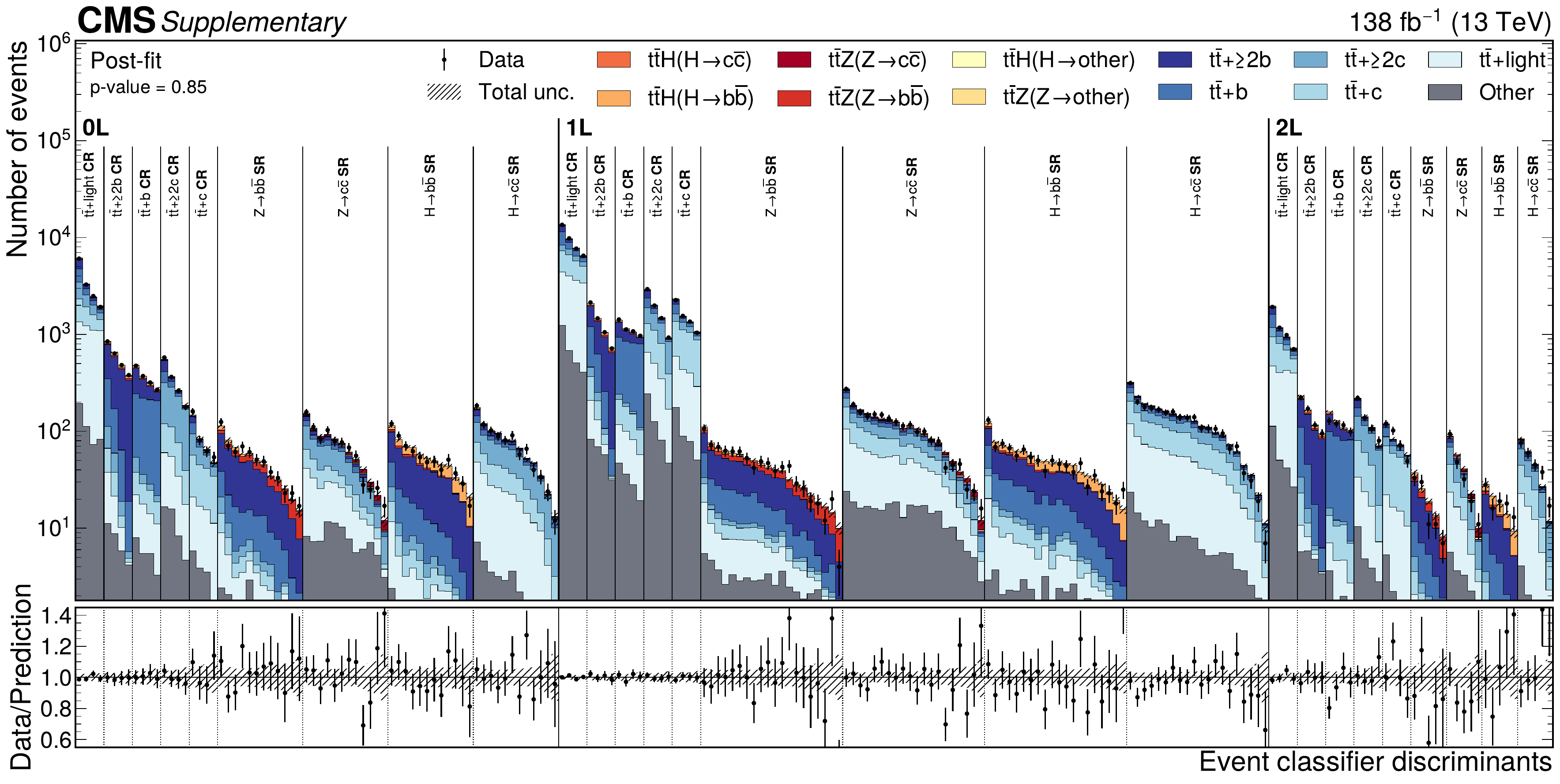}
\caption{%
    Distributions of the \parT discriminants in data (points) and predicted signal and backgrounds (colored histograms) after the fit to data.
    The vertical bars on the points represent the statistical uncertainties in data.
    The hatched band represents the total uncertainty in the sum of the signal and background predictions.
    The lower panel shows the ratio of the data to the sum of the signal and background predictions.
}
\label{fig:results:postfitsupp_noline}
\end{figure*}

Figure~\ref{fig:results:postfitsupp_noline} shows the fitted event yields in each bin for the CRs and SRs.

\subsection{The \texorpdfstring{\ttjets}{tt+jets} background model}
\label{endmatter:bkgttjets}

Simulated \ttjets events are divided into five mutually exclusive components: \ttbb, \ttbj, \ttcc, \ttcj, and \ttlight.
The \ttbb and \ttbj events are taken from a dedicated \ttbbbar sample, where the \ttbbbar matrix elements are calculated at NLO in QCD using the four-flavor scheme (\fourFS).
The \PQb quark mass is set to 4.75\GeV, and the \fourFS NNLO {NNPDF3.1} PDF set is used to describe the proton structure.
The \ttbbbar production cross section is computed at NLO in \fourFS.
The \ttcc, \ttcj, and \ttlight components are extracted from the inclusive \ttbar simulation after removing the \ttbb and \ttbj contributions, thereby avoiding double counting.
This inclusive \ttbar sample is generated at NLO in QCD with up to one additional parton included at the matrix-element level, while additional emissions are modeled via parton showering with \PYTHIA.
The matrix element calculation is performed in the five-flavor scheme (\fiveFS) using the \fiveFS NNLO {NNPDF3.1} PDF set, where the \PQb quarks are treated as massless.
A subcomponent of \ttbj (\ttcj), referred to as \ttDoubleB (\ttDoubleC), consists of events where the additional \PQb (\PQc) jet contains two or more \PQb (\PQc) hadrons.
These events predominantly arise from collinear \gbb (\gcc) splittings, which are not well modeled in simulation.
To account for this modeling limitation, an additional 50\% uncertainty is assigned to the \ttDoubleB and \ttDoubleC contributions in the background estimation.

Production of \ttbbbar via double parton scattering (DPS) also contributes to the \ttbb and \ttbj categories. This contribution is included in the inclusive \ttbar sample but not in the dedicated \ttbbbar sample.
It is therefore modeled separately by generating \ttbar production at NLO in QCD using \POWHEG, with additional \bbbar production simulated by \PYTHIA at LO in QCD using the \texttt{SecondHard} option.
The cross section for this process is estimated to be 8\pb, assuming an effective DPS cross section of 30\unit{mb}~\cite{ATLAS:2018zbr} and an inclusive \bbbar cross section of 0.3\unit{mb}, the latter estimated from simulation.
In the background estimation, the DPS contributions are included in the corresponding \ttbb and \ttbj categories, with a 50\% uncertainty assigned to the normalization of the DPS components.

Table~\ref{tab:rateParam} summarizes the best fit values of the \ttjets background normalization factors obtained from a simultaneous fit to all CRs and SRs.

\begin{table}[!ht]
\centering
\topcaption{%
    Best fit values of the \ttjets background normalization factors for each analysis category (Cat.).
}
\label{tab:rateParam}
\renewcommand{\arraystretch}{1.2}
\cmsNarrowTable{\begin{scotch}{lccccc}
    Cat. & \ttbb & \ttbj & \ttcc & \ttcj & \ttlight \\ \hline
    \DL/\SL & $0.88\pm0.07$ & $0.92\pm0.16$ & $1.39\pm0.23$ & $1.41\pm0.36$ & $0.89\pm0.11$ \\
    \FH & $1.12\pm0.11$ & $0.89\pm0.18$ & $2.03\pm0.41$ & $0.95\pm0.34$ & $0.89\pm0.17$ \\
\end{scotch}}
\end{table}

\subsection{Uncertainty breakdown}
\label{endmatter:uncertainties}

\begin{table}[!htp]
\centering
\topcaption{The absolute (relative) contributions to the total uncertainties, $\Delta\mu$ ($\Delta\mu/\Delta\mu_{\text{tot}}$).}
\label{tab:syst-breakdown}
\cmsNarrowTable{\begin{scotch}{ll@{\cmsExtraSep}r@{\cmsNormalSep}r@{\cmsExtraSep}r@{\cmsNormalSep}r}
    & & \multicolumn{4}{c}{$\Delta\mu$ ($\Delta\mu/\Delta\mu_{\text{tot}}$)} \\
    \multicolumn{2}{l}{Uncertainty source} & \multicolumn{2}{c@{\cmsExtraSep}}{\muHcc} & \multicolumn{2}{c}{\muHbb} \\[\cmsTabSkip] \hline
    \multicolumn{2}{l}{Statistical} & 3.3 & (74\%) & 0.14 & (57\%) \\
    & \ttjets normalizations & 1.4 & (32\%) & 0.06 & (26\%) \\
    & \ttZ normalizations & 0.4 & (8.4\%) & 0.06 & (30\%) \\[\cmsTabSkip]
    \multicolumn{2}{l}{Theory} & 2.1 & (47\%) & 0.18 & (75\%) \\
    & Signal & 0.7 & (15\%) & 0.11 & (47\%) \\
    & \ttbjets & 0.7 & (15\%) & 0.14 & (60\%) \\
    & \ttcjets & 1.4 & (32\%) & 0.01 & (5.8\%) \\
    & \ttlight & 1.3 & (29\%) & 0.01 & (5.2\%) \\
    & Minor backgrounds & 0.2 & (4.6\%) & 0.01 & (4.6\%) \\[\cmsTabSkip]
    \multicolumn{2}{l}{Experimental} & 2.0 & (47\%) & 0.07 & (31\%) \\
    & Jet flavor tagging & 1.7 & (39\%) & 0.07 & (28\%) \\
    & Size of the simulated samples & 1.1 & (24\%) & 0.05 & (21\%) \\
    & Jet energy scale and resolution & 0.8 & (18\%) & 0.02 & (8.6\%) \\
    & Lepton identification & 0.3 & (6.0\%) & 0.02 & (6.3\%) \\
    & Integrated luminosity & 0.1 & (2.0\%) & 0.02 & (6.2\%) \\[\cmsTabSkip]
    \multicolumn{2}{l}{Total} & 4.5 & (100\%) & 0.24 & (100\%) \\
\end{scotch}}
\end{table}

{\tolerance=800
The contributions of individual uncertainty sources to the total uncertainties in the fitted \muHcc and \muHbb are summarized in Table~\ref{tab:syst-breakdown}.
The values are determined by repeating the fit with the nuisance parameters associated with each category fixed to their best fit values, and then subtracting the resulting uncertainty in quadrature from the total uncertainty.
The total uncertainty differs from the quadrature sum of the individual components because of correlations among nuisance parameters in the fit.
\par}

\ifthenelse{\boolean{cms@external}}{
    \nocite{Artoisenet:2012st,Alwall:2007fs,Qu:2019gqs,Bols:2020bkb,CMS:LUM-17-003,CMS:LUM-17-004,CMS:LUM-18-002,CMS:JME-18-001,CMS:2016lmd,Kulesza:2020nfh,deFlorian:2016spz,Mrenna:2016sih,CMS:MLG-24-001,CMS:TOP-22-008,Buonocore:2023ljm,ATLAS:2018zbr,CMS:TOP-22-009,Heinemeyer:2013tqa,CMS:HIG-21-008,ATLAS:2024gth}
}{%
    \clearpage
    \numberwithin{figure}{section}
    \numberwithin{table}{section}
    \section{Supplemental Material}
    \label{supp}
    \newcommand{\ttSingleC}{\ensuremath{\smash[b]{\ttbar{+}\text{j}_{\PQc}}}\xspace}
\newcommand{\ttSingleB}{\ensuremath{\smash[b]{\ttbar{+}\text{j}_{\PQb}}}\xspace}
\newcommand{\qcd}{\ensuremath{\text{QCD}}\xspace}
\newcommand{\MadSpin}{\textsc{MadSpin}\xspace}
\newcommand{\fxfx}{\text{FxFx}\xspace}
\newcommand{\ttbbDPS}{\ensuremath{\ttbar\bbbar\,\text{DPS}}\xspace}
\newcommand{\ttHother}{\ensuremath{\ttH(\PH{\to}\text{Other})}\xspace}
\newcommand{\ttZother}{\ensuremath{\ttZ(\PZ{\to}\text{Other})}\xspace}
\newcommand{\tW}{\ensuremath{\PQt\PW}\xspace}
\newcommand{\bb}{\ensuremath{\PQb\PQb}\xspace}
\newcommand{\cc}{\ensuremath{\PQc\PQc}\xspace}
\newcommand{\mumu}{\ensuremath{\PGmp\PGmm}\xspace}
\newcommand{\ee}{\ensuremath{\Pep\Pem}\xspace}
\newcommand{\emu}{\ensuremath{\Pepm\PGmmp}\xspace}
\newcommand{\mll}{\ensuremath{m_{\Pell\Pell}}\xspace}
\newcommand{\muR}{\ensuremath{\mu_{\text{R}}}\xspace}
\newcommand{\muF}{\ensuremath{\mu_{\text{F}}}\xspace}
\newcommand{\hdamp}{\ensuremath{h_{\text{damp}}}\xspace}
\newcommand{\mtop}{\ensuremath{m_{\PQt}}\xspace}

\newcommand{\fullcorr}{\ensuremath{\checkmark}\xspace}
\newcommand{\partcorr}{\ensuremath{\sim}\xspace}
\newcommand{\nocorr}{\ensuremath{\times}\xspace}

\newcommand{\PQx}{{\HepParticle{x}{}{}}\xspace}
\newcommand{\gtogg}{\ensuremath{{\Pg{\to}\Pg\Pg}}\xspace}
\newcommand{\gtoqq}{\ensuremath{{\Pg{\to}\qqbar}}\xspace}
\newcommand{\qtoqg}{\ensuremath{{\PQq{\to}\PQq\Pg}}\xspace}
\newcommand{\xtoxg}{\ensuremath{{\PQx{\to}\PQx\Pg}}\xspace}

\newlength{\cmsFigWidthB}
\ifthenelse{\boolean{cms@external}}{\setlength{\cmsFigWidthB}{\columnwidth}}{\setlength{\cmsFigWidthB}{0.6\textwidth}}
\newlength{\cmsFigWidthC}
\ifthenelse{\boolean{cms@external}}{\setlength{\cmsFigWidthC}{\columnwidth}}{\setlength{\cmsFigWidthC}{0.6\textwidth}}

\subsection{Simulated samples}
\label{supp:samples}

Signal and background processes are modeled using Monte Carlo (MC) simulations.
The generators used for the \ttH, \ttZ, and \ttbar processes are listed in Table~\ref{tab:generatorsettings},
while those for the other background processes are summarized in Table~\ref{tab:generatorsettingsbkg}.
The decays of heavy particles in \ttW, \ttZ, and $t$-channel single top quark production are modeled using \MadSpin~\cite{Artoisenet:2012st}.
Details about the modeling parameters of the \ttbbbar sample in the four-flavor scheme (\fourFS) and of the inclusive \ttbar sample in the five-flavor scheme (\fiveFS) are provided in Table~\ref{tab:generatorsettings:ttjets}.

The \qcd multijet events are generated with \MGvATNLO at leading-order (LO) accuracy, employing the MLM scheme~\cite{Alwall:2007fs} for matrix element (ME) and parton shower (PS) matching, and are used only for training the neural network event classifier and are not included in the background estimation for the profile likelihood fits.

\begin{table*}[!ht]
\centering
\topcaption{%
    Generator settings for signal and major background samples.
    The ``Groups'' column refers to the grouping of processes in the maximum likelihood fits.
    Here, \ttHother and \ttZother refer to all Higgs boson and \PZ boson decay channels other than those to \PQb and \PQc quark pairs.
}
\label{tab:generatorsettings}
\renewcommand{\arraystretch}{1.1}
\cmsTable{\begin{scotch}{lcccc}
    Process & Groups & ME order & Generator & Notes  \\
    \hline
    \multirow{2}{*}{\ttH} & \ttHcc, \ttHbb, & \multirow{2}{*}{NLO} & \multirow{2}{*}{\POWHEG v2} &  \\
    & \ttHother & & &  \\[\cmsTabSkip]
    \multirow{2}{*}{\ttZ} & \ttZcc, \ttZbb, & \multirow{2}{*}{NLO} & \multirow{2}{*}{\MGvATNLO v2.6.1} & \multirow{2}{*}{\MadSpin for heavy particle decays} \\
    & \ttZother & & &  \\[\cmsTabSkip]
    \ttbar & \ttlight, \ttcc, \ttcj & NLO & \POWHEG v2 & \\[\cmsTabSkip]
    \ttbbbar & \ttbb, \ttbj & NLO & \POWHEGBOX & \\[\cmsTabSkip]
    \multirow{2}{*}{\ttbbDPS} & \multirow{2}{*}{\ttbb, \ttbj} & NLO / & \POWHEG v2 / & \ttbar NLO ME with \POWHEG / \\
    & & LO & \PYTHIA v8.240 & LO \bbbar DPS with \PYTHIA \\
\end{scotch}}
\end{table*}

\begin{table*}[!ht]
\centering
\topcaption{%
    Generator settings for minor background samples.
    The ``Group'' column refers to the grouping of processes in the maximum likelihood fits.
}
\label{tab:generatorsettingsbkg}
\renewcommand{\arraystretch}{1.1}
\cmsTable{\begin{scotch}{lcccc}
    Process & Group & ME order & Generator & Notes \\
    \hline
    \tW & & & \POWHEG v2 & \\
    $t$ channel & single \PQt & NLO & \POWHEG v2 & \MadSpin for heavy particle decays \\
    $s$ channel & & & \MGvATNLO v2.6.1 & \\[\cmsTabSkip]
    \multirow{2}{*}{\ttW} & \multirow{2}{*}{\ttW} & \multirow{2}{*}{NLO} & \multirow{2}{*}{\MGvATNLO v2.6.1} & \fxfx merging up to 1 additional jet \\
    & & & & \MadSpin for heavy particle decays \\[\cmsTabSkip]
    \tWZ & \tWZ & LO & \MGvATNLO v2.6.5 & \MadSpin for heavy particle decays \\[\cmsTabSkip]
    QCD & QCD & LO & \MGvATNLO v2.5.1 & MLM merging for ME-PS matching \\
\end{scotch}}
\end{table*}

\begin{table*}[!ht]
\centering
\topcaption{%
    Summary of generator settings used for the \ttbbbar and inclusive \ttbar samples. The transverse mass is defined as $\mT=\sqrt{\smash[b]{m^2 + \pt^2}}$.
}
\label{tab:generatorsettings:ttjets}
\renewcommand{\arraystretch}{1.1}
\begin{scotch}{lcc}
    & \ttbar NLO (\fiveFS) & \ttbbbar NLO (\fourFS) \\
    \hline
    ME generator & \POWHEG v2 & \POWHEGBOX \\
    Parton shower & \PYTHIA v8.230 & \PYTHIA v8.230 \\
    Flavor scheme & \fiveFS & \fourFS \\
    PDF set & \texttt{NNPDF31\_nnlo\_hessian\_pdfas} & \texttt{NNPDF31\_nnlo\_as\_0118\_nf\_4} \\
    \mtop & 172.5\GeV & 172.5\GeV \\
    $m_{\PQb}$ & 0 & 4.75\GeV \\
    \muR & $\sqrt{\tfrac{1}{2}\big(m^2_{
        rm{T},\PQt}+m^2_{\mathrm{T},\PAQt}\big)}$ & $\tfrac{1}{2}\sqrt[4]{m_{\PQt,\PQt}\,m_{\PQt,\PAQt}\,m_{\PQt,\PQb}\,m_{\PQt,\PAQb}\strut}$ \\
    \muF & \muR & $\tfrac{1}{4}\big[m_{\PQt,\PQt}+m_{\PQt,\PAQt}+m_{\PQt,\PQb}+m_{\PQt,\PAQb}+m_{\PQt,\Pg}\big]$ \\
    \hdamp & 1.379\mtop & 1.379\mtop \\
    Tune & CP5 & CP5 \\
\end{scotch}
\end{table*}

\subsection{Jet flavor identification}
\label{supp:flavor-tagging}

Jets containing \PQb or \PQc hadrons are identified using the \pNet algorithm~\cite{Qu:2019gqs}.
The \pNet algorithm outputs eight probabilities, trained to discriminate jets containing one ($p_{\PQb}$) or two ($p_{\bb}$) \PQb hadrons, one ($p_{\PQc}$) or two ($p_{\cc}$) \PQc hadrons, jets originating from \PQu, \PQd, or \PQs quarks ($p_{\PQu\PQd\PQs}$), gluons ($p_{\Pg}$), pileup, or other sources.
For this measurement, it is essential to distinguish jets originating from \PQb quarks, \PQc quarks, and those of other origins.
Therefore, two discriminants are constructed, based on the individual output probabilities of the \mbox{\pNet} tagger:
\begin{equation}\begin{aligned}
    \pBpC &= \frac{p_{\PQb}+p_{\bb}+p_{\PQc}+p_{\cc}}{p_{\PQb}+p_{\bb}+p_{\PQc}+p_{\cc}+p_{\PQu\PQd\PQs}+p_{\Pg}}\,, \\
    \pBvC &= \frac{p_{\PQb}+p_{\bb}}{p_{\PQb}+p_{\bb}+p_{\PQc}+p_{\cc}}\,.
\end{aligned}\end{equation}
The first discriminant, \pBpC, differentiates between heavy-flavor (HF) jets (\ie, of \PQb or \PQc origin) and light jets (\ie, of \PQu, \PQd, \PQs, and \Pg origin).
The second discriminant, \pBvC, further separates HF jets into \PQb and \PQc jets.
Based on these two discriminants, 11 mutually exclusive tagging categories are defined: 5 categories for \PQb tagging (\tagBin{B0}--\tagBin{B4}), 5 categories for \PQc tagging (\tagBin{C0}--\tagBin{C4}), and one untagged category (\tagBin{L0}).

In Fig.~\ref{fig:rejection}, the rejection factors, \ie, the inverse of the efficiency, for the subdominant jet flavors in each of the categories are compared to the \deepJet tagger~\cite{Bols:2020bkb} used in previous \hcc searches.
For this purpose, similar working points were derived for the \deepJet tagger, mimicking the tagging efficiencies of the dominant jet flavor in each of the tagging bins, \ie, requiring the same light-jet efficiency in \tagBin{L0}, the same \PQc jet efficiency in \tagBin{C0}--\tagBin{C4}, and the same \PQb jet efficiency in \tagBin{B1}--\tagBin{B4}. The bin \tagBin{B0} is not shown as no comparable \PQb tagging efficiency for \deepJet can be achieved for all bins at the same time due to its lower overall tagging efficiency.
In summary, the tagging layout of the \pNet algorithm improves the rejection factors of the subdominant jet flavors by factors of 1--2 in all tagging bins, compared to \deepJet.

\begin{figure*}[!ht]
\centering
\includegraphics[width=\textwidth]{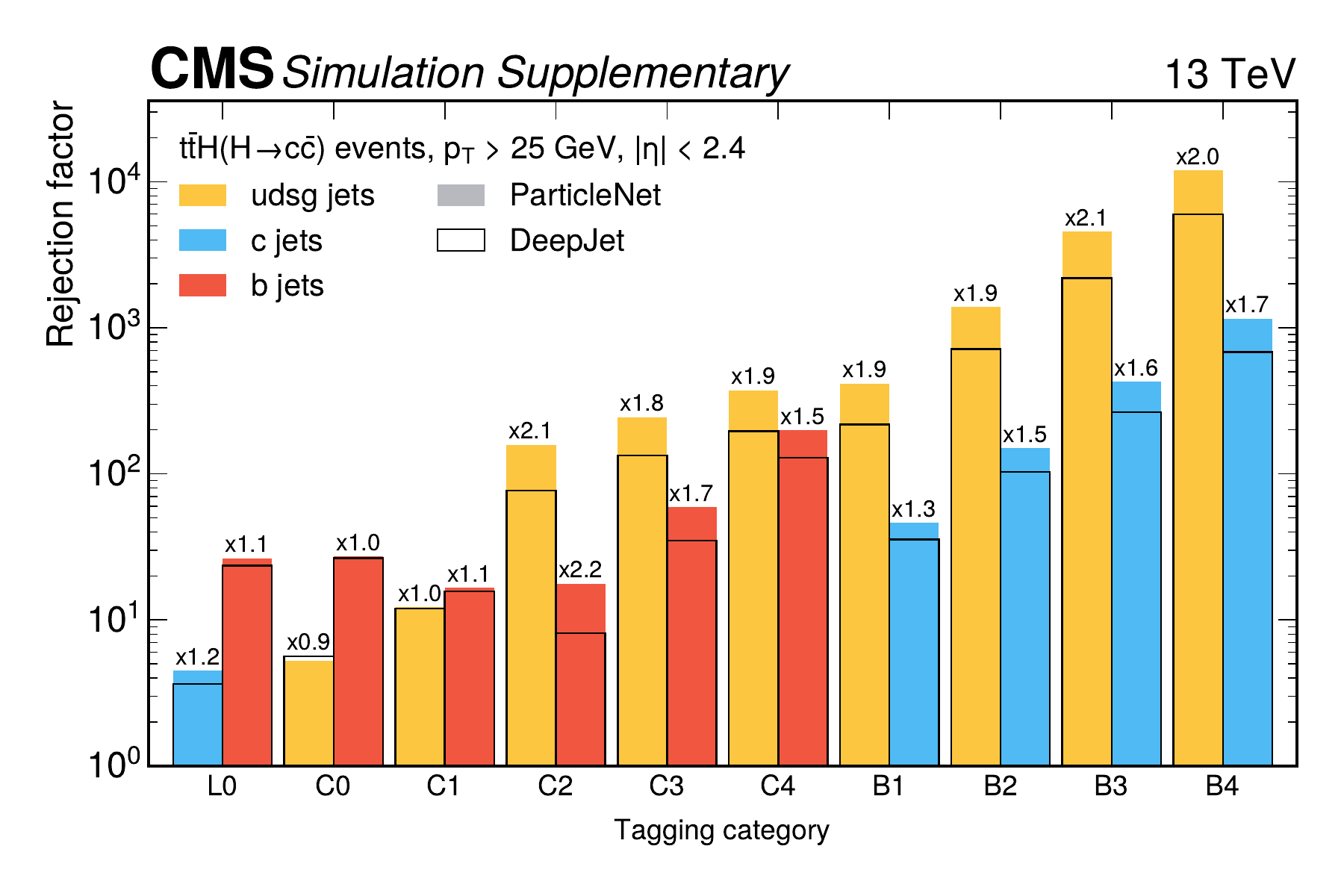}
\caption{Rejection factors for the subdominant jet flavors in each of the tagging bins. The filled bars represent the rejection factors achieved with the \pNet tagger and the corresponding working point definitions. The black bars represent the rejection factors achieved with the \deepJet tagger with working points mimicking the dominant flavor tagging efficiencies. Each bin is labeled with the relative improvement of the \pNet tagger compared to the \deepJet tagger. All rejection factors and tagging efficiencies are evaluated using simulated \ttHcc events with 2018 detector conditions.}
\label{fig:rejection}
\end{figure*}

Differences in the flavor tagging efficiencies between simulation and data are measured using three data samples, enriched each in dileptonic \ttbar, \Wjets, and \Zjets events, focusing on \PQb, \PQc, and light jets, respectively.
\begin{itemize}
\item The \ttbar region is defined by selecting events that contain an \emu pair. In addition, the transverse mass $\mT(\Pe,\PGm,\ptmiss)$ is required to be larger than 100\GeV to reduce contributions of \Zjets events.
\item For the \Wjets region, events with one isolated lepton are selected. Events further need to have exactly one jet that contains a nonisolated muon with a \pt larger than 5\GeV inside the jet cone and with a charge opposite to that of the selected isolated lepton.
The requirement of a non-isolated muon inside the jet cone enriches the selection in events with semileptonic \PQc hadron decays in the jets, and thereby $\PW{+}\PQc$ signatures.
A set of additional kinematic requirements are imposed to reduce contributions from other background processes and enrich the selection with \Wjets events.
\item Finally, for the \Zjets region, events with two opposite-charge same-flavor (\ee, \mumu) lepton candidates are selected. The leading lepton is required to have $\pt>25\GeV$, and \mll has to be within the range of 81{--}101\GeV. Exactly one additional jet in the event that balances the reconstructed \PZ boson \pt is required.
\end{itemize}
For all regions, the jets are selected using the same criteria as used in the present \ttHcc analysis.

Simulation-to-data scale factors (SFs) are then derived as a function of jet \pt with the following procedure.
Tagging SFs are derived by simultaneously fitting all three flavor-enriched regions (\Zjets, \Wjets, \ttbar) in a profile likelihood fit.
The tagging SFs are discretized into the eleven aforementioned tagging bins \tagBin{L0}, \tagBin{C0}--\tagBin{C4}, and \tagBin{B0}--\tagBin{B4}.
For each of the ten exclusive \tagBin{C0}--\tagBin{C4} and \tagBin{B0}--\tagBin{B4} bins, one parameter of interest (POI) is defined per jet flavor (\ie, in total 30 independent POIs per jet \pt bin), which corresponds to the jet tagging scale factor for the exclusive tagging bin and jet flavor.
In order to preserve unitarity, the SFs for the tagger bin \tagBin{L0} are not associated with an independent POI but are constructed such that the sum of tagging probabilities per jet flavor is retained, \ie, requiring
\ifthenelse{\boolean{cms@external}}{\begin{multline}
    \sum_{\text{WP}\in[\tagBin{L0},\tagBin{C0}-\tagBin{C4},\tagBin{B0}-\tagBin{B4}]} \epsilon_{\text{MC}}\big(\text{WP}\big|\text{flav}\big) \\
    =\sum_{\text{WP}}\epsilon_{\text{data}}\big(\text{WP}\big|\text{flav}\big)\equiv1.
\end{multline}}{\begin{equation}
    \sum_{\text{WP}\in[\tagBin{L0},\tagBin{C0}-\tagBin{C4},\tagBin{B0}-\tagBin{B4}]} \epsilon_{\text{MC}}\big(\text{WP}\big|\text{flav}\big)=\sum_{\text{WP}}\epsilon_{\text{data}}\big(\text{WP}\big|\text{flav}\big)\equiv1.
\end{equation}}
This constraint guarantees that the overall normalization of each type (\PQb, \PQc, and light) component is retained in the fits, and the resulting SFs correct for only the efficiencies but do not alter the normalizations. This helps preserve the normalization of each flavor component when the SFs are applied to a different phase space region.
To be insensitive to the overall normalization of the total background prediction as obtained by the simulation and to fit only the relative tagging efficiencies, the overall expected yields are scaled to match the observed data in the scale factor derivation regions, inclusive in jet flavor.
The fits are performed in six independent jet \pt ranges, to introduce a \pt dependence on the tagging SFs:
\begin{equation}
    \pt(\text{jet})\in[25,35,50,70,90,120,\infty]\GeV.
\end{equation}
The normalization and derivation of SFs is completely independent for each of the \pt ranges.

In some cases, where the contribution of a certain jet flavor to a certain tagging bin is very small, the SFs cannot be determined well.
For example, in the high \PQb tagging bins \tagBin{B3} or \tagBin{B4}, the contamination of \PQc and LF jets is almost negligible.
In cases where the fits for these parameters do not lead to a well-converged result, \ie, where the expected uncertainty in the SF exceeds 50\%, the SF value is set to unity in the fit to data, and the corresponding SF is assigned an uncertainty covering the range $[0.3,3]$.
Changes in these SFs have negligible impact on the analysis results, as the contamination of jets of the affected flavors in the corresponding tagging bins is negligible.

In the fit to data, a range of experimental and modeling uncertainties are considered.
The same set of experimental uncertainties as described in Section~\ref{sec:syst:experimental} is considered for the calibration measurement.
Modeling uncertainties due to renormalization scale \muR and factorization scale \muF choices, and uncertainties due to initial-state and final-state radiation scales (ISR and FSR) are applied, following the same procedures as described in Section~\ref{sec:syst:modelling}. These uncertainties are treated as uncorrelated between major processes in the SF derivation regions (\ttbar, \Wjets, \Zjets, \etc).
Normalization uncertainties are included to account for the limited knowledge of the cross sections of the relevant processes. A 5\% normalization uncertainty is applied to the minor \ttbar background in the \Wjets and \Zjets regions.
In the \ttbar channel, the \ttbar normalization is fitted to data.
A 10\% normalization uncertainty is applied for \tW production, and 15\% for single top $t$-channel production.
Diboson production is associated with a 15\% normalization uncertainty.
For the production of a \PW or \PZ boson with HF jets, dedicated normalization uncertainties are applied, as the modeling of these processes is known to be insufficient. An 8\% normalization uncertainty is applied to $\PW{+}\PQc$ production. The $\PW{+}\PQb$, $\PZ{+}\PQb$, and $\PZ{+}\PQc$ processes are each associated with a free parameter in the fit to determine the normalization in-situ.

The impacts of systematic uncertainties in the final tagging SFs are derived from the impacts of the corresponding nuisance parameters.
Variations of a systematic source are considered fully correlated between all SFs, \pt ranges, and years when applied in the \ttHcc analysis.
The statistical uncertainties are derived by freezing all nuisance parameters to their best fit values and refitting the tagging SFs. The resulting uncertainty in the SFs is considered the statistical uncertainty.
In cases where the SFs could not be determined, \ie, cases where the SFs are fixed to a value of 1 and the uncertainties are inflated to the range $[0.3,3]$, the full uncertainty is treated as a statistical uncertainty.
Statistical uncertainties of different jet flavors, tagging bins, and years are all treated as uncorrelated parameters when applied in the \ttHcc analysis.

The large majority of SFs for \PQb, \PQc, and light jets are within the range $[0.75, 1.5]$, with only few SFs outside this range.
Uncertainties in the \PQb jet SFs in the \PQb jet enriched regions are generally small (of order 1--5\%), and show continuous trends (\ie, SF values appear smooth for neighboring bins). Similarly, \PQb jet SFs are well described in the \PQc jet enriched bins \tagBin{C0}--\tagBin{C4}, but have somewhat larger uncertainties (of order 50\%) in the \tagBin{L0} bin.
SFs for \PQc jets converge mostly well in \PQc jet and light jet enriched bins. In higher \PQb jet tagging bins (\tagBin{B1}--\tagBin{B4}), \PQc jet SFs often cannot be constrained in the fit and are assigned a default value of one with large uncertainties.
Similarly, light jet SFs cannot be well constrained in most \PQc and \PQb jet enriched bins, because of the low light-jet mistagging efficiencies.

\subsection{Baseline event selection}
\label{supp:baseline-selection}

The baseline event selection of this analysis is summarized in Table~\ref{tab:baselineselection}.
Corrections are applied to the simulated samples to account for differences in the jet energy scale and resolution, lepton selection efficiencies, and trigger efficiencies.
Corrections for the variations in jet flavor identification efficiencies, discussed above, are also applied.
Different selection criteria for different data-taking periods are motivated by changes in the trigger selection.

\begin{table*}[!ht]
\centering
\topcaption{Baseline selection criteria in the \FH, \SL, and \DL channels. Where the selection criteria differ per year, they are quoted as 2016/2017/2018.}
\label{tab:baselineselection}
\cmsTable{\begin{scotch}{lccc}
    Selection criteria & \FH channel & \SL channel & \DL channel \\
    \hline
    Number of leptons & 0 & 1 & 2 \\
    Sign and flavor of leptons & \NA & $\Pepm$, $\PGmpm$ & \EE, $\PGmpm\Pemp$, \MM \\[\cmsTabSkip]
    Min.\ \pt of \pt-leading electron [{\GeVns}] & \NA & 29/30/30 & 25 \\
    Electron identification & \NA & Tight (80\% efficiency) & Loose (90\% efficiency) \\
    Electron isolation & \NA & Tight (80\% efficiency) & Loose (90\% efficiency) \\[\cmsTabSkip]

    Min.\ \pt of \pt-leading muon [{\GeVns}] & \NA & 26/29/26 & 25 \\
    Max.\ muon relative isolation & \NA & 0.15 & 0.25 \\[\cmsTabSkip]
    Min.\ $m_{\Pe\Pe/\PGm\PGm}$ [{\GeVns}] & \NA & \NA & 20 \\
    Range of $m_{\Pe\Pe/\PGm\PGm}$ [{\GeVns}] & \NA & \NA & $<$76, $>$106 \\[\cmsTabSkip]
    Min.\ \pt of jets [{\GeVns}] & 25 & 25 & 25 \\
    Min.\ \pt of 6\textsuperscript{th} jet [{\GeVns}] & 40 & \NA & \NA \\
    Max.\ \abseta of jets & 2.4 & 2.4 & 2.4 \\
    Min.\ number of jets & 7 & 5 & 4 \\
    Min.\ number of \PQb-tagged jets & 1 & 1 & 1 \\
    Min.\ number of HF (\PQb- or \PQc-tagged) jets & 3 & 3 & 3 \\[\cmsTabSkip]

    Min.\ \HT [{\GeVns}] & 500 & \NA & \NA \\
    Min.\ \ptmiss [{\GeVns}] & \NA & 20 & 20 \\
\end{scotch}}
\end{table*}

\subsection{Neural network event classifier}
\label{supp:event-classifier}

The performance of the \parT event classifier is summarized via the confusion matrices shown in Fig.~\ref{fig:confusion_matrix_baseline}.

\begin{figure*}[!htp]
\centering
\includegraphics[width=0.48\textwidth]{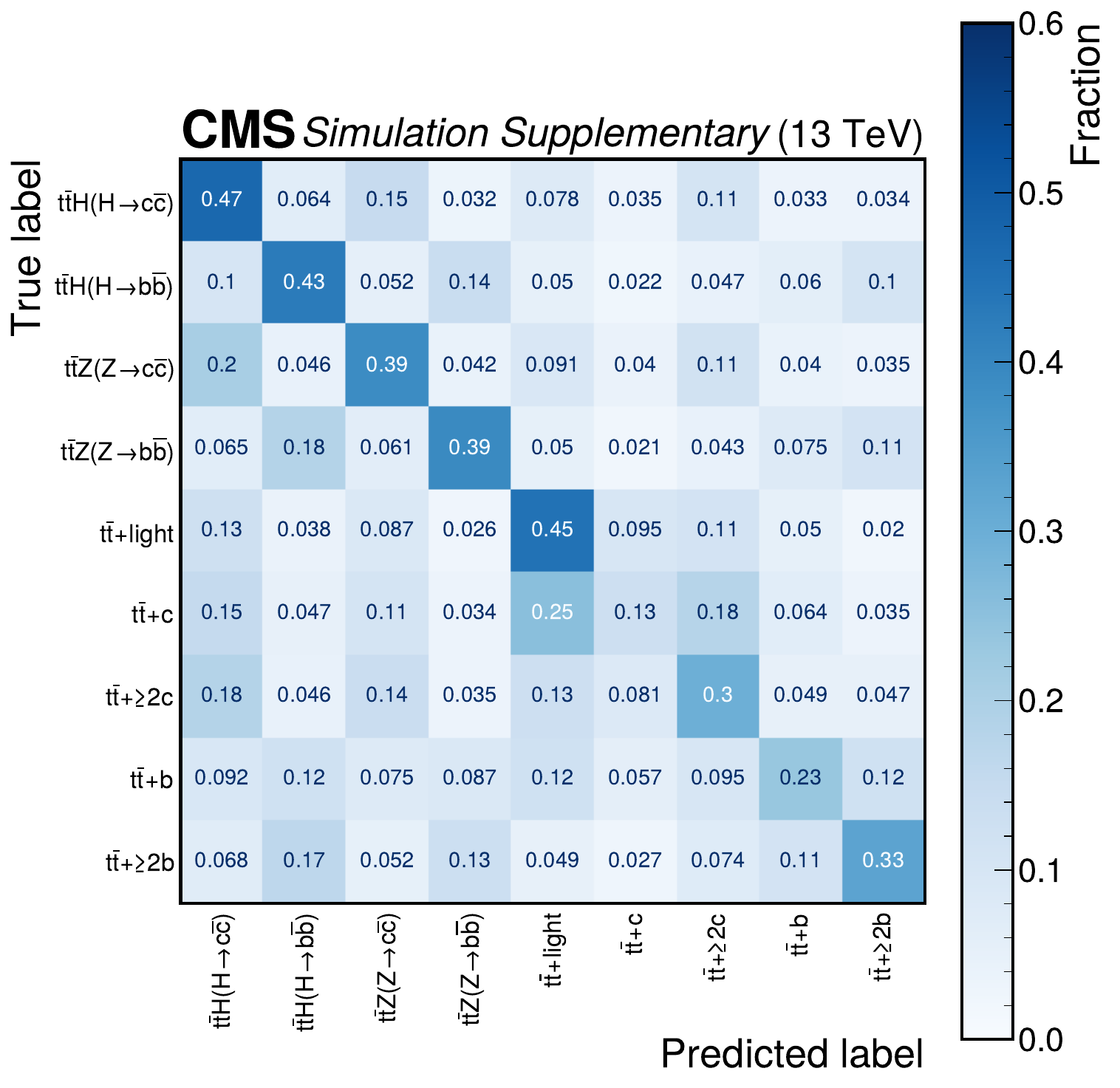} \\
\includegraphics[width=0.48\textwidth]{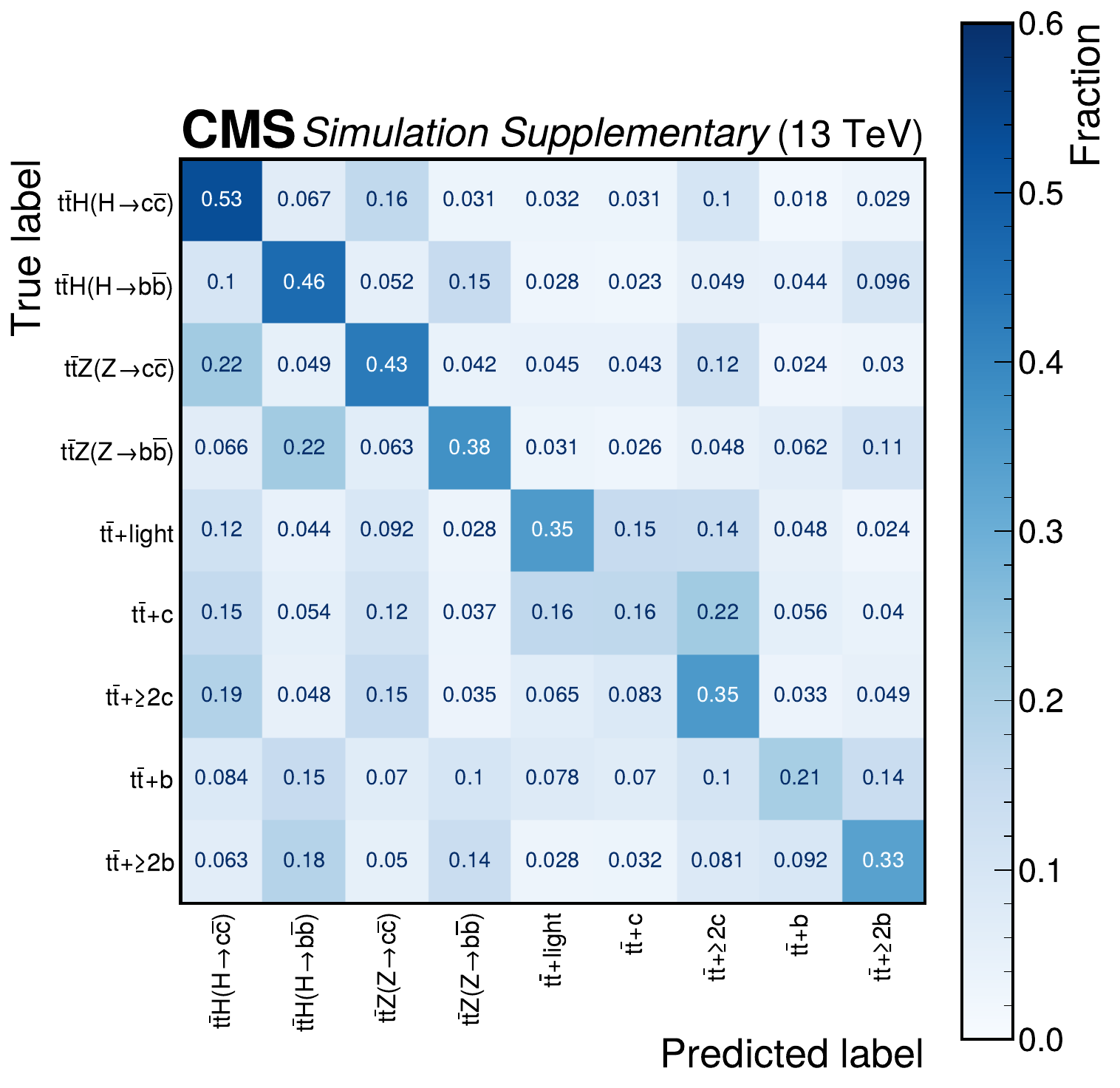}%
\hfill%
\includegraphics[width=0.48\textwidth]{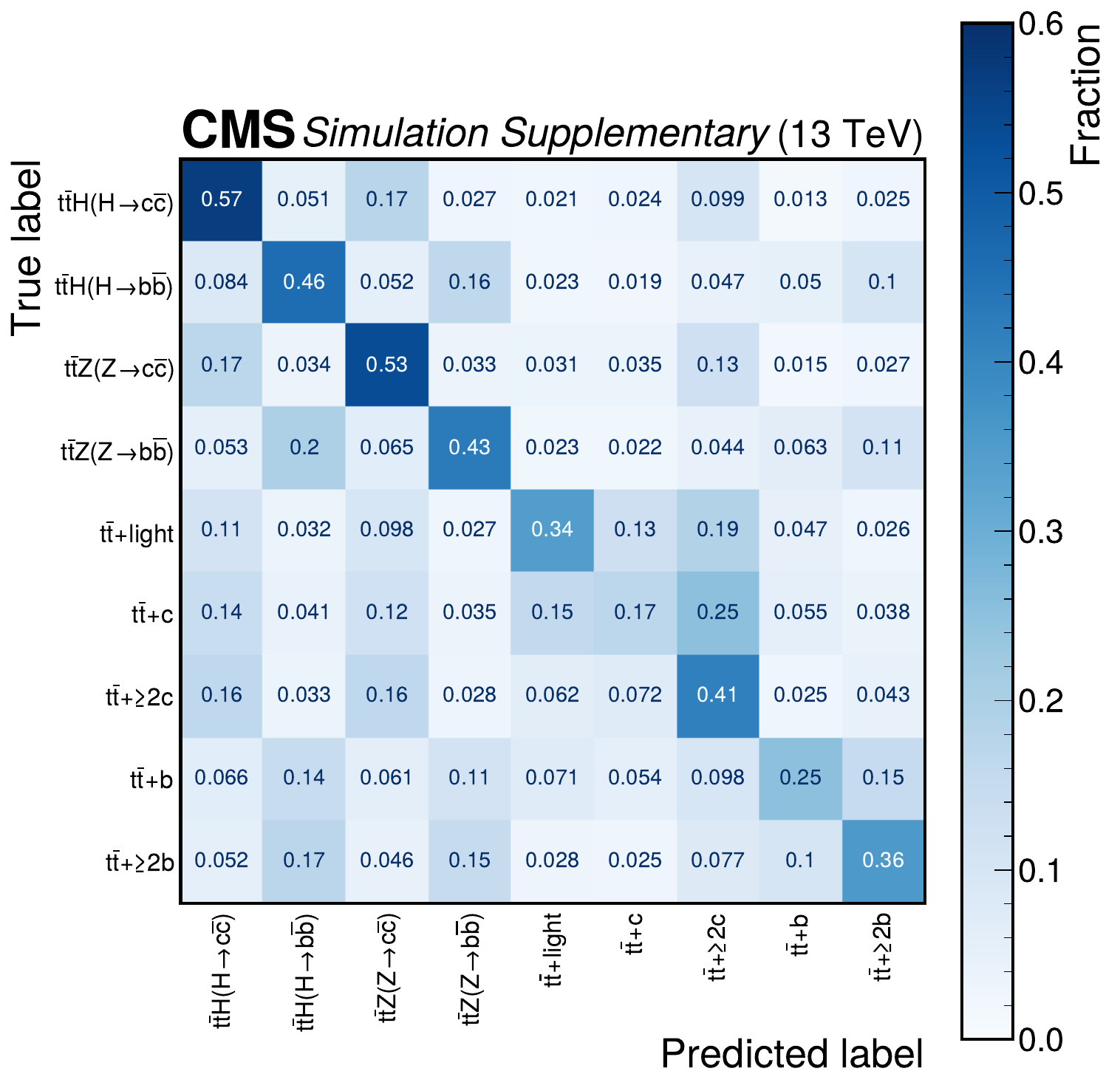}
\caption{Confusion matrices of the \parT event classifier in the \FH (upper), \SL (lower left), and \DL (lower right) channels after the baseline selection. For each event, the predicted label is the process with the highest output discriminant. The event yield fractions are normalized per true label such that each row sums up to unity.}
\label{fig:confusion_matrix_baseline}
\end{figure*}

\begin{table}[!htp]
\centering
\topcaption{Definition of the \parT event classifier discriminant for each category in the maximum likelihood fit.}
\label{tab:fit-variables}
\renewcommand{\arraystretch}{1.5}
\begin{scotch}{lc}
    Category & Discriminant \\
    \hline
    \hcc SR & \score{\ttHcc} \\
    \hbb SR & \score{\ttHbb} \\
    \zcc SR & \score{\ttZcc} \\
    \zbb SR & \score{\ttZbb} \\
    \ttlight CR & $\score{\ttlight}/\score{\ttjets}$ \\
    \ttcj CR & $\score{\ttcj}/\score{\ttjets}$ \\
    \ttcc CR & $\score{\ttcc}/\score{\ttjets}$ \\
    \ttbj CR & $\score{\ttbj}/\score{\ttjets}$ \\
    \ttbb CR & $\score{\ttbb}/\score{\ttjets}$ \\
\end{scotch}
\end{table}

The fitted variable is the \parT event classifier discriminant for the corresponding category as summarized in Table~\ref{tab:fit-variables}.
Several selections on the classifier discriminants define the analysis and validation regions (AR and VR), summarized in Fig.~\ifthenelse{\boolean{cms@external}}{5 of the paper}{\ref{fig:event-categorization}}.
In the \FH channel the classifier discriminant for the QCD multijet is used to remove the multijet background.
The discriminant and the corresponding cut are shown in Fig.~\ref{fig:scores:qcd}.
In all three channels, a cut on \score{\ttlight} is applied to reduce the overwhelming background from \ttlight production, shown in Fig.~\ref{fig:scores:ttlf}.
After these selections, the AR and VR, and the SRs and CRs are defined via requirements on \score{\ttX}, summarized in Fig.~\ref{fig:scores:ttX}.

The event yields in the AR after the fit to data are summarized in Tables~\ref{tab:yields:DL}--\ref{tab:yields:FH} for the three individual channels, and in Tables~\ref{tab:yields:total}--\ref{tab:yields:channels} for the full analysis.

\begin{figure}[!htp]
\centering
\includegraphics[width=\cmsFigWidth]{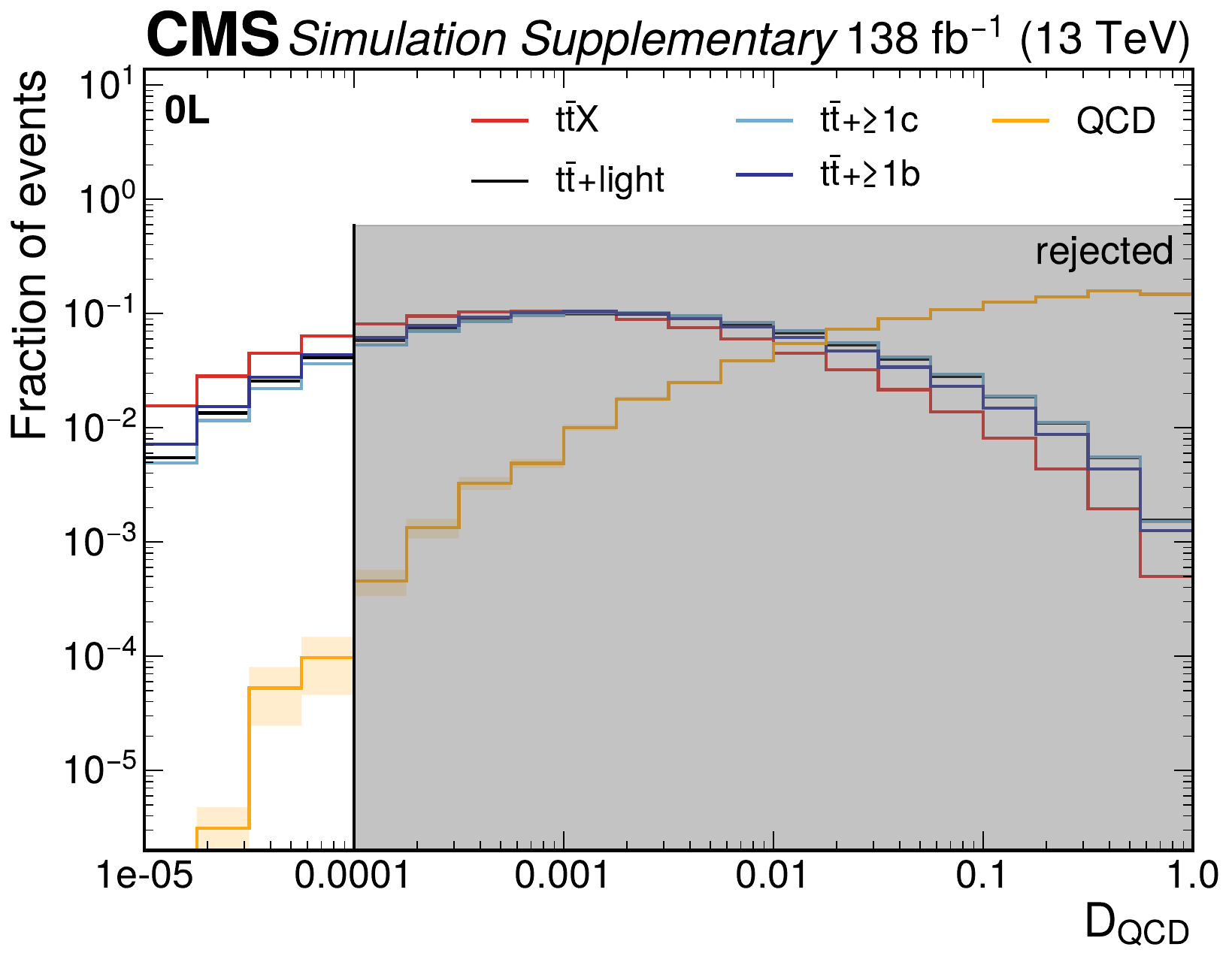}
\caption{%
    Distribution of the \parT \score{\qcd} discriminant used in the \FH channel to remove the \qcd background.
    The gray area indicates the region that is rejected in the analysis.
    The shaded band indicates the uncertainty in the \qcd prediction due to limited size of simulated \qcd multijet samples.
    All contributions are normalized to unity.
}
\label{fig:scores:qcd}
\end{figure}

\begin{figure}[!htp]
\centering
\includegraphics[width=\cmsFigWidth]{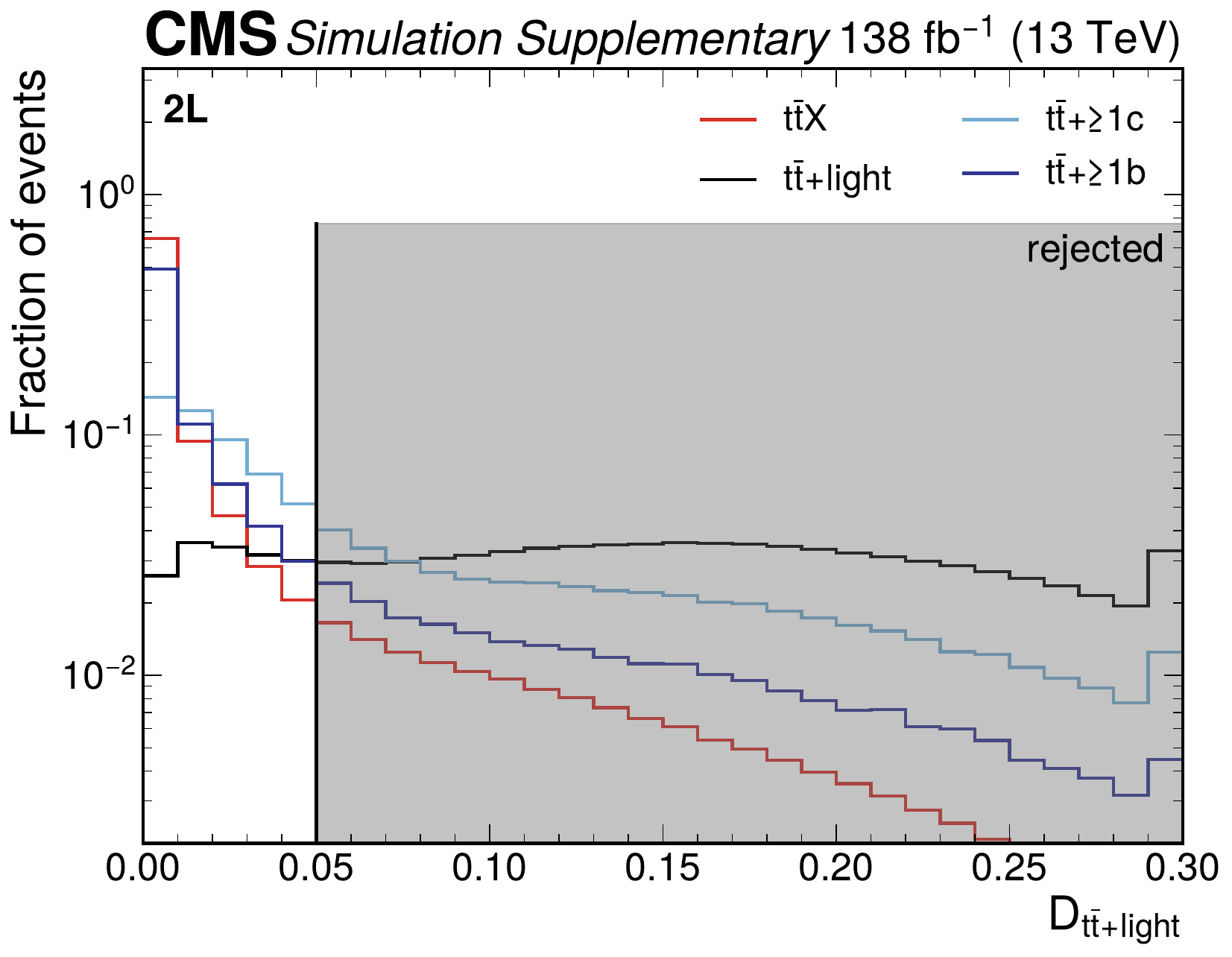}
\caption{%
    Distribution of the \parT \score{\ttlight} discriminant used to reduce the \ttlight background in the \DL channel.
    The gray area indicates the region that is rejected in the analysis.
    All contributions are normalized to unity.
    The last bin includes the overflow.
}
\label{fig:scores:ttlf}
\end{figure}

\begin{figure}[!htp]
\centering
\includegraphics[width=\cmsFigWidth]{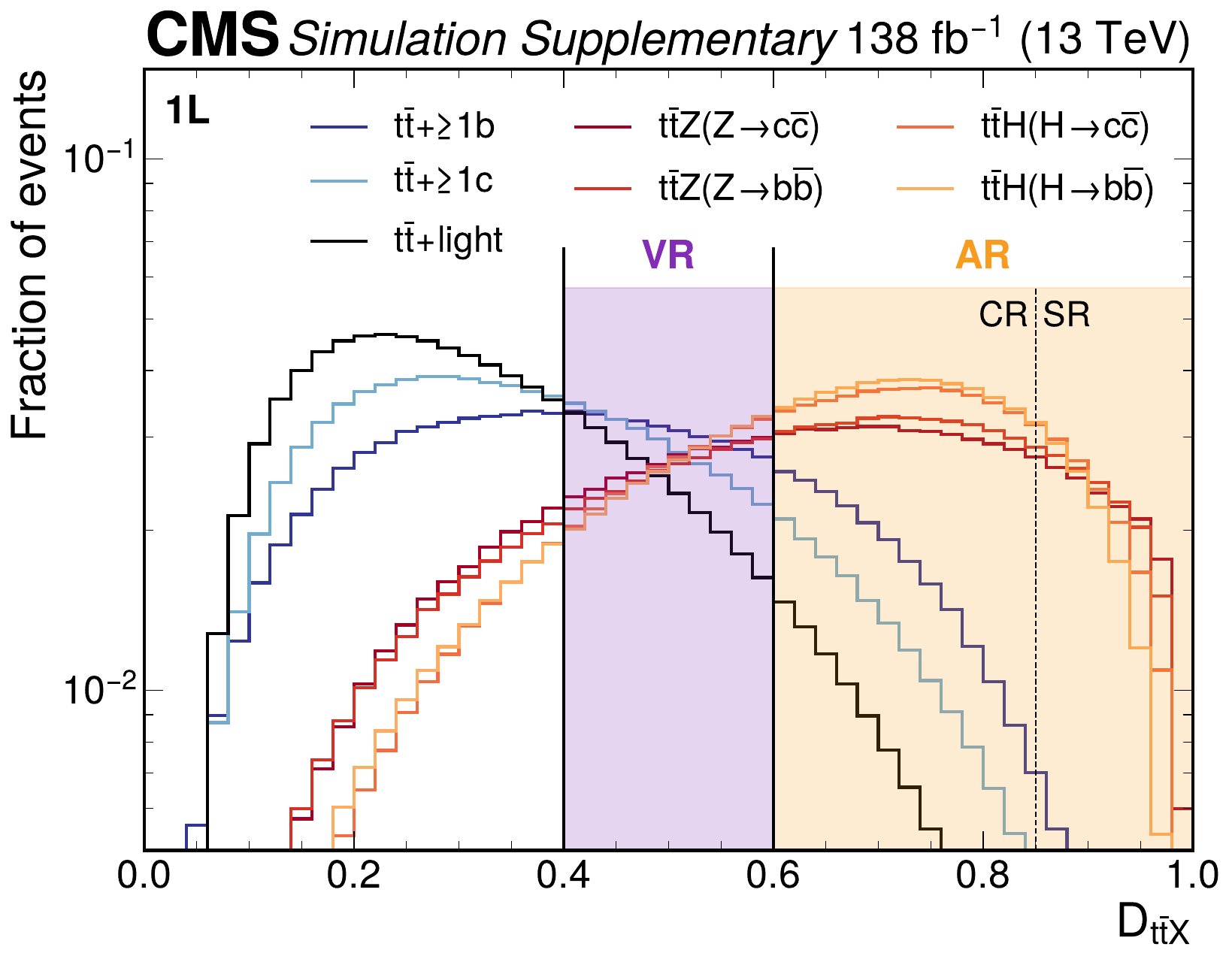}
\caption{%
    Distribution of the \parT \score{\ttX} discriminant used to define the different regions in the \SL channel.
    The purple (yellow) area indicates the region that is used for the validation (analysis).
    The dashed line indicates the separation of SRs and CRs in the analysis.
    All contributions are normalized to unity.
}
\label{fig:scores:ttX}
\end{figure}

\begin{table*}[!p]
\centering
\topcaption{%
    Event yields in the \DL channel.
    Values in brackets correspond to the pre-fit expectations.
}
\label{tab:yields:DL}
\renewcommand{\arraystretch}{1.1}
\cmsWideTable{\scriptsize\begin{scotch}{l*{9}{r@{\,$\pm$\,}l}}
    \multirow{2}{*}{Process} & \multicolumn{2}{c}{SR} & \multicolumn{2}{c}{SR} & \multicolumn{2}{c}{SR} & \multicolumn{2}{c}{SR} & \multicolumn{2}{c}{CR} & \multicolumn{2}{c}{CR} & \multicolumn{2}{c}{CR} & \multicolumn{2}{c}{CR} & \multicolumn{2}{c}{CR} \\
    & \multicolumn{2}{c}{\ttHcc} & \multicolumn{2}{c}{\ttHbb} & \multicolumn{2}{c}{\ttZcc} & \multicolumn{2}{c}{\ttZbb} & \multicolumn{2}{c}{\ttbb} & \multicolumn{2}{c}{\ttbj} & \multicolumn{2}{c}{\ttcc} & \multicolumn{2}{c}{\ttcj} & \multicolumn{2}{c}{\ttlight} \\
    \hline
    \multirow{2}{*}{\ttHcc} & $-$2.5 & 4.4 & $-$0.06 & 0.62 & $-$0.5 & 1.8 & $-$0.02 & 0.35 & $-$0.13 & 0.83 & $-$0.04 & 0.51 & $-$1.2 & 2.8 & $-$0.4 & 1.5 & $-$4.4 & 5.2 \\
    & (1.5 & 2.8) & (0.03 & 0.37) & (0.3 & 1.1) & (0.01 & 0.21) & (0.08 & 0.50) & (0.02 & 0.31) & (0.7 & 1.7) & (0.24 & 0.97) & (2.7 & 3.2) \\[\cmsTabSkip]
    \multirow{2}{*}{\ttHbb} & 2.4 & 3.1 & 20.2 & 9.6 & 1.3 & 2.1 & 6.4 & 4.7 & 37 & 12 & 18.6 & 8.1 & 4.1 & 3.4 & 1.8 & 2.4 & 45 & 11 \\
    & (2.7 & 3.4) & (22 & 11) & (1.4 & 2.2) & (7.1 & 5.2) & (40 & 13) & (20.8 & 9.0) & (4.4 & 3.6) & (2.1 & 2.6) & (50 & 13) \\[\cmsTabSkip]
    \ttZcc & 1.9 & 2.7 & 0.11 & 0.64 & 9.9 & 7.1 & 0.21 & 0.90 & 0.4 & 1.1 & 0.12 & 0.71 & 4.3 & 4.2 & 1.2 & 2.2 & 14.2 & 7.4 \\
    \ttZbb & 0.7 & 2.1 & 3.8 & 4.7 & 1.1 & 2.7 & 16 & 11 & 19 & 11 & 9.1 & 7.2 & 2.1 & 3.1 & 0.9 & 2.2 & 23 & 10 \\[\cmsTabSkip]
    \ttbb & 12.7 & 7.9 & 45 & 16 & 11.8 & 7.4 & 52 & 17 & 370 & 42 & 121 & 24 & 52 & 15 & 14.3 & 8.0 & 385 & 40 \\
    \ttbj & 19 & 14 & 14 & 13 & 16 & 12 & 17 & 14 & 120 & 32 & 303 & 56 & 54 & 22 & 37 & 19 & 727 & 81 \\
    \ttcc & 67 & 21 & 1.9 & 3.7 & 63 & 20 & 2.4 & 3.3 & 16.0 & 7.7 & 5.5 & 5.6 & 214 & 35 & 44 & 15 & 524 & 52 \\
    \ttcj & 61 & 28 & 1.7 & 4.0 & 49 & 24 & 3.4 & 6.8 & 16 & 11 & 23 & 16 & 121 & 37 & 166 & 43 & 1290 & 120 \\
    \ttlight & 50 & 14 & 1.7 & 2.7 & 44 & 12 & 1.8 & 2.7 & 8.1 & 4.7 & 11.2 & 6.1 & 60 & 14 & 54 & 13 & 1470 & 74 \\[\cmsTabSkip]
    Other & 15 & 18 & 2.8 & 7.3 & 24 & 24 & 5.4 & 9.7 & 22 & 17 & 8.5 & 8.1 & 32 & 25 & 13 & 14 & 313 & 75 \\[\cmsTabSkip]
    Total & 228 & 14 & 90.7 & 7.9 & 220 & 14 & 103.6 & 9.1 & 608 & 22 & 499 & 19 & 541 & 27 & 332 & 18 & 4790 & 110 \\[\cmsTabSkip]
    Data & \multicolumn{2}{c}{236} & \multicolumn{2}{c}{93} & \multicolumn{2}{c}{204} & \multicolumn{2}{c}{92} & \multicolumn{2}{c}{605} & \multicolumn{2}{c}{463} & \multicolumn{2}{c}{546} & \multicolumn{2}{c}{343} & \multicolumn{2}{c}{4784} \\
\end{scotch}}
\end{table*}

\begin{table*}[!p]
\centering
\topcaption{%
    Event yields in the \SL channel.
    Values in brackets correspond to the pre-fit expectations.
}
\label{tab:yields:SL}
\renewcommand{\arraystretch}{1.1}
\cmsWideTable{\begin{scotch}{l*{9}{r@{\,$\pm$\,}l}}
    \multirow{2}{*}{Process} & \multicolumn{2}{c}{SR} & \multicolumn{2}{c}{SR} & \multicolumn{2}{c}{SR} & \multicolumn{2}{c}{SR} & \multicolumn{2}{c}{CR} & \multicolumn{2}{c}{CR} & \multicolumn{2}{c}{CR} & \multicolumn{2}{c}{CR} & \multicolumn{2}{c}{CR} \\
    & \multicolumn{2}{c}{\ttHcc} & \multicolumn{2}{c}{\ttHbb} & \multicolumn{2}{c}{\ttZcc} & \multicolumn{2}{c}{\ttZbb} & \multicolumn{2}{c}{\ttbb} & \multicolumn{2}{c}{\ttbj} & \multicolumn{2}{c}{\ttcc} & \multicolumn{2}{c}{\ttcj} & \multicolumn{2}{c}{\ttlight} \\
    \hline
    \multirow{2}{*}{\ttHcc} & $-$16 & 22 & $-$0.5 & 3.6 & $-$3.1 & 8.9 & $-$0.2 & 1.9 & $-$1.2 & 2.6 & $-$0.5 & 1.8 & $-$11.7 & 8.8 & $-$5.4 & 5.9 & $-$24 & 13 \\
    & (10 & 14) & (0.3 & 2.1) & (1.9 & 5.4) & (0.1 & 1.1) & (0.7 & 1.5) & (0.3 & 1.0) & (7.0 & 5.3) & (3.3 & 3.6) & (14.8 & 7.7) \\[\cmsTabSkip]
    \multirow{2}{*}{\ttHbb} & 21 & 18 & 166 & 55 & 9 & 11 & 43 & 25 & 264 & 31 & 139 & 22 & 52 & 12 & 35 & 10 & 327 & 32 \\
    & (23 & 19) & (181 & 60) & (10 & 12) & (47 & 27) & (284 & 34) & (153 & 25) & (55 & 13) & (38 & 11) & (356 & 35) \\[\cmsTabSkip]
    \ttZcc & 16 & 16 & 1.1 & 4.1 & 59 & 35 & 1.5 & 4.9 & 3.5 & 3.6 & 1.6 & 2.5 & 37 & 12 & 18.1 & 8.6 & 78 & 18 \\
    \ttZbb & 6 & 12 & 32 & 28 & 8 & 14 & 114 & 58 & 136 & 28 & 70 & 20 & 25 & 11 & 19.2 & 9.7 & 166 & 29 \\[\cmsTabSkip]
    \ttbb & 109 & 48 & 387 & 91 & 85 & 41 & 360 & 88 & 2760 & 120 & 957 & 67 & 633 & 51 & 293 & 34 & 2830 & 110 \\
    \ttbj & 198 & 96 & 205 & 97 & 168 & 88 & 165 & 87 & 1300 & 110 & 2700 & 170 & 841 & 88 & 719 & 81 & 5760 & 240 \\
    \ttcc & 470 & 110 & 26 & 24 & 389 & 99 & 19 & 21 & 153 & 26 & 60 & 18 & 2080 & 110 & 718 & 63 & 3400 & 140 \\
    \ttcj & 700 & 190 & 61 & 54 & 470 & 150 & 45 & 46 & 256 & 48 & 291 & 59 & 1910 & 150 & 2680 & 180 & 9220 & 330 \\
    \ttlight & 780 & 110 & 73 & 33 & 605 & 94 & 69 & 33 & 205 & 24 & 226 & 29 & 1180 & 65 & 1326 & 70 & 12620 & 220 \\[\cmsTabSkip]
    Other & 190 & 120 & 43 & 48 & 320 & 160 & 58 & 55 & 249 & 50 & 136 & 33 & 550 & 86 & 423 & 70 & 3080 & 190 \\[\cmsTabSkip]
    Total & 2470 & 92 & 993 & 52 & 2115 & 80 & 875 & 48 & 5330 & 110 & 4580 & 110 & 7300 & 200 & 6230 & 190 & 37450 & 530 \\[\cmsTabSkip]
    Data & \multicolumn{2}{c}{2411} & \multicolumn{2}{c}{983} & \multicolumn{2}{c}{2108} & \multicolumn{2}{c}{879} & \multicolumn{2}{c}{5372} & \multicolumn{2}{c}{4594} & \multicolumn{2}{c}{7317} & \multicolumn{2}{c}{6209} & \multicolumn{2}{c}{37497} \\
\end{scotch}}
\end{table*}

\begin{table*}[!p]
\centering
\topcaption{%
    Event yields in the \FH channel.
    Values in brackets correspond to the pre-fit expectations.
}
\label{tab:yields:FH}
\renewcommand{\arraystretch}{1.1}
\cmsWideTable{\begin{scotch}{l*{9}{r@{\,$\pm$\,}l}}
    \multirow{2}{*}{Process} & \multicolumn{2}{c}{SR} & \multicolumn{2}{c}{SR} & \multicolumn{2}{c}{SR} & \multicolumn{2}{c}{SR} & \multicolumn{2}{c}{CR} & \multicolumn{2}{c}{CR} & \multicolumn{2}{c}{CR} & \multicolumn{2}{c}{CR} & \multicolumn{2}{c}{CR} \\
    & \multicolumn{2}{c}{\ttHcc} & \multicolumn{2}{c}{\ttHbb} & \multicolumn{2}{c}{\ttZcc} & \multicolumn{2}{c}{\ttZbb} & \multicolumn{2}{c}{\ttbb} & \multicolumn{2}{c}{\ttbj} & \multicolumn{2}{c}{\ttcc} & \multicolumn{2}{c}{\ttcj} & \multicolumn{2}{c}{\ttlight} \\
    \hline
    \multirow{2}{*}{\ttHcc} & $-$9 & 13 & $-$0.4 & 2.5 & $-$1.6 & 5.0 & $-$0.1 & 1.4 & $-$0.7 & 1.9 & $-$0.3 & 1.4 & $-$3.4 & 4.7 & $-$0.5 & 1.9 & $-$17 & 10 \\
    & (5.5 & 8.1) & (0.2 & 1.4) & (1.0 & 3.0) & (0.08 & 0.79) & (0.4 & 1.1) & (0.17 & 0.79) & (2.0 & 2.8) & (0.3 & 1.1) & (10.2 & 6.2) \\[\cmsTabSkip]
    \multirow{2}{*}{\ttHbb} & 12 & 10 & 128 & 38 & 5.4 & 6.5 & 31 & 16 & 140 & 23 & 62 & 15 & 14.9 & 6.5 & 2.9 & 2.8 & 185 & 23 \\
    & (13 & 11) & (137 & 41) & (5.9 & 7.1) & (34 & 17) & (151 & 24) & (68 & 16) & (15.8 & 6.9) & (3.1 & 3.0) & (201 & 25) \\[\cmsTabSkip]
    \ttZcc & 8.0 & 9.0 & 0.8 & 2.9 & 36 & 21 & 1.3 & 3.6 & 1.9 & 2.6 & 1.0 & 2.0 & 10.4 & 6.7 & 2.0 & 2.9 & 51 & 14 \\
    \ttZbb & 3.1 & 6.5 & 22 & 18 & 4.7 & 8.3 & 87 & 40 & 62 & 19 & 28 & 13 & 6.2 & 5.4 & 1.5 & 2.7 & 75 & 19 \\[\cmsTabSkip]
    \ttbb & 73 & 32 & 336 & 73 & 55 & 28 & 320 & 70 & 1548 & 97 & 452 & 50 & 178 & 30 & 28 & 11 & 1677 & 89 \\
    \ttbj & 86 & 47 & 96 & 49 & 70 & 43 & 88 & 50 & 405 & 60 & 742 & 86 & 159 & 37 & 45 & 20 & 2040 & 130 \\
    \ttcc & 296 & 81 & 22 & 21 & 260 & 75 & 20 & 19 & 74 & 22 & 30 & 15 & 653 & 72 & 85 & 25 & 2410 & 130 \\
    \ttcj & 163 & 59 & 21 & 20 & 113 & 48 & 16 & 17 & 49 & 18 & 44 & 19 & 180 & 39 & 113 & 30 & 2520 & 140 \\
    \ttlight & 213 & 42 & 27 & 15 & 165 & 36 & 21 & 13 & 44 & 11 & 43 & 12 & 122 & 20 & 50 & 13 & 4310 & 120 \\[\cmsTabSkip]
    Other & 39 & 40 & 13 & 20 & 106 & 67 & 18 & 23 & 32 & 19 & 24 & 15 & 52 & 29 & 17 & 16 & 565 & 88 \\[\cmsTabSkip]
    Total & 883 & 50 & 665 & 42 & 814 & 47 & 602 & 40 & 2357 & 74 & 1426 & 63 & 1372 & 64 & 343 & 26 & 13800 & 310 \\[\cmsTabSkip]
    Data & \multicolumn{2}{c}{898} & \multicolumn{2}{c}{668} & \multicolumn{2}{c}{831} & \multicolumn{2}{c}{622} & \multicolumn{2}{c}{2345} & \multicolumn{2}{c}{1435} & \multicolumn{2}{c}{1382} & \multicolumn{2}{c}{357} & \multicolumn{2}{c}{13733} \\
\end{scotch}}
\end{table*}

\begin{table*}[!p]
\centering
\topcaption{%
    Event yields in the full analysis, separated in the SRs and CRs.
    Values in brackets correspond to the pre-fit expectations.
}
\label{tab:yields:total}
\renewcommand{\arraystretch}{1.1}
\cmsWideTable{\begin{scotch}{l*{9}{r@{\,$\pm$\,}l}}
    \multirow{2}{*}{Process} & \multicolumn{2}{c}{SR} & \multicolumn{2}{c}{SR} & \multicolumn{2}{c}{SR} & \multicolumn{2}{c}{SR} & \multicolumn{2}{c}{CR} & \multicolumn{2}{c}{CR} & \multicolumn{2}{c}{CR} & \multicolumn{2}{c}{CR} & \multicolumn{2}{c}{CR} \\
    & \multicolumn{2}{c}{\ttHcc} & \multicolumn{2}{c}{\ttHbb} & \multicolumn{2}{c}{\ttZcc} & \multicolumn{2}{c}{\ttZbb} & \multicolumn{2}{c}{\ttbb} & \multicolumn{2}{c}{\ttbj} & \multicolumn{2}{c}{\ttcc} & \multicolumn{2}{c}{\ttcj} & \multicolumn{2}{c}{\ttlight} \\
    \hline
    \multirow{2}{*}{\ttHcc} & $-$27 & 40 & $-$0.9 & 6.7 & $-$5 & 16 & $-$0.3 & 3.7 & $-$2.0 & 5.4 & $-$0.8 & 3.6 & $-$16 & 16 & $-$6.3 & 9.3 & $-$46 & 28 \\
    & (17 & 25) & (0.5 & 3.8) & (3.2 & 9.5) & (0.2 & 2.1) & (1.2 & 3.1) & (0.5 & 2.1) & (9.8 & 9.8) & (3.9 & 5.7) & (28 & 17) \\[\cmsTabSkip]
    \multirow{2}{*}{\ttHbb} & 35 & 31 & 310 & 100 & 16 & 20 & 81 & 45 & 441 & 65 & 220 & 45 & 71 & 22 & 40 & 15 & 557 & 67 \\
    & (38 & 34) & (340 & 110) & (17 & 21) & (88 & 49) & (475 & 71) & (242 & 50) & (75 & 24) & (43 & 17) & (607 & 73) \\[\cmsTabSkip]
    \ttZcc & 25 & 28 & 2.0 & 7.6 & 105 & 63 & 2.9 & 9.4 & 5.8 & 7.3 & 2.7 & 5.3 & 51 & 23 & 21 & 14 & 143 & 40 \\
    \ttZbb & 10 & 21 & 57 & 51 & 14 & 25 & 220 & 110 & 218 & 58 & 108 & 40 & 34 & 20 & 22 & 15 & 263 & 58 \\[\cmsTabSkip]
    \ttbb & 194 & 88 & 770 & 180 & 152 & 76 & 730 & 170 & 4680 & 260 & 1530 & 140 & 862 & 95 & 335 & 53 & 4890 & 240 \\
    \ttbj & 300 & 160 & 320 & 160 & 250 & 140 & 270 & 150 & 1830 & 200 & 3740 & 310 & 1050 & 150 & 800 & 120 & 8520 & 450 \\
    \ttcc & 830 & 210 & 50 & 49 & 710 & 190 & 41 & 43 & 244 & 56 & 96 & 39 & 2950 & 220 & 850 & 100 & 6330 & 320 \\
    \ttcj & 920 & 270 & 84 & 78 & 640 & 230 & 64 & 70 & 321 & 77 & 358 & 94 & 2210 & 230 & 2960 & 250 & 13030 & 590 \\
    \ttlight & 1050 & 160 & 102 & 51 & 810 & 140 & 92 & 49 & 257 & 39 & 280 & 47 & 1362 & 99 & 1430 & 96 & 18390 & 410 \\[\cmsTabSkip]
    Other & 240 & 180 & 58 & 76 & 450 & 250 & 82 & 88 & 303 & 87 & 169 & 56 & 630 & 140 & 453 & 100 & 3950 & 350 \\[\cmsTabSkip]
    Total & 3580 & 160 & 1750 & 100 & 3150 & 140 & 1580 & 97 & 8300 & 210 & 6500 & 200 & 9210 & 290 & 6910 & 240 & 56040 & 950 \\[\cmsTabSkip]
    Data & \multicolumn{2}{c}{3545} & \multicolumn{2}{c}{1744} & \multicolumn{2}{c}{3143} & \multicolumn{2}{c}{1593} & \multicolumn{2}{c}{8322} & \multicolumn{2}{c}{6492} & \multicolumn{2}{c}{9245} & \multicolumn{2}{c}{6909} & \multicolumn{2}{c}{56014} \\
\end{scotch}}
\end{table*}

\begin{table*}[!htp]
\centering
\topcaption{%
    Event yields in the full analysis, separated per lepton channel.
    Values in brackets correspond to the pre-fit expectations.
}
\label{tab:yields:channels}
\renewcommand{\arraystretch}{1.1}
\begin{scotch}{l*{4}{r@{\,$\pm$\,}l}}
    Process & \multicolumn{2}{c}{\DL} & \multicolumn{2}{c}{\SL} & \multicolumn{2}{c}{\FH} & \multicolumn{2}{c}{Total} \\
    \hline
    \multirow{2}{*}{\ttHcc} & $-$9 & 18 & $-$62 & 69 & $-$33 & 42 & $-$100 & 130 \\
    & (6 & 11) & (38 & 42) & (20 & 25) & (64 & 78) \\[\cmsTabSkip]
    \multirow{2}{*}{\ttHbb} & 137 & 56 & 1060 & 220 & 580 & 140 & 1770 & 410 \\
    & (151 & 62) & (1150 & 240) & (630 & 150) & (1930 & 450) \\[\cmsTabSkip]
    \ttZcc & 32 & 27 & 210 & 110 & 112 & 65 & 360 & 200 \\
    \ttZbb & 75 & 53 & 580 & 210 & 290 & 130 & 940 & 400 \\[\cmsTabSkip]
    \ttbb & 1060 & 180 & 8420 & 650 & 4670 & 480 & 14100 & 1300 \\
    \ttbj & 1310 & 260 & 12100 & 1000 & 3730 & 530 & 17100 & 1800 \\
    \ttcc & 940 & 160 & 7320 & 610 & 3850 & 460 & 12100 & 1200 \\
    \ttcj & 1730 & 290 & 15600 & 1200 & 3220 & 390 & 20600 & 1900 \\
    \ttlight & 1700 & 140 & 17080 & 670 & 4990 & 280 & 23800 & 1100 \\[\cmsTabSkip]
    Other & 430 & 200 & 5040 & 810 & 870 & 320 & 6300 & 1300 \\[\cmsTabSkip]
    Total & 7410 & 240 & 67300 & 1400 & 22270 & 720 & 97000 & 2400 \\[\cmsTabSkip]
    Data & \multicolumn{2}{c}{7366} & \multicolumn{2}{c}{67370} & \multicolumn{2}{c}{22271} & \multicolumn{2}{c}{97007} \\
\end{scotch}
\end{table*}

\clearpage
\subsection{Systematic uncertainties}
\label{supp:systematics}

Systematic uncertainties are evaluated by appropriate variations of the signal and background simulations.
The uncertainty sources affect either the rate or the discriminant shape of the signal or background processes, or both.
They are taken into account via nuisance parameters in the final profile likelihood fit.
The effects from the same source are treated as fully correlated among the different categories.

The considered sources of uncertainties are listed in Table~\ref{tab:systematics} and described in detail in the following sections, grouped by experimental uncertainties in Section~\ref{sec:syst:experimental} and modeling uncertainties in Section~\ref{sec:syst:modelling}.
Their impact on the final result is discussed in Section~\ref{supp:results}.

As for the final statistical inference, the templates corresponding to individual years are merged to obtain only one template.
We then consider the appropriate correlation scheme in the following way.
For correlated systematic uncertainties the templates are varied simultaneously in all years, while for decorrelated systematics the template of the corresponding year(s) are varied, and for all other years the nominal template is considered. This way, partial effects are correctly taken into account as partial variations.

\begin{table*}[!ht]
\centering
\topcaption{%
    Summary of the systematic uncertainty sources in the measurement.
    The first column lists the source of the uncertainty.
    The second (third) column indicates the treatment of correlations of the uncertainties between different data-taking periods (processes), where \fullcorr means fully correlated, \partcorr means partially correlated (\ie, contains sub-sources that are either fully correlated or uncorrelated), and \nocorr means uncorrelated. The last column indicates whether the uncertainties are applied to all processes or only a subset.
}
\label{tab:systematics}
\renewcommand{\arraystretch}{1.1}
\cmsTable{\begin{scotch}{llccc}
    & Source & Corr. (period) & Corr. (process) & Processes \\
    \hline
    \multirow{9}{*}{\rotatebox{90}{Experimental}}
    & Integrated luminosity & \partcorr & \fullcorr & all \\
    & Electron reconstruction and identification & \fullcorr & \fullcorr & all \\
    & Muon reconstruction and identification & \fullcorr & \fullcorr & all \\
    & Trigger efficiencies & \nocorr & \fullcorr & all \\
    & Pileup reweighting & \fullcorr & \fullcorr & all \\
    & Jet energy scale (11 sources) & \partcorr & \fullcorr & all \\
    & Jet energy resolution & \nocorr & \fullcorr & all \\
    & Unclustered \ptmiss & \nocorr & \fullcorr & all \\
    & Jet flavor tagging & \partcorr & \fullcorr & all \\[\cmsTabSkip]
    \multirow{8}{*}{\rotatebox{90}{Modeling}}
    & Incl. cross sections \& normalization & \fullcorr & \partcorr & all \\
    & \muR scale & \fullcorr & \nocorr & \ttjets, \ttX, single \PQt \\
    & \muF scale & \fullcorr & \nocorr & \ttjets, \ttX, single \PQt \\
    & PDF shape & \fullcorr & \nocorr & \ttjets, \ttX \\
    & PS scales: ISR & \fullcorr & \nocorr & \ttjets, \ttX, single \PQt \\
    & PS scales: FSR & \fullcorr & \nocorr & \ttjets, \ttX, single \PQt \\
    & ME-PS matching (\hdamp) & \fullcorr & \nocorr & \ttjets \\
    & ME (4 \vs 5 flavor scheme) & \fullcorr & \nocorr & \ttbjets \\
\end{scotch}}
\end{table*}

\subsection{Experimental uncertainties}
\label{sec:syst:experimental}

\begin{itemize}
\item \textbf{Luminosity:} The integrated luminosities of the data-taking periods are individually measured with uncertainties of 1.2, 2.3, and 2.5\% for 2016, 2017, and 2018 data-taking periods, respectively~\cite{CMS:LUM-17-003, CMS:LUM-17-004, CMS:LUM-18-002}.
The uncertainty in the integrated luminosity of the combined data set is 1.6\%, when taking into account the correlations between the periods.
\item \textbf{Lepton scale factors:} Separate \pt and $\eta$ dependent uncertainties are assigned to the reconstruction and identification efficiencies for electrons in the simulation.
Similarly, the muon tracking, identification, and isolation efficiencies are estimated.
The associated uncertainties are propagated to the final discriminant distributions as a total shape uncertainty estimated from the sum of all sources and of the statistical and systematic components, separately for electrons and muons.
The impact of the lepton calibration uncertainties in the final result is negligible.
\item \textbf{Trigger scale factors:} The trigger efficiency scale factors are varied within their uncertainties, separately for each lepton number and flavor combination and the hadronic trigger selection.
\item \textbf{Pileup reweighting:} The prediction of number of pileup interactions in simulation is performed with a total inelastic \pp cross section of 69.2\unit{mb}.
Changes in the assumed pileup multiplicity are estimated by varying the total inelastic cross section by $\pm$4.6\%~\cite{CMS:JME-18-001}.
This uncertainty is treated as fully correlated between data-taking periods.
\item \textbf{L1 prefiring issue:} During the 2016--2017 data-taking periods, a gradual shift in the timing of the inputs of the ECAL L1 trigger in the forward endcap region ($\abseta>2.4$) led to a specific inefficiency, known as ``prefiring''.
A similar effect is present for the muon system because of its limited time resolution, which was most pronounced in 2016 but also impacted data collected in 2017--2018.
Corrections for this effect are applied to simulated events, and 20\% of the corrections are assigned as the associated uncertainties.
\item \textbf{Jet energy scale:} Uncertainties in the determination of the jet energy scale are taken into account by shifting the jet energy scale applied to the reconstructed jets in the simulation up and down by one standard deviation, separately for each of several sources of uncertainty, such as the overall energy scale, differences in flavor response, and residual differences between energy scale measurements~\cite{CMS:2016lmd}.
Some of these sources are treated separately per data-taking period, while some are correlated for all periods.
\item \textbf{Jet energy resolution:} Uncertainties in the jet energy resolution are evaluated by increasing or decreasing the difference between jet energies at reconstructed and particle level, or by smearing the measured jet energy in case no matching particle-level jet could be found~\cite{CMS:2016lmd}.
This uncertainty is uncorrelated between data-taking periods.
\item \textbf{Unclustered \ptmiss:}  An additional uncertainty in \ptmiss\ is derived by varying the energies of reconstructed
particles not clustered into jets.
\item \textbf{Jet flavor tagging:} Uncertainties in the flavor tagging calibration described in Section~\ref{supp:flavor-tagging} are taken into account. The flavor tagging uncertainties include statistical uncertainties in the derived SFs, effects due to the choice of the \muF and \muR scales, variations in PS ISR and FSR in \PYTHIA, uncertainties in the fraction of \PQb and \PQc jets in the simulated \Wjets and \Zjets events, jet energy scale and resolution uncertainties, and pileup uncertainties. The statistical uncertainties in the scale factors are treated as uncorrelated, while all the other uncertainties in the scale factors are treated as fully correlated.
\end{itemize}

\subsection{Modeling uncertainties}
\label{sec:syst:modelling}

This measurement is strongly affected by uncertainties in the modeling of the \ttjets background processes, especially the \ttbjets processes.
Variations applied to the \ttjets background processes are defined such that the predicted yields in the inclusive phase space region before acceptance effects or any event selection remain constant for the individual processes.
This procedure ensures that the modeling uncertainties for the \ttjets processes only have an effect on the selection efficiency, and not on the cross section of the processes.
The normalization of the individual processes are degrees of freedom in the binned profile likelihood fit.
Furthermore, the uncertainty in the overall normalization of the \ttbar process is covered with separate systematic uncertainties, as detailed below.

All modeling uncertainties are correlated between the data-taking years. Some uncertainties are correlated between processes.

\begin{itemize}
\item \textbf{Inclusive cross sections:} The expectation for the inclusive signal and background cross sections are derived from theoretical predictions of at least NLO accuracy.
Uncertainties affecting these normalizations are split into a component due to the \muR and \muF scale uncertainties in the ME calculation and a component due to the PDF set and choice of the strong coupling constant \alpS (PDF+\alpS) for the \ttH signal and the \ttZ process.
These uncertainties are taken into account as rate-changing uncertainties in the final fit and are on the order of 9\% for \ttZ~\cite{Kulesza:2020nfh} and are split into three components for \ttH production~\cite{deFlorian:2016spz}, taking into account the effects from variations of the scales ($^{+5.8\%}{-9.2\%}$), \alpS (2\%), and the PDFs (3\%).
\item \textbf{Branching fractions:} For the \ttH process, several rate-changing uncertainties are assigned to the signal and background contributions, depending on the final state. The uncertainties are taken from Ref.~\cite{deFlorian:2016spz} and denote about 1.3\% (10\%) in total for \hbb (\hcc) decays.
\item \textbf{Renormalization and factorization scales:} Uncertainties due to the renormalization scale \muR and the factorization scale \muF in the ME generators are modeled by shifting the scales independently up and down by a factor of two.
These uncertainties are treated separately for each process and for each of the seven \ttjets background components.
\item \textbf{PDF shape:} The NNPDF3.1 NNLO PDF set is used for the description of the substructure of the colliding protons in simulation.
The shape variations of the fit distribution are evaluated by reweighting the simulated events with the replicas of the PDF set.
We calculate the root-mean-square of all residuals for the \fourFS PDF set, used for the \ttbjets backgrounds.
For the other processes, using the \fiveFS PDF set, the variations correspond to the eigenvectors of the PDF fit covariance matrix, and individual uncertainties are obtained from each eigenvector. Each eigenvector variation is treated as one separate uncertainty in the fit.
An additional uncertainty from the value of \alpS, used in the PDFs, is included.
These uncertainties are treated as correlated for all processes of common PDF flavor schemes.
\item \textbf{PS scales ISR/FSR:} In order to estimate the impact of the choice of scale at which \alpS is evaluated in the PS simulation, the scales in the shower simulation are varied up and down by a factor of two, independently for ISR and FSR.
For both ISR and FSR, the uncertainties are split into 8 separate components.
We individually vary the scales and nonsingular terms for the splitting kernels \gtogg, \gtoqq, \qtoqg, \xtoxg, where $\PQx=\PQt,\PQb$, as implemented in \PYTHIA v8~\cite{Mrenna:2016sih}.
These uncertainties are considered as uncorrelated for all processes.
\item \textbf{ME-PS matching (\hdamp):} In the \POWHEG generator, the scale that separates the phase space of the first QCD emission into soft and hard parts is controlled via the \hdamp parameter.
For the CP5 tune, when \POWHEG is matched to the \PYTHIA v8~\cite{Mrenna:2016sih} PS, the central value of \hdamp is set to $\hdamp=1.379\mtop$, and the uncertainties are estimated with varied values of $\hdamp=2.305\mtop$ and 0.874\mtop for the \ttbar and \ttbjets simulations.
The alternative predictions with varied \hdamp values are obtained via an ML-reweighting approach~\cite{CMS:MLG-24-001}.
This uncertainty is treated as uncorrelated between the different \ttjets processes.
\item \textbf{Background normalization:} To account for the known issues in the modeling of different background processes, we assign the following normalization uncertainties: 25\% for single top quark, 100\% for \tWZ due to the discrepancies between prediction and measurements~\cite{CMS:TOP-22-008}, 6.8\% for \ttW~\cite{Buonocore:2023ljm}, and 50\% for the relative \ttDoubleB and \ttDoubleC  contributions with respect to \ttSingleB and \ttSingleC, respectively, to account for the limited knowledge of the rate of collinear gluon splitting.
\item \textbf{\ttbbDPS normalization:} The \ttbbDPS cross section is not precisely known and an estimate of the effective cross section for DPS is used, which is process dependent. Different measurements of this effective cross section range between 10--30\unit{mb}~\cite{ATLAS:2018zbr}, while the \PYTHIA reference value is 30\unit{mb}. Given the limited knowledge of the effective cross section value for \ttbbDPS, a 50\% normalization uncertainty is applied to the \ttbbDPS contributions.
{\tolerance=800
\item \textbf{\ttbjets modeling differences:} The nominal model for the estimation of the \ttbjets background relies on the NLO ME description of \ttbb using \POWHEGBOX, interfaced with \PYTHIA for parton showering and hadronization.
Dedicated measurements of \ttbjets production show that this simulation approach is able to describe the most relevant observables in data~\cite{CMS:TOP-22-009}, when taking into account the systematic modeling uncertainties described previously.
Nevertheless, in order to provide additional freedom for the background model, the difference between the nominal model and the description of \ttbjets with the \ttbar-inclusive NLO \POWHEG model, interfaced with \PYTHIA, is used as a systematic uncertainty. The variations are normalized to the nominal event yield after baseline selection.
This alternative \ttbjets model does not describe relevant physics objects significantly differently than the nominal background model. This uncertainty is considered decorrelated for the \ttSingleB processes and decorrelated among all four SRs.
\par}
\end{itemize}

\subsection{Validation of the background model}

A signal-depleted sideband validation region (VR), with a partition into a control-like region for $0.40<\score{\ttX}<0.58$ and a signal-like region for $0.58<\score{\ttX}<0.60$, analogous to the AR partition into a CR and a SR, is used to validate the background estimation strategy, and good agreement is observed between data and estimated backgrounds, as shown in Fig.~\ref{fig:results:postfitVR}.
Each region in the VR is defined such that the purity of the background components is as close as possible to those in the CRs and SRs.

\begin{figure*}[!p]
\centering
\includegraphics[width=0.95\textwidth]{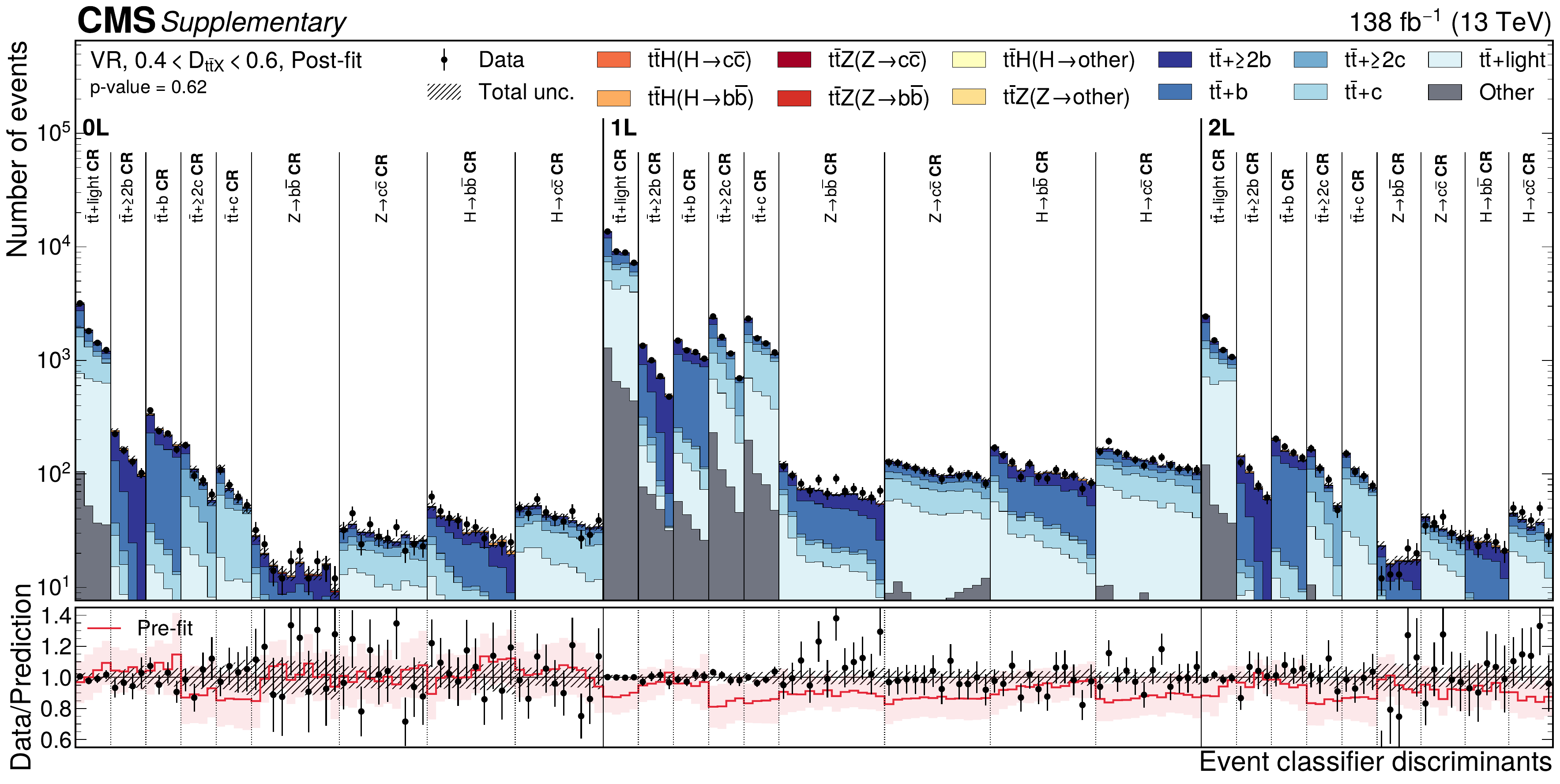}
\caption{%
    Distributions of the \parT discriminants in data (points) and predicted signal and backgrounds (colored histograms) after the maximum likelihood fit to data in the VR, defined by $0.4<\score{\ttX}<0.6$.
    The vertical bars on the points represent the statistical uncertainties in data.
    The hatched band represents the total uncertainty in the sum of the signal and background predictions.
    The lower panel shows the ratio of the data to the sum of the signal and background predictions.
    The ratio of the pre-fit expectation to the sum of the signal and background predictions after the fit is shown as a red line in the lower panel, including the pre-fit uncertainties as a shaded band.
}
\label{fig:results:postfitVR}
\end{figure*}

\subsection{Additional results}
\label{supp:results}

\begin{figure*}[!tp]
\centering
\includegraphics[width=0.95\textwidth]{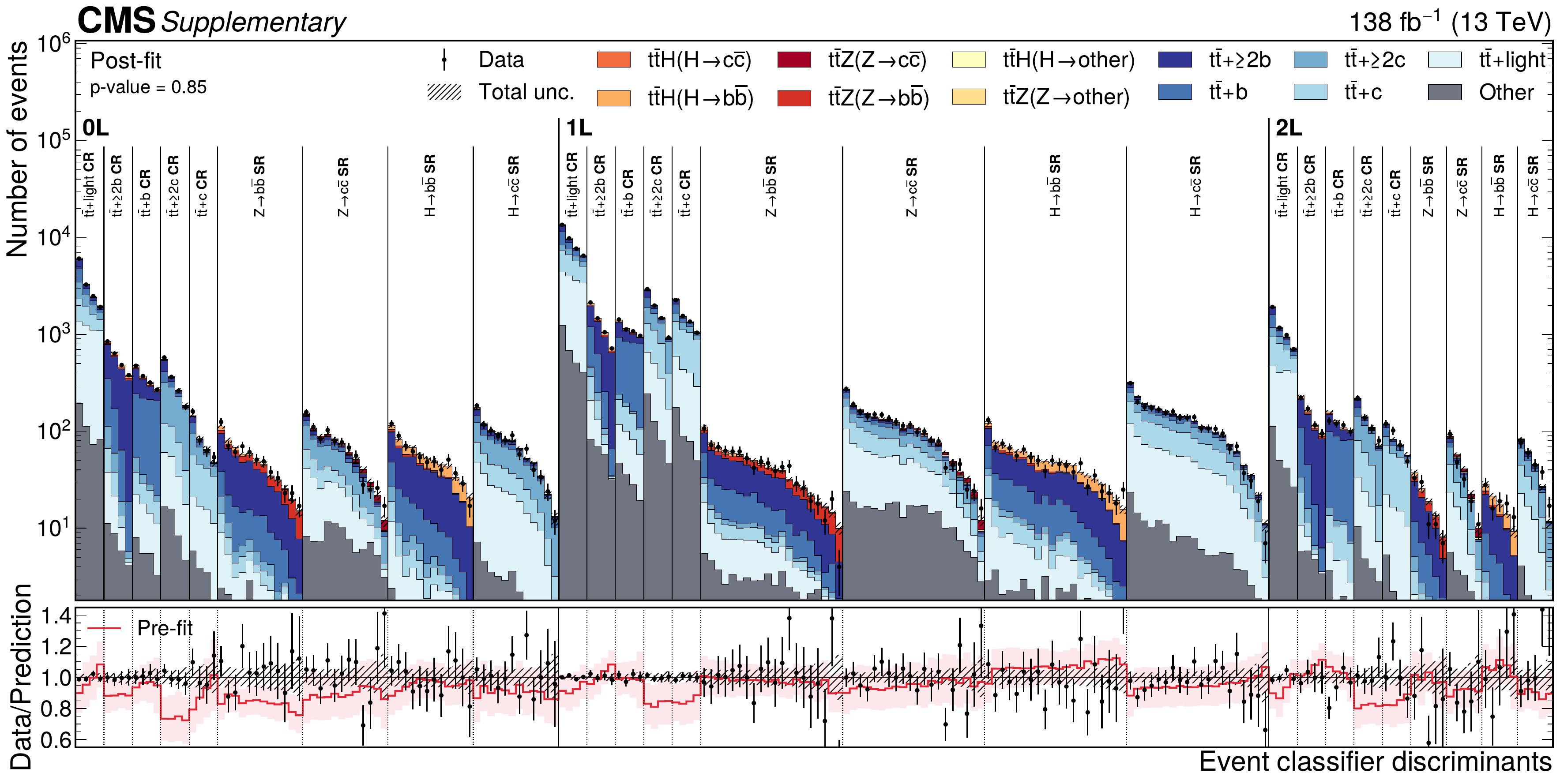}
\caption{%
    Distributions of the \parT discriminants in data (points) and predicted signal and backgrounds (colored histograms) after the maximum likelihood fit to data.
    The vertical bars on the points represent the statistical uncertainties in data.
    The hatched band represents the total uncertainty in the sum of the signal and background predictions.
    The lower panel shows the ratio of the data to the sum of the signal and background predictions.
    The ratio of the pre-fit expectation to the sum of the signal and background predictions after the fit is shown as a red line in the lower panel, including the pre-fit uncertainties as a shaded band.
}
\label{fig:results:postfitsupp}
\end{figure*}

The production rates of the signal processes and the normalization factors of the \ttbar background components are determined via a binned profile likelihood fit to data in all the SRs and CRs.
The fitted observables are described in Section~\ref{supp:event-classifier}.
Figure~\ref{fig:results:postfitsupp} shows the observed yields in each bin of the event classifier discriminant for all CRs and SRs, along with the fitted signal and background yields.
Figures~\ref{fig:results:ttZcc}--\ref{fig:results:ttZbb} summarize the results of the measurements of the \ttZcc and \ttZbb signal strengths, respectively, obtained from the simultaneous fit of all four signal strength parameters and all SRs and CRs.
Figure~\ref{fig:results:ttHbb} shows the contributions from the individual channels for the best fit values of \muHbb.

\begin{figure*}[!htp]
\centering
\includegraphics[width=0.405\textwidth]{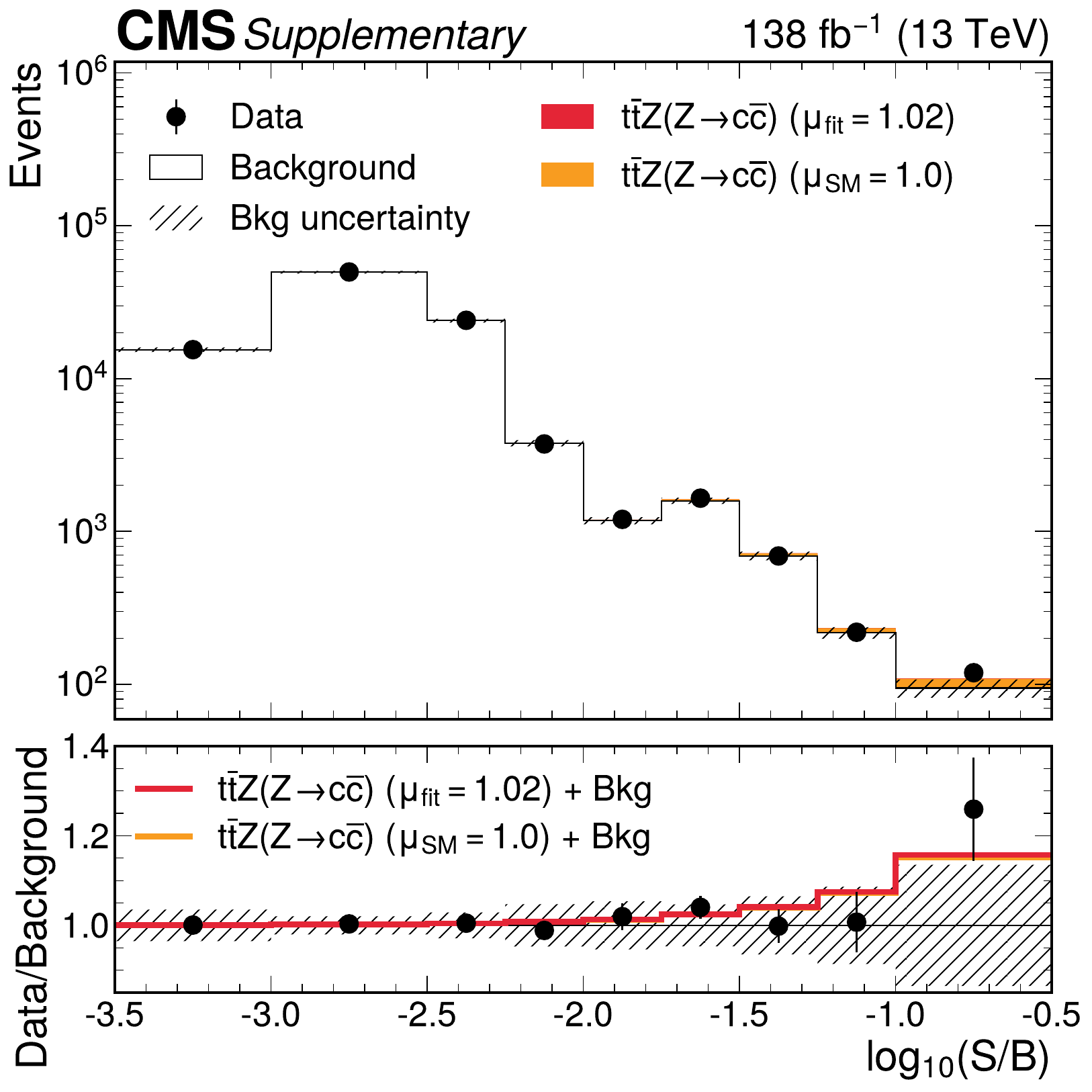}%
\hfill%
\includegraphics[width=0.575\textwidth]{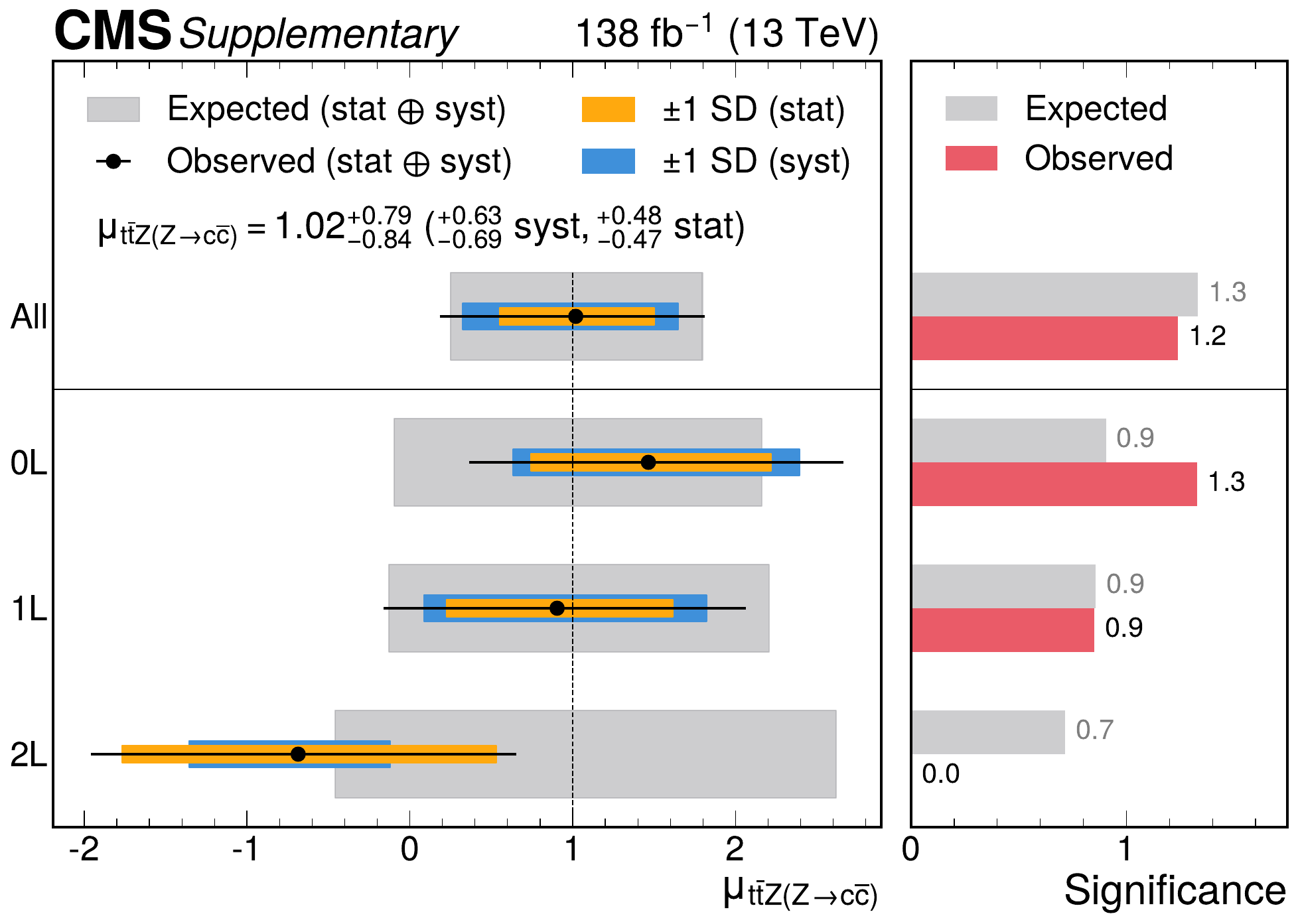}
\caption{%
    On the left, observed and expected event yields from all SRs and CRs as a function of $\log_{10}(S/B)$, where $S$ are the observed \ttZcc yields, and $B$ are the total background yields in the combined fit to data.
    The signals are shown for the best fit signal strength (red), and the SM prediction, $\mu=1$ (orange).
    The lower panel shows the ratio of the data to the post-fit background prediction, compared to the signal-plus-background predictions.
    On the right, fit results of \muZcc in the combination of channels (first row) and the channels individually (lower rows).
    The left panel shows the observed signal strength, compared to the expected results.
    The right panel shows the expected and observed significance.
}
\label{fig:results:ttZcc}
\end{figure*}

\begin{figure*}[!ht]
\centering
\includegraphics[width=0.405\textwidth]{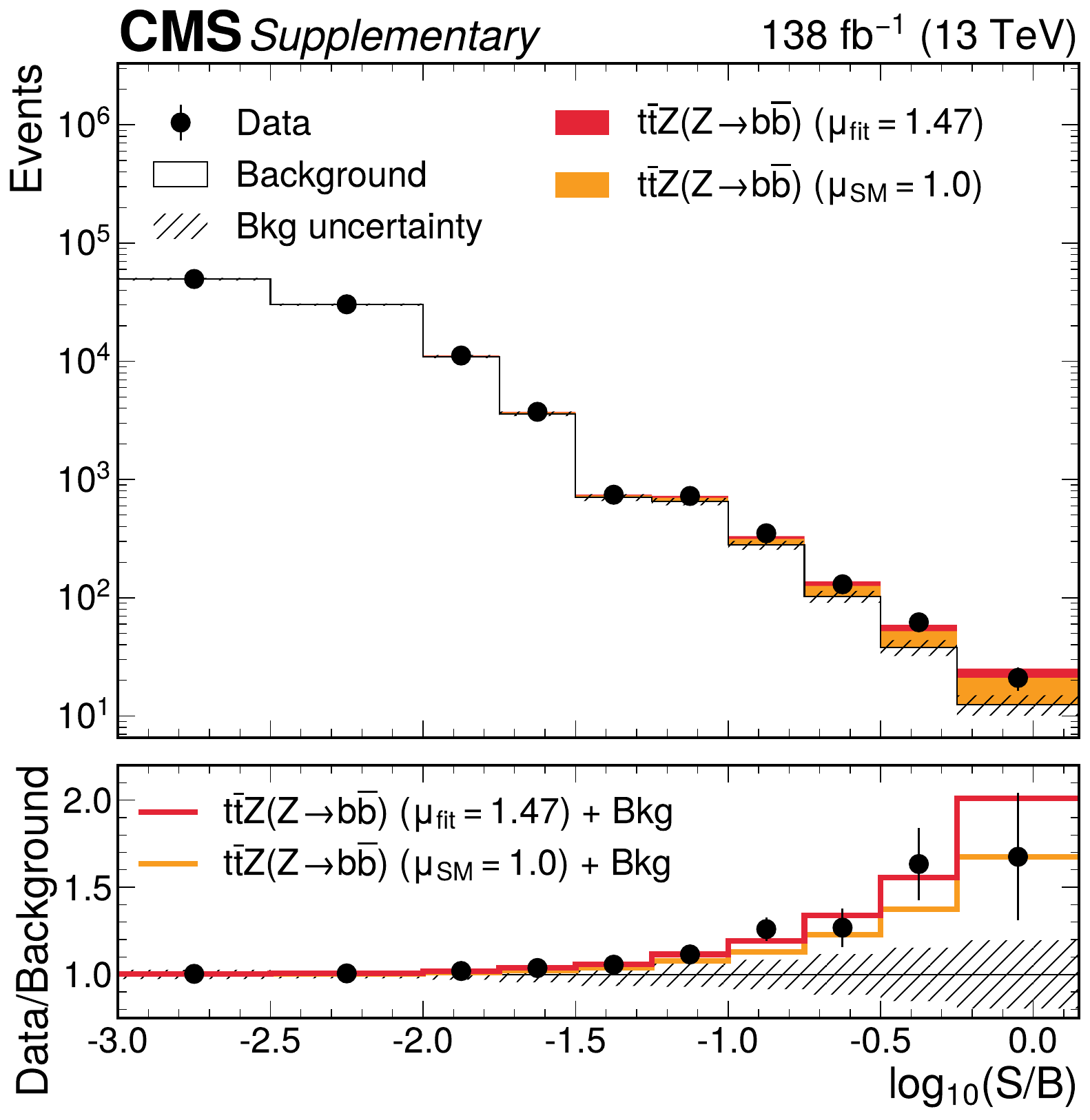}%
\hfill%
\includegraphics[width=0.575\textwidth]{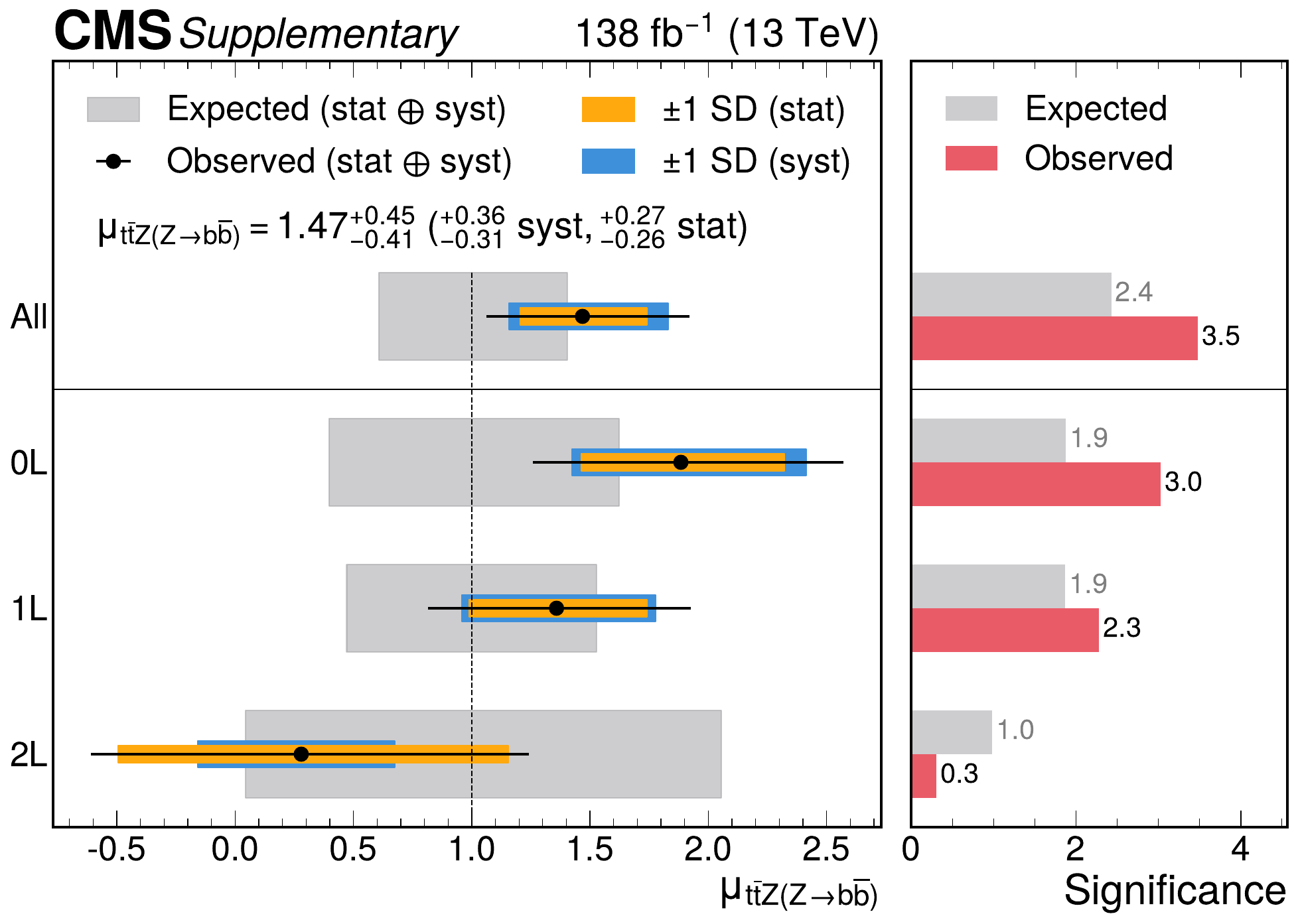}
\caption{%
    On the left, observed and expected event yields from all SRs and CRs as a function of $\log_{10}(S/B)$, where $S$ are the observed \ttZbb yields, and $B$ are the total background yields in the combined fit to data.
    The signals are shown for the best fit signal strength (red), and the SM prediction, $\mu = 1$ (orange).
    The lower panel shows the ratio of the data to the post-fit background prediction, compared to the signal-plus-background predictions.
    On the right, fit results of \muZbb in the combination of channels (first row) and the channels individually (lower rows).
    The left panel shows the observed signal strength, compared to the expected results.
    The right panel shows the expected and observed significance.
}
\label{fig:results:ttZbb}
\end{figure*}

\begin{figure}[!ht]
\centering
\includegraphics[width=\cmsFigWidth]{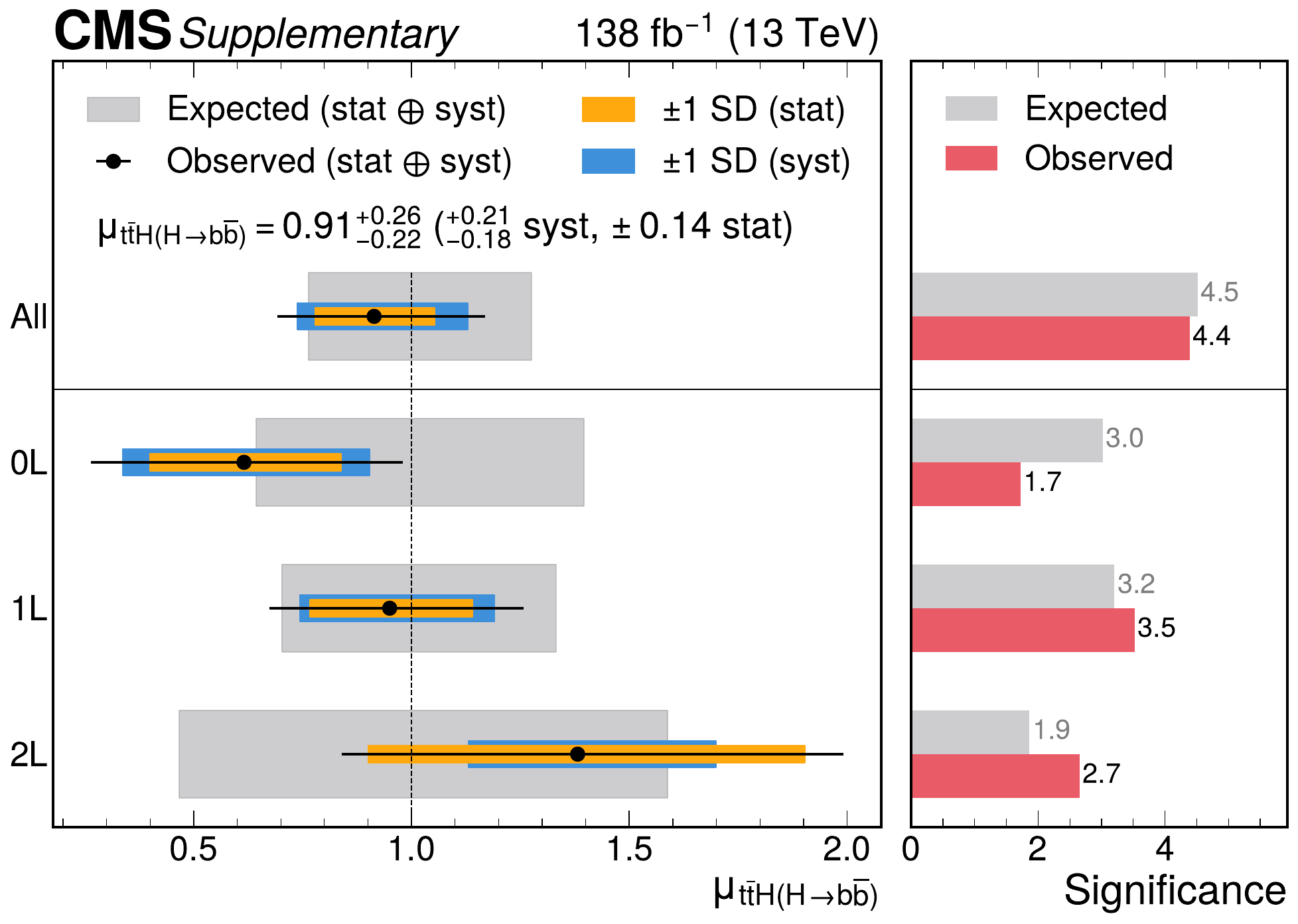}
\caption{%
    Fit results of \muHbb.
    The left panel shows the observed signal strength, compared to the expected results.
    The right panel shows the expected and observed significance.
}
\label{fig:results:ttHbb}
\end{figure}

The result is interpreted in the $\kappa$-framework ~\cite{Heinemeyer:2013tqa,deFlorian:2016spz}.
The one-dimensional likelihood scan for \kappaC is shown in Fig.~\ref{fig:results:kappa1d}, both for a fixed and a floating \kappaB value.

\begin{figure}[!ht]
\centering
\ifthenelse{\boolean{cms@external}}{%
    \includegraphics[width=0.82\columnwidth]{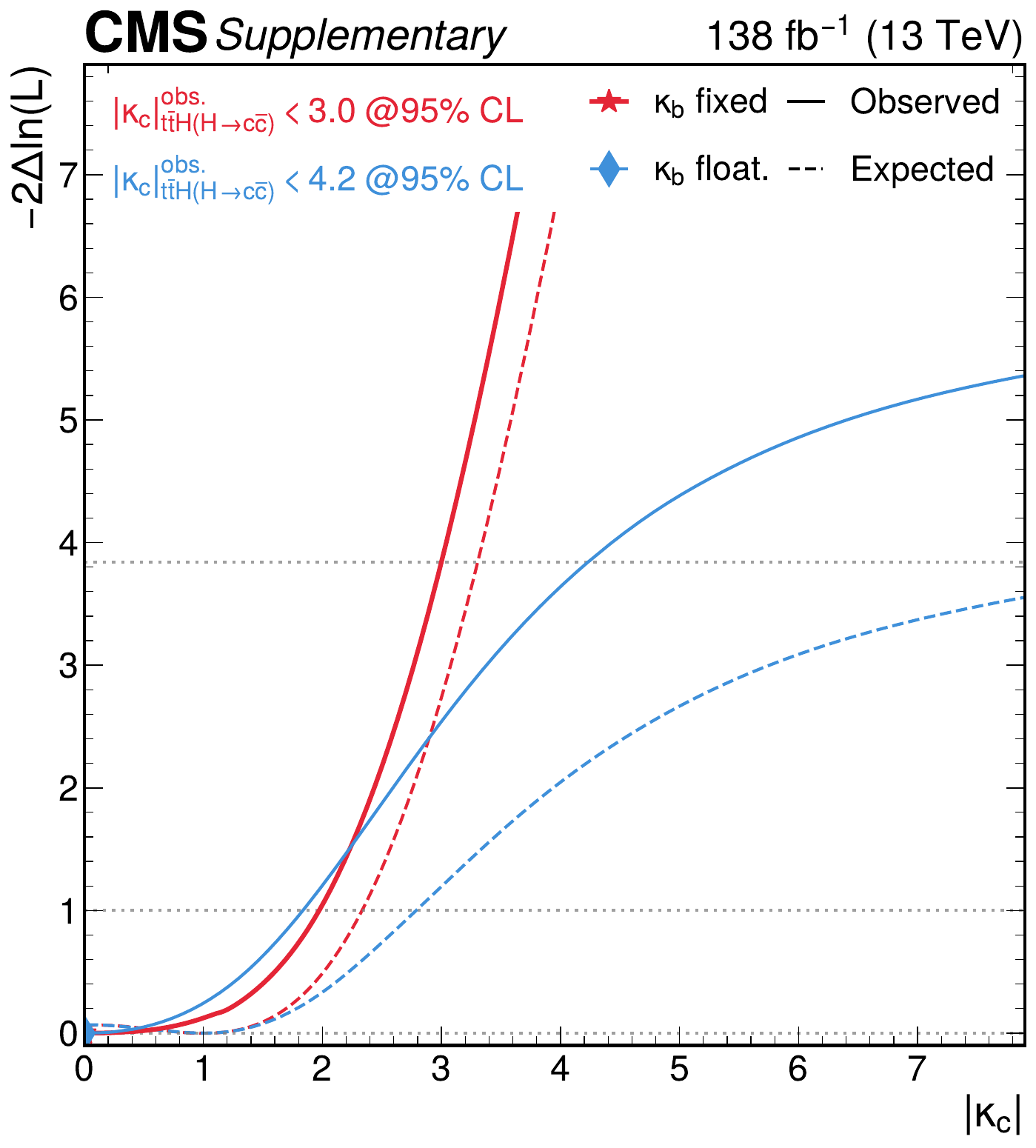}
}{%
    \includegraphics[width=0.4\textwidth]{Figure_J005.pdf}
}
\caption{%
    Likelihood scans of \kappaC with fixed $\kappaB=1$ (red) and floating \kappaB (blue).
    The 68\% and 95\% \CL intervals are indicated by the horizontal dotted lines.
}
\label{fig:results:kappa1d}
\end{figure}

\begin{figure}[!tp]
\centering
    \includegraphics[width=\cmsFigWidth]{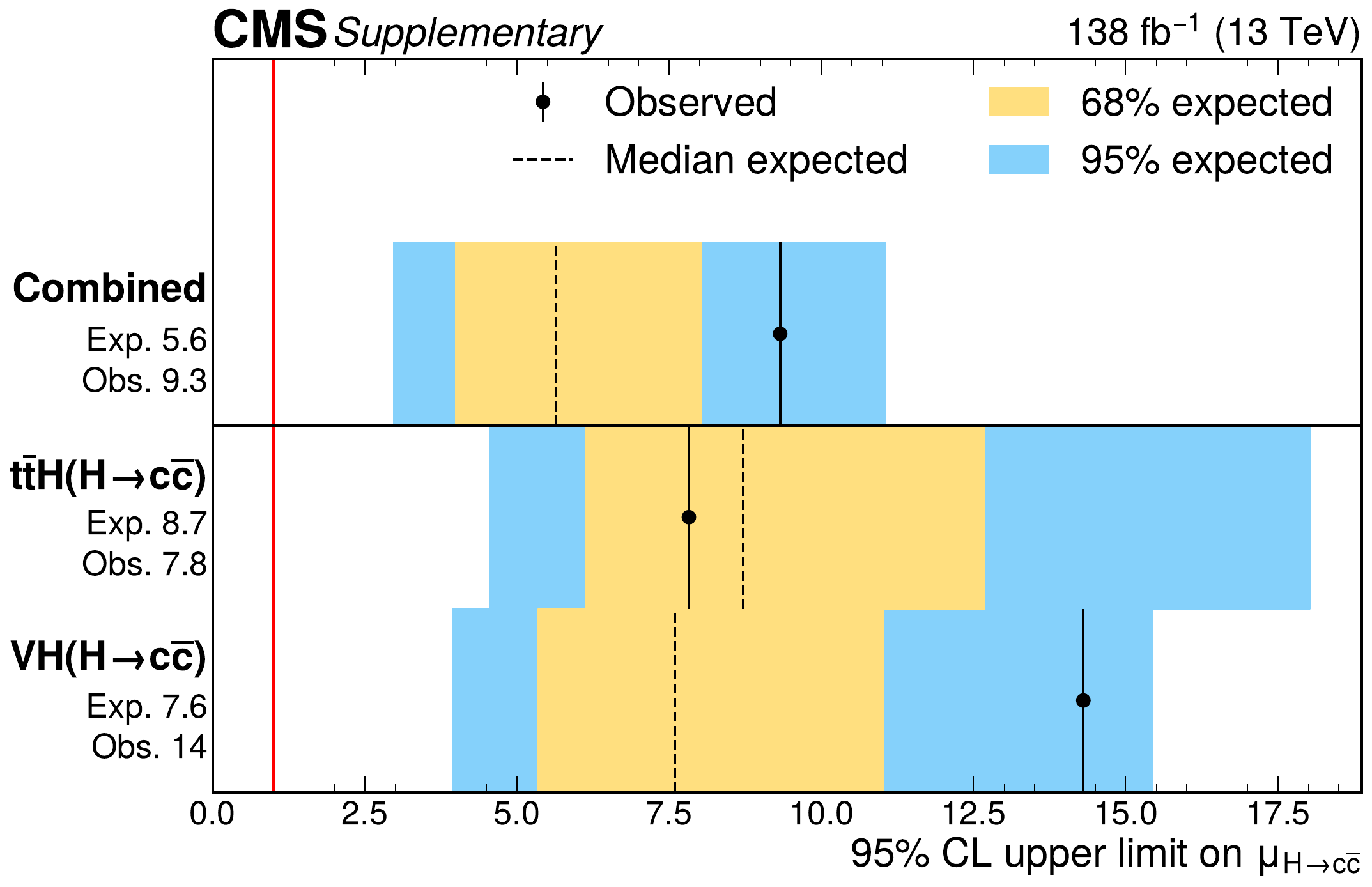}

\caption{%
    The 95\% \CL upper limits on \muJustHcc.
    The blue and yellow bands indicate the expected 68\% and 95\% \CL regions, respectively, under the background-only hypothesis.
    The vertical red line indicates the SM value $\muJustHcc=1$.
}
\label{fig:results:limitsJustHcc}
\end{figure}

\begin{figure}[!tp]
\centering
\ifthenelse{\boolean{cms@external}}{%
    \includegraphics[width=0.82\columnwidth]{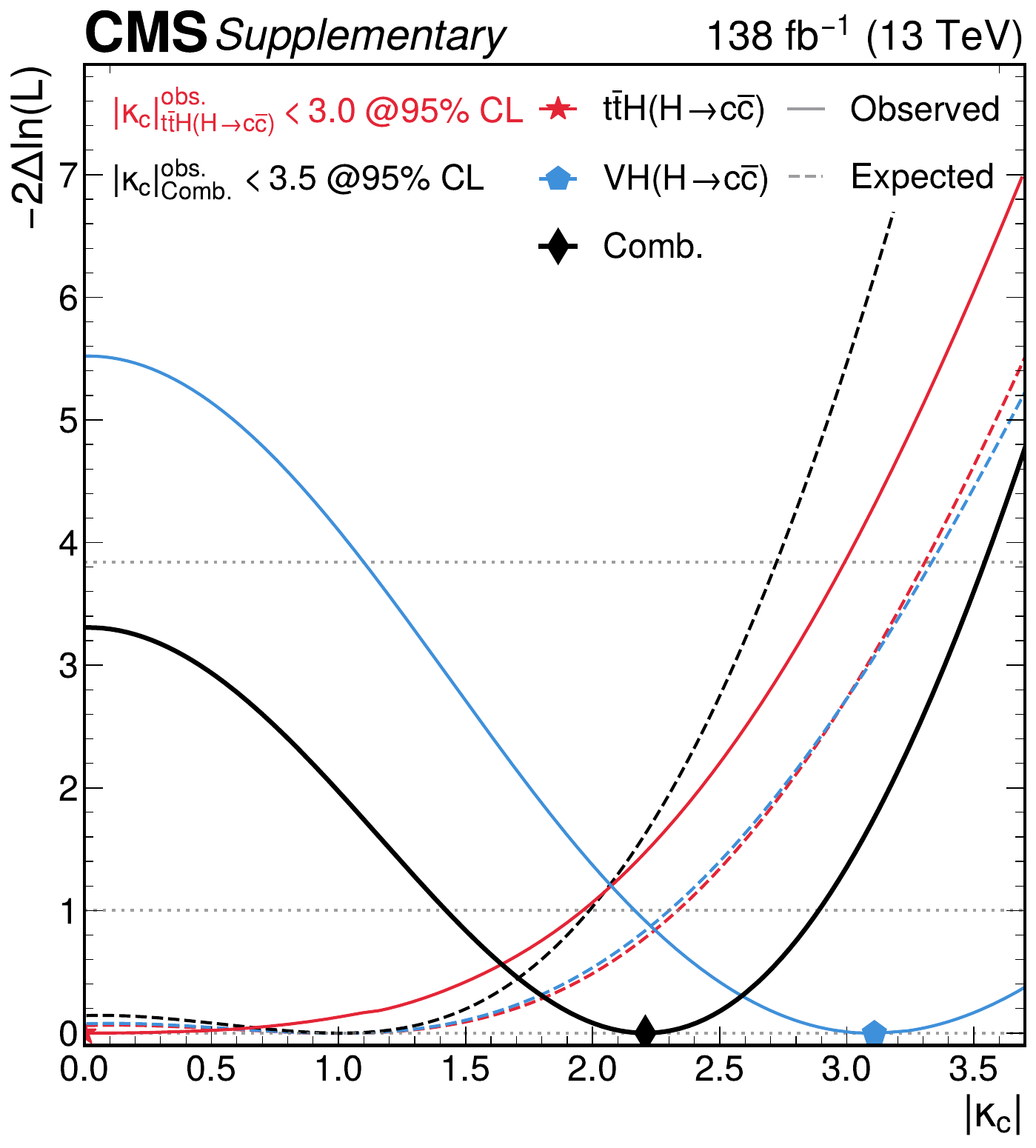}
}{%
    \includegraphics[width=0.4\textwidth]{Figure_J007.pdf}
}
\caption{%
    Likelihood scans of \kappaC with fixed $\kappaB=1$ in the individual \ttH (red) and \VH (blue) channels and their combination (black).
    The 68\% and 95\% \CL intervals are indicated by the horizontal dotted lines.
}
\label{fig:results:kappa_comb}
\end{figure}

The search is combined with a previous search in the \VH channel~\cite{CMS:HIG-21-008}.
All uncertainties in experimental effects, such as jet energy scale and resolution, lepton identification efficiencies, the modeling of detector effects, pileup and \ptmiss modeling, and the integrated luminosity are correlated.
The flavor (mis-)tagging scale factors of the merged and resolved \VH analyses are treated as uncorrelated with this measurement because of the differences in jet tagging algorithms and jet cone sizes, as well as in calibration methods.
All uncertainties in the background and signal modeling are considered uncorrelated as well, since the main background in the \VH analysis, \Zjets and \Wjets production in association with heavy-flavor jets, and the \VH signal contribution are negligible for this analysis.
The dominant systematic uncertainties in each measurement are the normalization and modeling of the background components, which are negligible in the respective other measurement.
Theoretical uncertainties such as in the Higgs boson branching fraction and production cross section are correlated.
Each input measurement is limited by statistical uncertainties.
The observed and expected 95\% \CL upper limit on \muJustHcc, the signal strength for the \hcc decay channel assuming SM production rates for \ttH and \VH, is displayed in Fig.~\ref{fig:results:limitsJustHcc}, along with the contributions from individual analyses.
Figure~\ref{fig:results:kappa_comb} shows the profile likelihood scan of $\abs{\kappaC}$ in the combined analysis.

Figures~\ref{fig:results:scan_mu2d}--\ref{fig:results:scan_sf} show likelihood scans for the simultaneous fit of two parameters each, \eg, for pairs of the four signal strength parameters \muHcc, \muHbb, \muZbb, and \muZcc, and for pairs of the normalization scale factors of the \ttjets background components \ttbb, \ttbj, \ttcc, and \ttcj.

\begin{figure*}[!htp]
\centering
\includegraphics[width=0.45\textwidth]{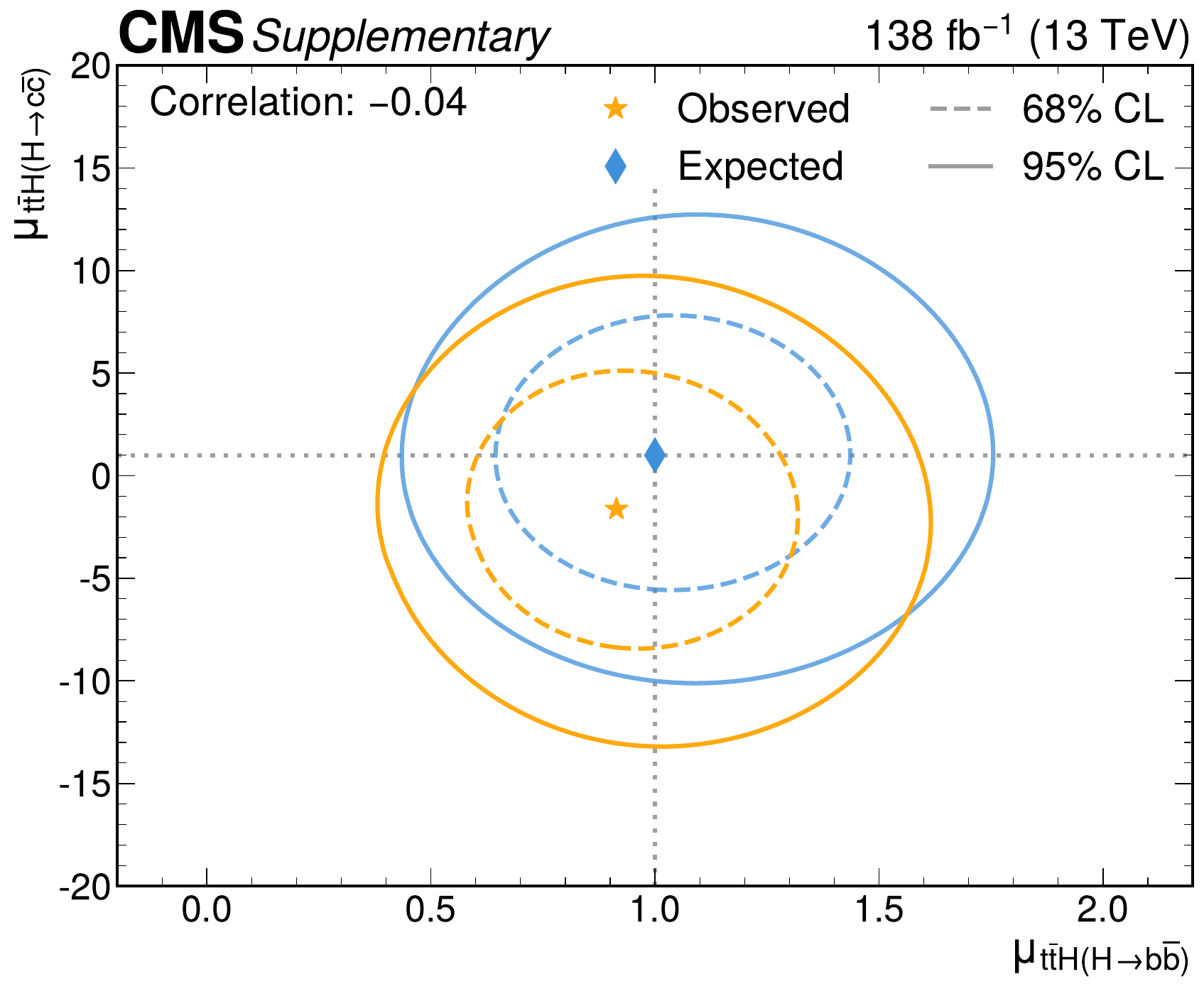}%
\hfill%
\includegraphics[width=0.45\textwidth]{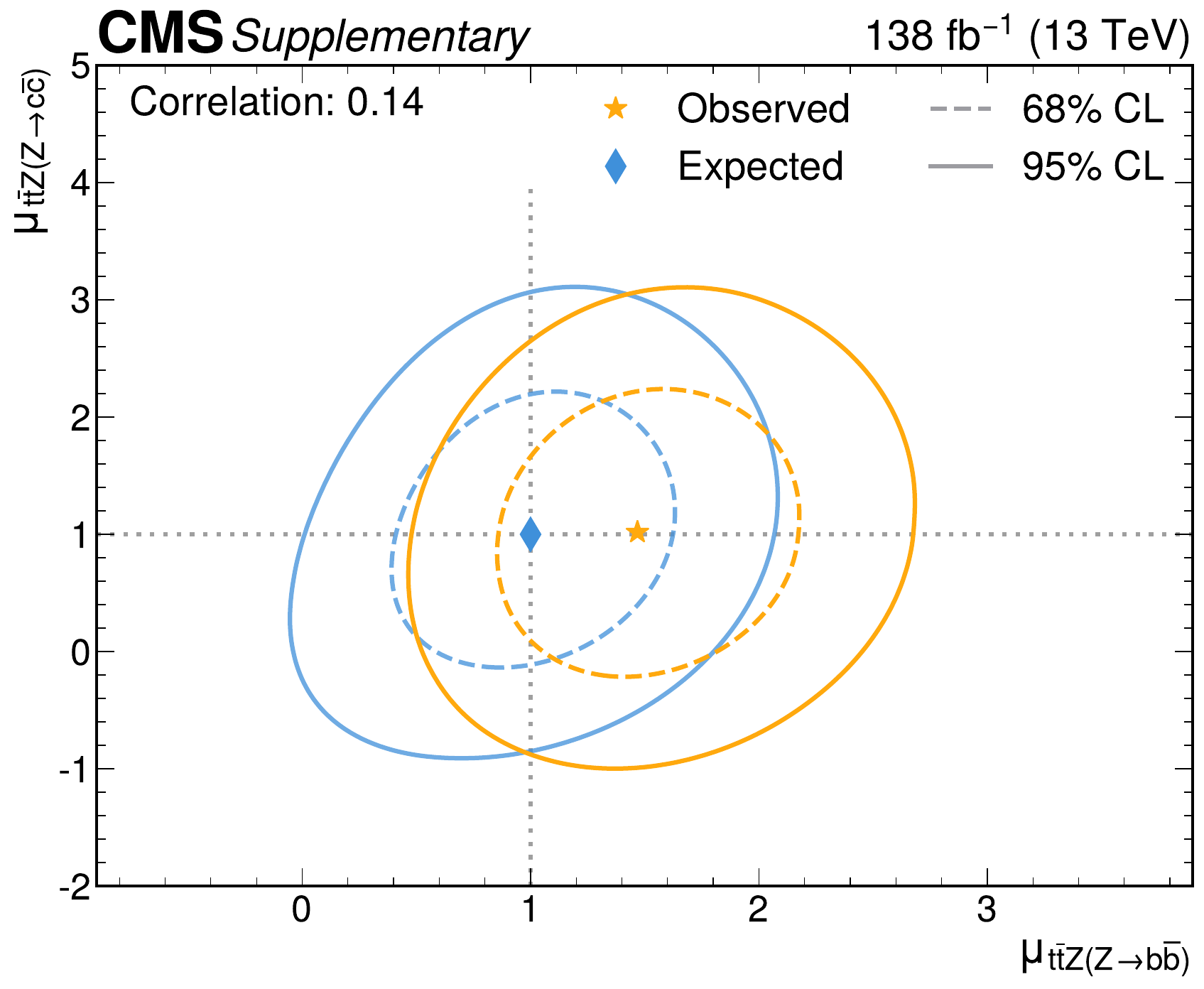} \\
\includegraphics[width=0.45\textwidth]{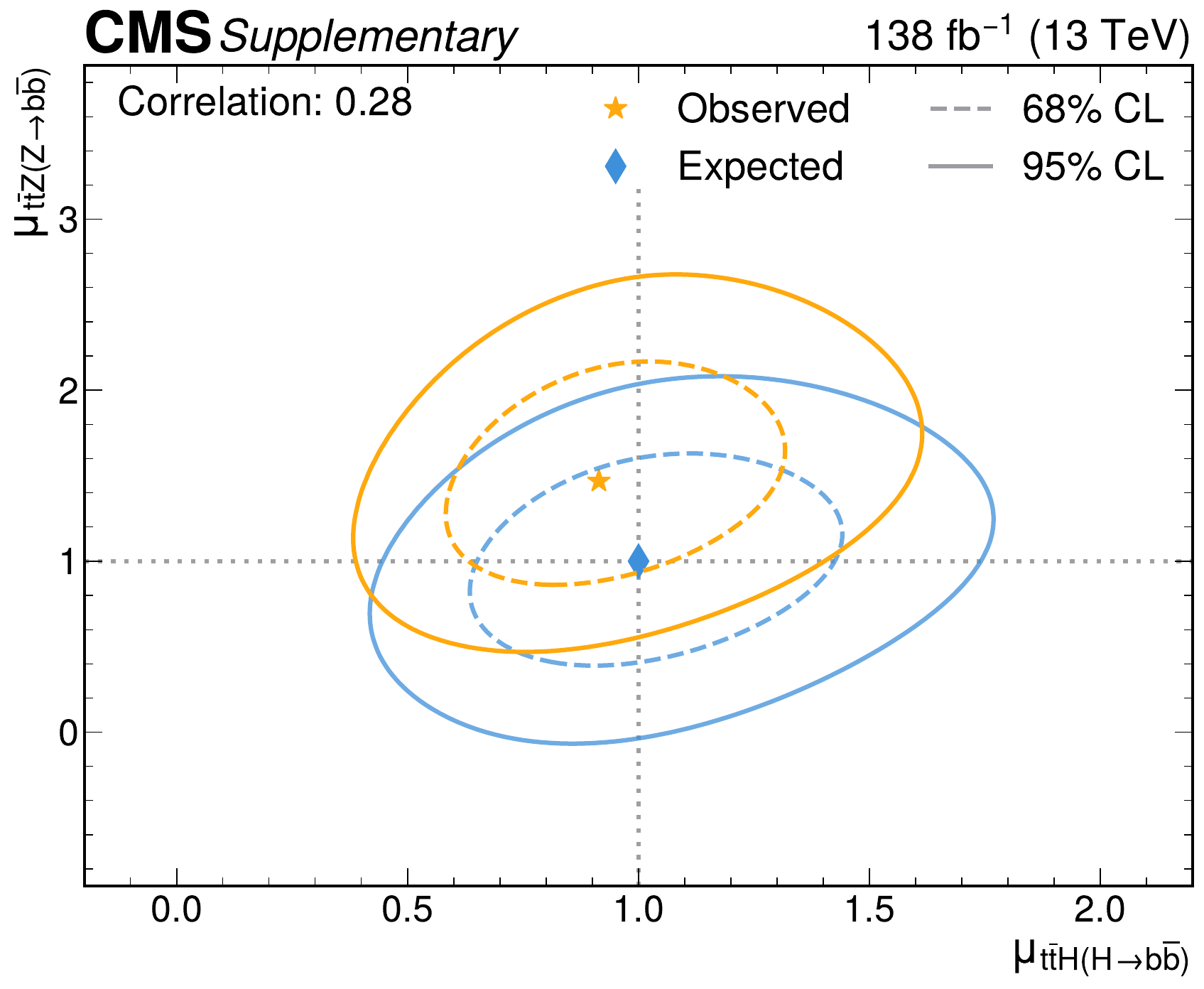}%
\hfill%
\includegraphics[width=0.45\textwidth]{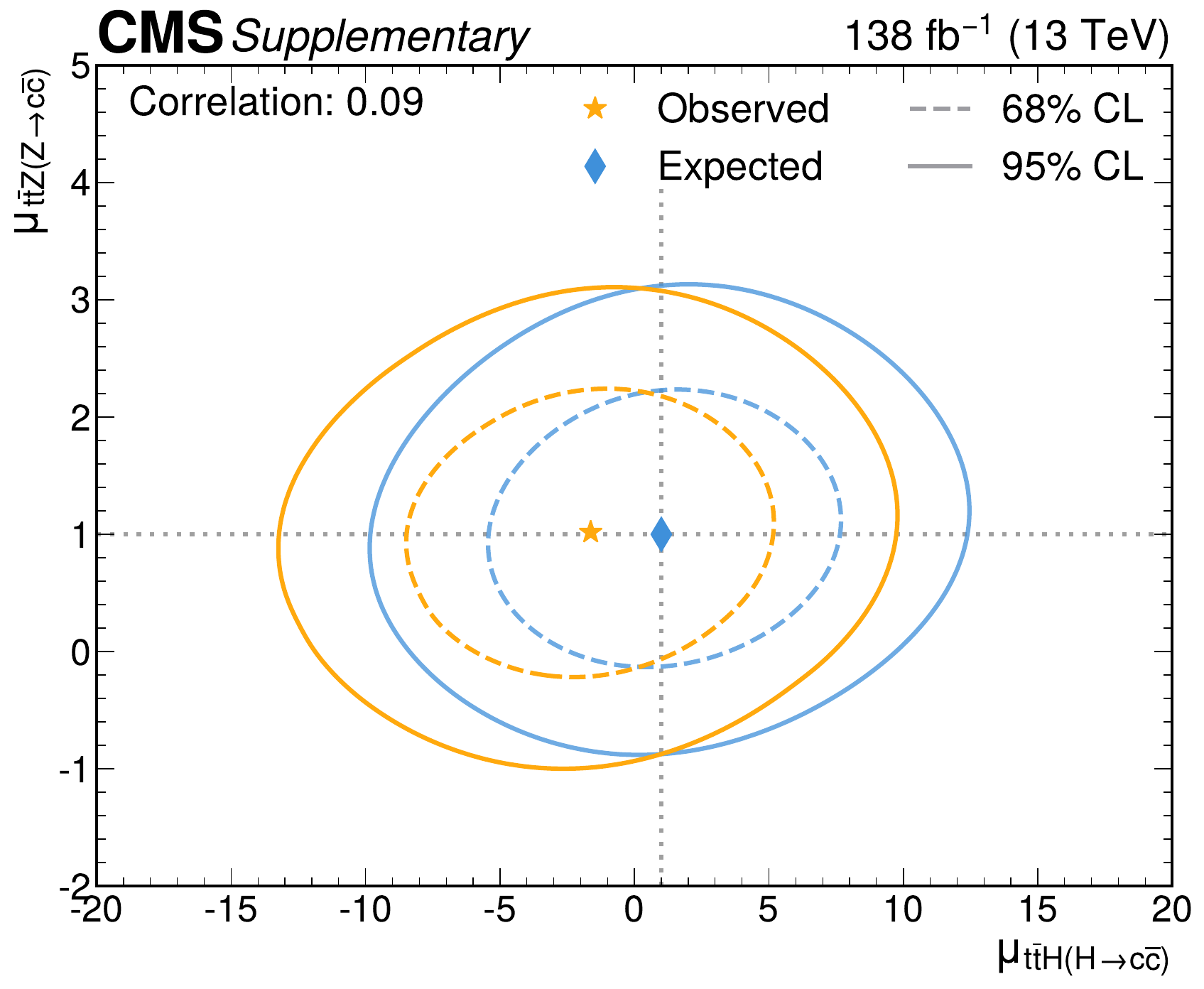}
\caption{%
    Likelihood scans for the simultaneous fit of \muHbb and \muHcc (upper left), \muZbb and \muZcc (upper right), \muHbb and \muZbb (lower left), and \muHcc and \muZcc (lower right).
    The 68\% (95\%) \CL intervals are indicated by the dotted (solid) lines.
    The observed (expected) best fit values are shown by the orange (blue) markers.
}
\label{fig:results:scan_mu2d}
\end{figure*}

\begin{figure*}[!htp]
\centering
\includegraphics[width=0.45\textwidth]{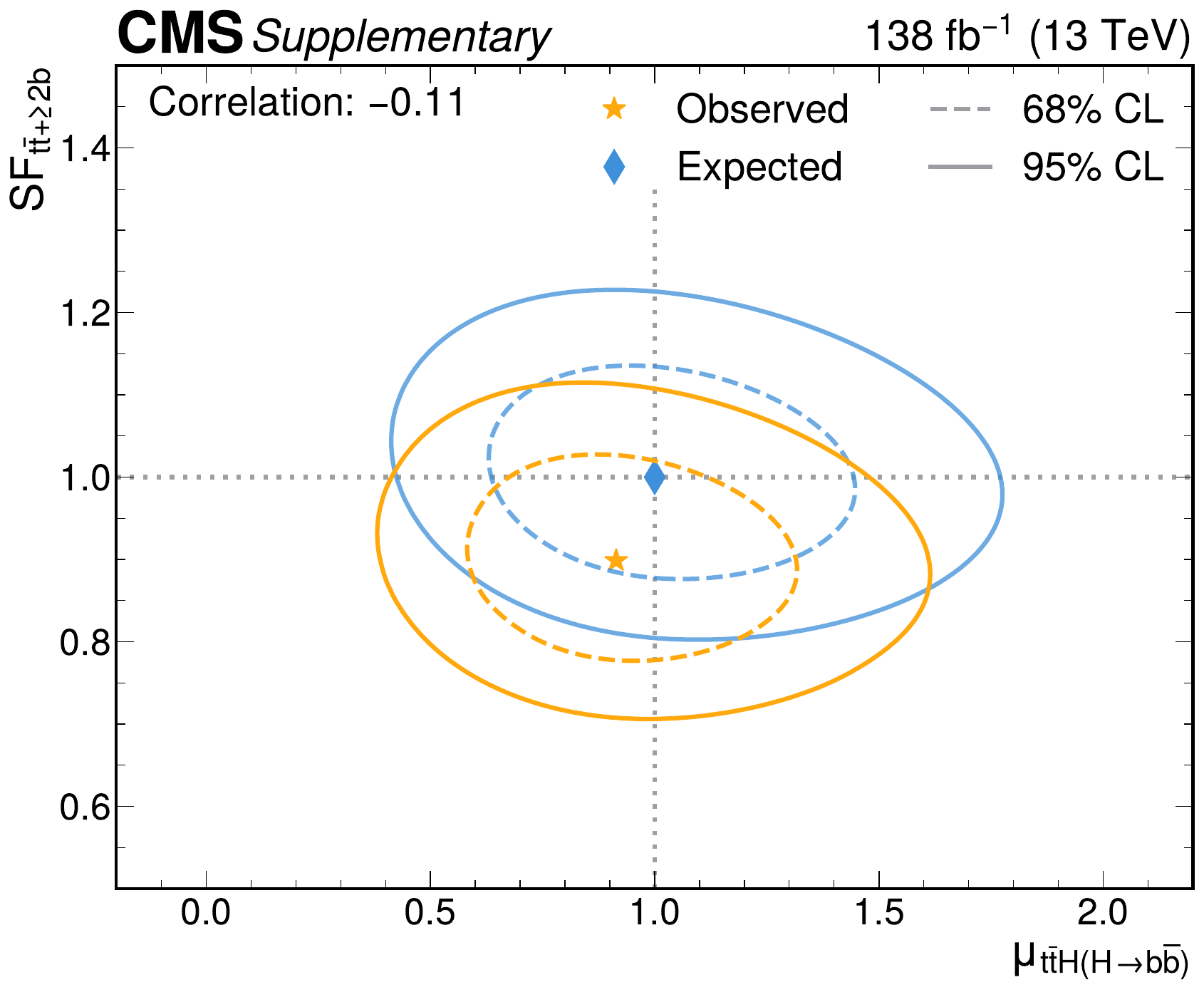}%
\hfill%
\includegraphics[width=0.45\textwidth]{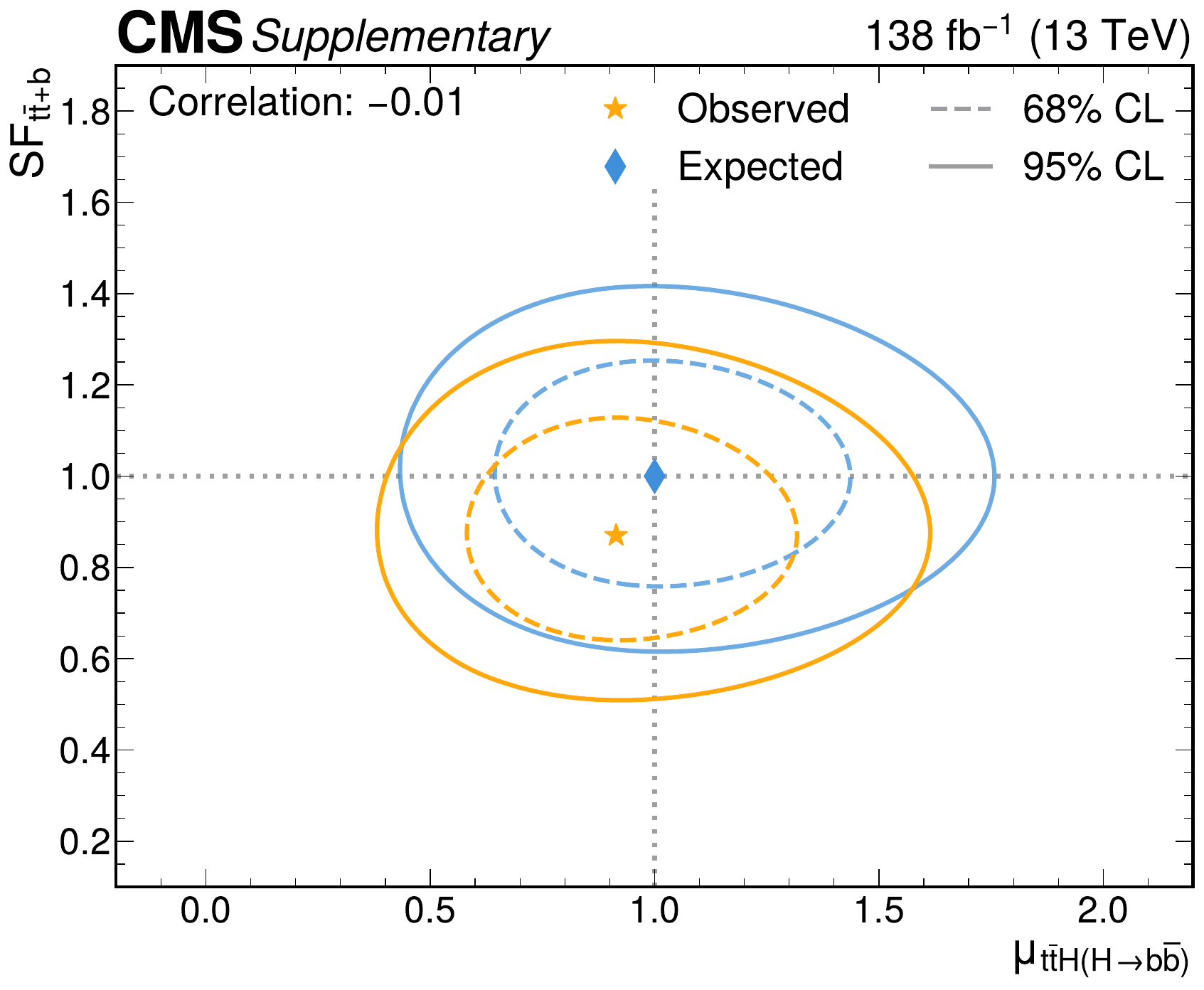} \\
\includegraphics[width=0.45\textwidth]{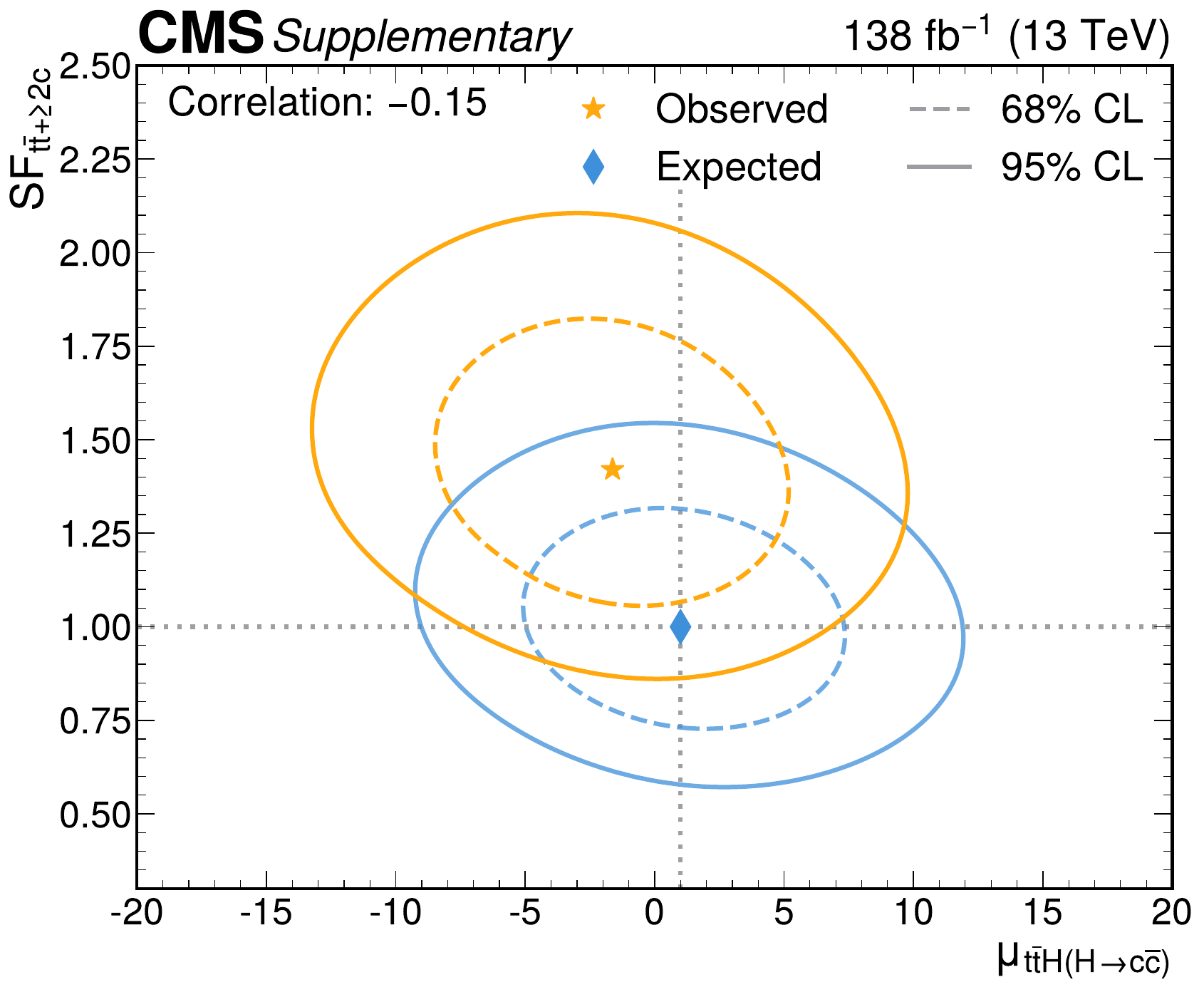}%
\hfill%
\includegraphics[width=0.45\textwidth]{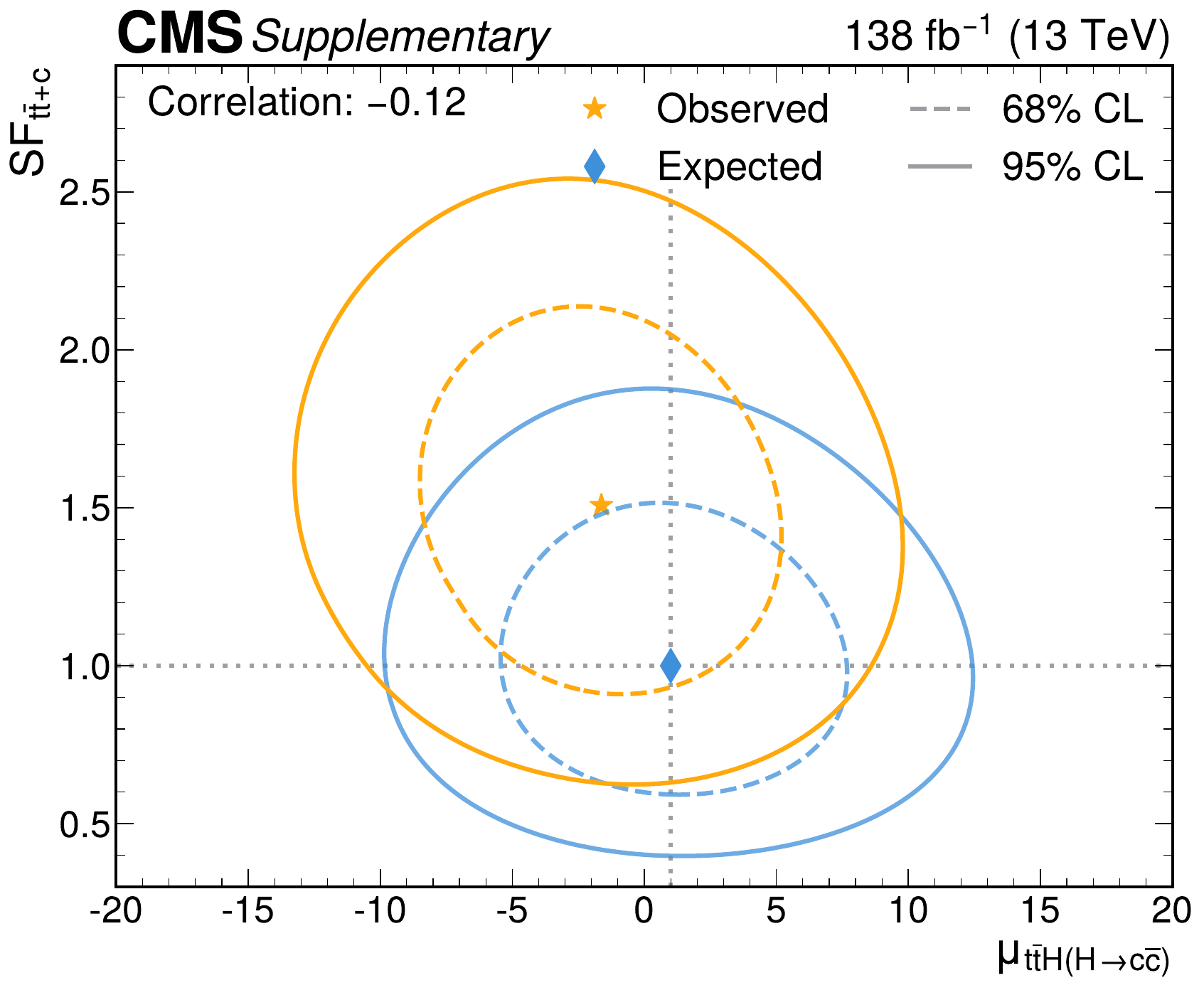}
\caption{%
    Likelihood scans for the simultaneous fit of \muHbb and the \ttbb scale factor (upper left), \muHbb and the \ttbj scale factor (upper right), \muHcc and the \ttcc scale factor (lower left), and \muHcc and the \ttcj scale factor (lower right).
    The 68\% (95\%) \CL intervals are indicated by the dotted (solid) lines.
    The observed (expected) best fit values are shown by the orange (blue) markers.
}
\label{fig:results:scan_mu_vs_sf}
\end{figure*}

\begin{figure*}[!htp]
\centering
\includegraphics[width=0.45\textwidth]{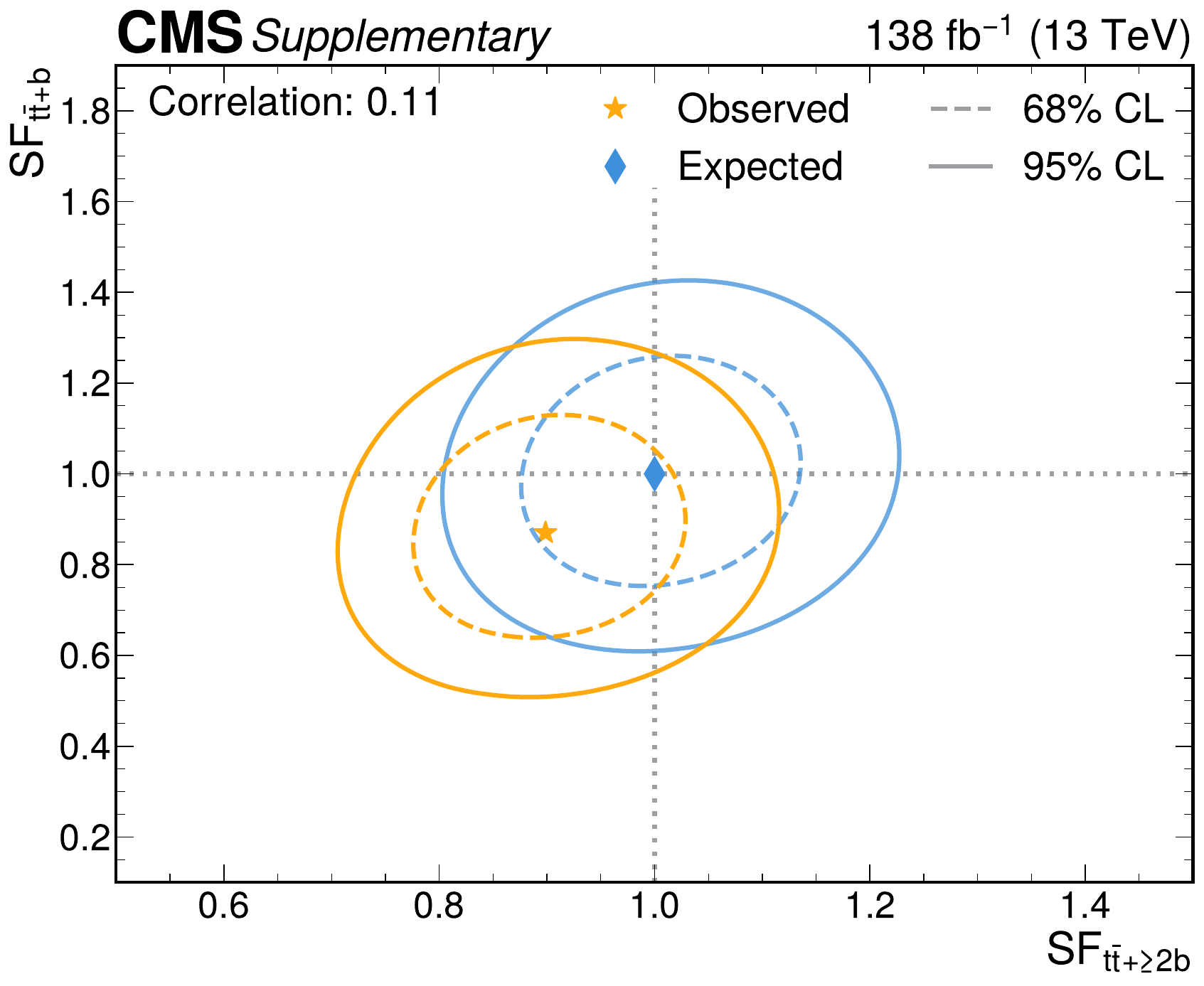}%
\hfill%
\includegraphics[width=0.45\textwidth]{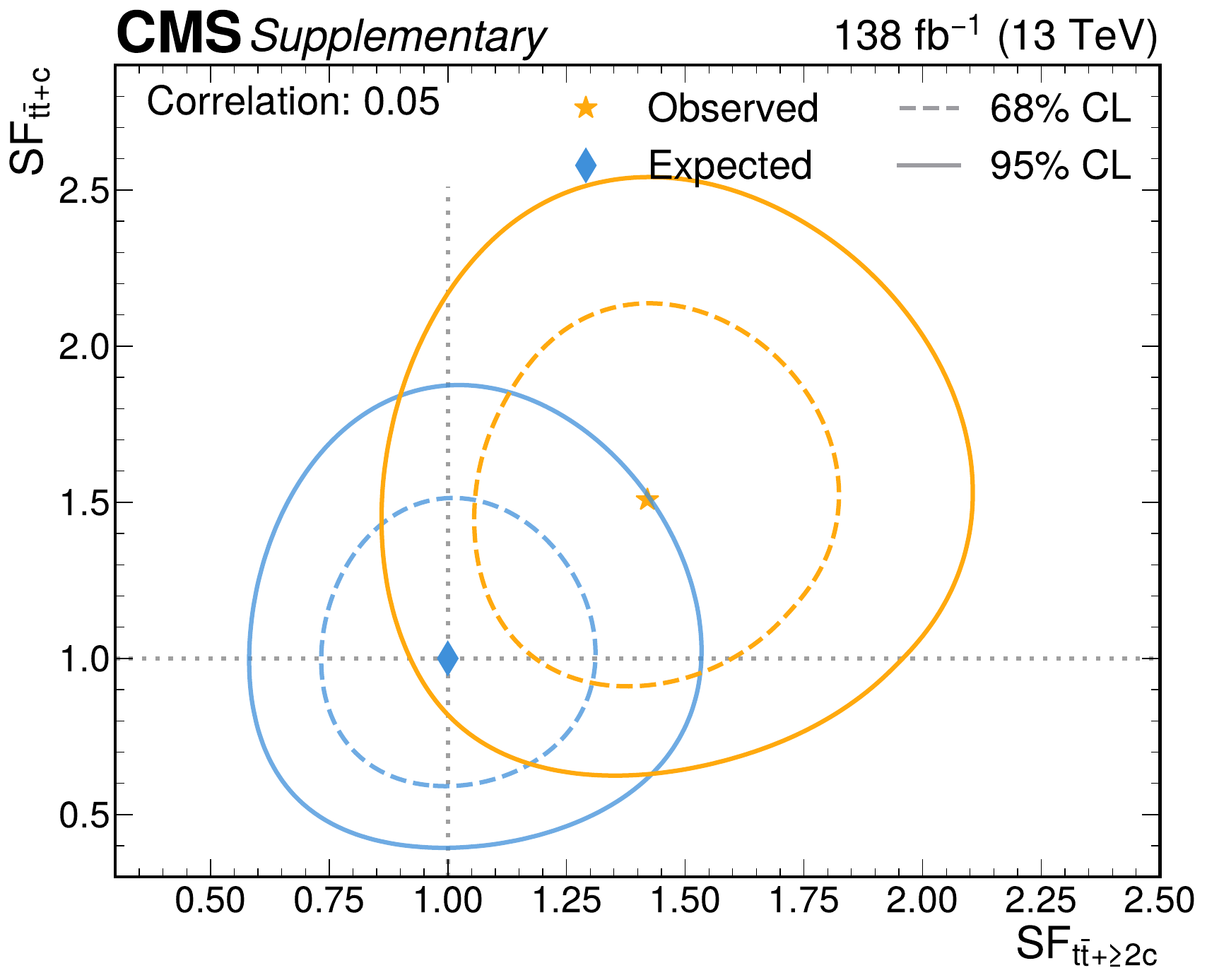}
\caption{%
    Likelihood scans for the simultaneous fit of the \ttbb and \ttbj scale factors (left), and of the \ttcc and \ttcj scale factors (right).
    The 68\% (95\%) \CL intervals are indicated by the dotted (solid) lines.
    The observed (expected) best fit values are shown by the orange (blue) markers.
}
\label{fig:results:scan_sf}
\end{figure*}

\ifthenelse{\boolean{cms@external}}{}{\clearpage}
\subsection{Cross checks with alternative \texorpdfstring{\ttbjets}{tt+b} background models}
\label{supp:ttbb-check}

The robustness of the results is assessed by performing the fit for the signal strength parameters using alternative background models, focusing on variations of the \ttbjets background model:
\begin{enumerate}
\renewcommand*\labelenumi{(\theenumi)}
\item \textbf{Simple PS:} Using a simplified scheme of the PS uncertainties with only one ISR and FSR scale uncertainty instead of the 16 individual components.
\item \textbf{$\muF\times2$:} Changing \muF by a factor two in the \ttbjets simulation and otherwise using the same \ttbjets uncertainty model. This alternative \muF and \muR choice corresponds to the setup used in the recent measurement of \ttHbb production by the ATLAS Collaboration~\cite{ATLAS:2024gth}.
\item \textbf{$\muF\times2$ \& Simple PS:} As described in (2), but using the simplification of (1).
\item \textbf{\fiveFS \ttbjets:} Using the \fiveFS \ttbjets simulation as the nominal background model.
\item \textbf{\fiveFS \ttbjets \& Simple PS:} As described in (4), but using the simplification of (1).
\end{enumerate}
The results for the nominal fit setup as well as for variations are given in Table~\ref{tab:altResultsMu} for the best fit signal strength parameters, the observed upper limits and the significance on the signal strengths, and in Table~\ref{tab:altResultsRateParam} for the \ttbjets background normalization parameters.

\begin{table*}[!!htp]
\centering
\topcaption{%
    Observed signal strengths for \ttHcc, \ttHbb, \ttZcc, and \ttZbb as obtained by the fits using alternative background models.
    In brackets are the observed upper limits for \ttHcc and the significance values for \ttHbb, \ttZcc, and \ttZbb, respectively.
}
\label{tab:altResultsMu}
\renewcommand{\arraystretch}{1.2}
\cmsWideTable{\begin{scotch}{lcc@{\hspace*{2em}}cc@{\hspace*{2em}}cc@{\hspace*{2em}}cc}
    \ttbjets bkg.\ model & \muHcc & (95\% \CL) & \muHbb & (Sign.\ [s.d.])& \muZcc & (Sign.\ [s.d.])& \muZbb & (Sign.\ [s.d.]) \\
    \hline
    Default & $-1.6\pm4.5$ & ($<$7.8) & 0.91\pmasym{+0.26}{-0.22} & (4.4) & 1.02\pmasym{+0.79}{-0.84} & (1.2) & 1.47\pmasym{+0.45}{-0.41} & (3.5) \\[\cmsTabSkip]
    Simple PS & $-0.4\pm4.4$ & ($<$8.8) & 0.91\pmasym{+0.27}{-0.22} & (4.4) & 0.94\pmasym{+0.82}{-0.73} & (1.3) & 1.47\pmasym{+0.44}{-0.39} & (3.6) \\
    $\muF\times2$ & $-$1.7\pmasym{+4.2}{-4.3} & ($<$8.0) & 0.94\pmasym{+0.26}{-0.23} & (4.4) & 1.06\pmasym{+0.77}{-0.75} & (1.4) & 1.62\pmasym{+0.43}{-0.38} & (4.1) \\
    $\muF\times2$ \& Simple PS & $-1.0\pm4.4$ & ($<$8.5) & 0.97\pmasym{+0.28}{-0.23} & (4.5) & 1.04\pmasym{+0.77}{-0.78} & (1.3) & 1.67\pmasym{+0.42}{-0.37} & (4.2) \\
    \fiveFS & $-1.8\pm4.3$ & ($<$7.7) & 0.97\pmasym{+0.24}{-0.20} & (5.7) & 1.17\pmasym{+0.77}{-0.71} & (1.6) & 1.14\pmasym{+0.41}{-0.38} & (3.0) \\
    \fiveFS \& Simple PS & $-0.9\pm4.5$ & ($<$7.6) & 1.02\pmasym{+0.26}{-0.21} & (6.0) & 1.19\pmasym{+0.77}{-0.71} & (1.6) & 1.10\pmasym{+0.40}{-0.37} & (2.9) \\
\end{scotch}}
\end{table*}

\begin{table*}[!htb]
\centering
\topcaption{%
    Observed normalization scale factors for the background components as obtained by the fits using alternative background models.
    The background normalization scale factors are given for the \DL \& \SL (\FH) channels.
}
\label{tab:altResultsRateParam}
\renewcommand{\arraystretch}{1.2}
\begin{scotch}{lcc}
    \ttbjets background model & \ttbb & \ttbj \\
    \hline
    Default & $0.90\pm0.08$ ($1.14\pm0.12$) & $0.87\pm0.15$ ($0.83\pm0.16$) \\[\cmsTabSkip]
    Simple PS & $0.88\pm0.07$ ($1.12\pm0.11$) & $0.92\pm0.16$ ($0.89\pm0.18$) \\
    $\muF\times2$ & $0.89\pm0.06$ ($1.37\pm0.10$) & $0.94\pm0.17$ ($1.05\pm0.21$) \\
    $\muF\times2$ \& Simple PS & $0.90\pm0.05$ ($1.41\pm0.10$) & $0.95\pm0.16$ ($1.06\pm0.22$) \\
    \fiveFS & $1.54\pm0.11$ ($2.06\pm0.19$) & $0.91\pm0.36$ ($0.62\pm0.64$) \\
    \fiveFS \& Simple PS & $1.57\pm0.09$ ($2.25\pm0.21$) & $0.83\pm0.34$ ($0.60\pm0.62$) \\
\end{scotch}
\end{table*}

}
\cleardoublepage \section{The CMS Collaboration \label{app:collab}}\begin{sloppypar}\hyphenpenalty=5000\widowpenalty=500\clubpenalty=5000\cmsinstitute{Yerevan Physics Institute, Yerevan, Armenia}
{\tolerance=6000
A.~Hayrapetyan, V.~Makarenko\cmsorcid{0000-0002-8406-8605}, A.~Tumasyan\cmsAuthorMark{1}\cmsorcid{0009-0000-0684-6742}
\par}
\cmsinstitute{Institut f\"{u}r Hochenergiephysik, Vienna, Austria}
{\tolerance=6000
W.~Adam\cmsorcid{0000-0001-9099-4341}, J.W.~Andrejkovic, L.~Benato\cmsorcid{0000-0001-5135-7489}, T.~Bergauer\cmsorcid{0000-0002-5786-0293}, M.~Dragicevic\cmsorcid{0000-0003-1967-6783}, C.~Giordano, P.S.~Hussain\cmsorcid{0000-0002-4825-5278}, M.~Jeitler\cmsAuthorMark{2}\cmsorcid{0000-0002-5141-9560}, N.~Krammer\cmsorcid{0000-0002-0548-0985}, A.~Li\cmsorcid{0000-0002-4547-116X}, D.~Liko\cmsorcid{0000-0002-3380-473X}, M.~Matthewman, I.~Mikulec\cmsorcid{0000-0003-0385-2746}, J.~Schieck\cmsAuthorMark{2}\cmsorcid{0000-0002-1058-8093}, R.~Sch\"{o}fbeck\cmsAuthorMark{2}\cmsorcid{0000-0002-2332-8784}, D.~Schwarz\cmsorcid{0000-0002-3821-7331}, M.~Shooshtari, M.~Sonawane\cmsorcid{0000-0003-0510-7010}, W.~Waltenberger\cmsorcid{0000-0002-6215-7228}, C.-E.~Wulz\cmsAuthorMark{2}\cmsorcid{0000-0001-9226-5812}
% ...
\par}
\vskip\cmsinstskip
\dag:~Deceased\\
$^{1}$Also at Yerevan State University, Yerevan, Armenia\\
$^{2}$Also at TU Wien, Vienna, Austria\\
% ...
\end{sloppypar}
\end{document}